\PassOptionsToPackage{unicode}{hyperref}
\PassOptionsToPackage{hyphens}{url}
\documentclass[11pt, ]{article}

\usepackage{mathtools}
\usepackage{amsmath}
\usepackage{amsthm}
\usepackage{amssymb}
\mathtoolsset{showonlyrefs}

\usepackage{iftex}
\ifPDFTeX
  \usepackage[OT1,T1]{fontenc}
  \usepackage[utf8]{inputenc}
  \usepackage{textcomp} 
    \usepackage[p,osf,swashQ]{cochineal}
  \usepackage[cochineal,vvarbb]{newtxmath}
      \usepackage[scale=0.95]{biolinum}
    \usepackage[scale=0.95,varl]{inconsolata}
\else 
  \usepackage[scale=0.95,varl]{inconsolata}
  \usepackage{newpxtext}
  \usepackage{mathpazo}
    \usepackage[scale=0.95]{biolinum}
  \fi
\ifLuaTeX
  \usepackage{selnolig}  
\fi
\IfFileExists{microtype.sty}{
  \usepackage[]{microtype}
  \UseMicrotypeSet[protrusion]{basicmath} 
}{}

\allowdisplaybreaks
\usepackage[dvipsnames,svgnames,x11names]{xcolor}
\usepackage[lmargin=1in,rmargin=1in,tmargin=1in,bmargin=1in]{geometry}
\usepackage[format=plain,
  labelfont={bf,sf,small,singlespacing},
  textfont={sf,small,singlespacing},
  justification=justified,
  margin=0.25in]{caption}

\usepackage{sectsty}
\usepackage[compact]{titlesec}
\makeatletter
\newcommand\@shorttitle{}
\newcommand\shorttitle[1]{\renewcommand\@shorttitle{#1}}
\usepackage{fancyhdr}
\fancyheadoffset{0pt}
\makeatother
\renewenvironment{abstract}{
  \centerline
  {\large\sffamily\bfseries Abstract}\vspace{-1em}
  \begin{quote}\small
}{
  \end{quote}
}

\makeatletter
\newcommand{\assumplabel}[2]{%
   \protected@write \@auxout {}{\string\newlabel{#1}{{#2}{\thepage}{#2}{#1}{}}}%
   \hypertarget{#1}{#2}%
}
\makeatother

\providecommand{\tightlist}{%
  \setlength{\itemsep}{0pt}\setlength{\parskip}{0pt}}\usepackage{longtable,booktabs,array}
\usepackage{calc} 
\usepackage{etoolbox}
\makeatletter
\patchcmd\longtable{\par}{\if@noskipsec\mbox{}\fi\par}{}{}
\makeatother
\IfFileExists{footnotehyper.sty}{\usepackage{footnotehyper}}{\usepackage{footnote}}
\makesavenoteenv{longtable}
\usepackage{graphicx}
\makeatletter
\newsavebox\pandoc@box
\newcommand*\pandocbounded[1]{
  \sbox\pandoc@box{#1}%
  \Gscale@div\@tempa{\textheight}{\dimexpr\ht\pandoc@box+\dp\pandoc@box\relax}%
  \Gscale@div\@tempb{\linewidth}{\wd\pandoc@box}%
  \ifdim\@tempb\p@<\@tempa\p@\let\@tempa\@tempb\fi
  \ifdim\@tempa\p@<\p@\scalebox{\@tempa}{\usebox\pandoc@box}%
  \else\usebox{\pandoc@box}%
  \fi%
}
\def\fps@figure{htbp}
\makeatother

\usepackage{placeins}
\usepackage[italicdiff]{physics}

\newcommand{\half}{\frac{1}{2}}

\DeclareMathOperator{\E}{\mathbb{E}}

\DeclareMathOperator{\Expo}{Expo}

\newcommand{\Beta}{\mathrm{Beta}\qty}

\newcommand{\Norm}{\mathcal{N}\qty}

\usepackage{booktabs}
\usepackage{longtable}
\usepackage{array}
\usepackage{multirow}
\usepackage{wrapfig}
\usepackage{float}
\usepackage{colortbl}
\usepackage{pdflscape}
\usepackage{tabu}
\usepackage{threeparttable}
\usepackage{threeparttablex}
\usepackage[normalem]{ulem}
\usepackage{makecell}
\usepackage{xcolor}
\makeatletter
\@ifpackageloaded{caption}{}{\usepackage{caption}}
\AtBeginDocument{%
\ifdefined\contentsname
  \renewcommand*\contentsname{Table of contents}
\else
  \newcommand\contentsname{Table of contents}
\fi
\ifdefined\listfigurename
  \renewcommand*\listfigurename{List of Figures}
\else
  \newcommand\listfigurename{List of Figures}
\fi
\ifdefined\listtablename
  \renewcommand*\listtablename{List of Tables}
\else
  \newcommand\listtablename{List of Tables}
\fi
\ifdefined\figurename
  \renewcommand*\figurename{Figure}
\else
  \newcommand\figurename{Figure}
\fi
\ifdefined\tablename
  \renewcommand*\tablename{Table}
\else
  \newcommand\tablename{Table}
\fi
}
\@ifpackageloaded{float}{}{\usepackage{float}}
\floatstyle{ruled}
\@ifundefined{c@chapter}{\newfloat{codelisting}{h}{lop}}{\newfloat{codelisting}{h}{lop}[chapter]}
\floatname{codelisting}{Listing}

\makeatother
\makeatletter
\@ifpackageloaded{caption}{}{\usepackage{caption}}
\@ifpackageloaded{subcaption}{}{\usepackage{subcaption}}
\makeatother

\usepackage[]{natbib}
\newenvironment{CSLReferences}[2]{
\bibliography{references.bib}
\clearpage
}{}

\usepackage{hyperref}
\usepackage{url}
\hypersetup{
  pdftitle={Redistricting Reforms Reduce Gerrymandering},
  pdfauthor={Cory McCartan; Christopher T. Kenny; Tyler Simko; Emma Ebowe; Michael Y. Zhao; Kosuke Imai},
  pdfkeywords={redistricting, differences-in-differences, formal
modeling, continuous treatment},
  colorlinks=true,
  linkcolor={black},
  filecolor={Maroon},
  citecolor={VioletRed4},
  urlcolor={DodgerBlue4},
  pdfcreator={LaTeX via pandoc}}

\title{\sffamily\bfseries\huge\parfillskip=0pt
\rightskip=0pt plus .5\textwidth
\leftskip=0pt plus .5\textwidth
\emergencystretch=.3\textwidth Redistricting Reforms Reduce
Gerrymandering by Constraining Partisan Actors}
\shorttitle{Redistricting Reforms Reduce Gerrymandering}
\author{\textbf{Cory McCartan}\footnote{
To whom correspondence should be addressed.
Email: \texttt{\href{mailto:mccartan@psu.edu}{mccartan@psu.edu}}.
Website: \url{https://corymccartan.com/}.
Address:
461 Pollock Road, University Park, PA 16801.
We acknowledge helpful comments from Christian Fong, Shiro Kuriwaki,
Jacob Montgomery, Daniel Thompson, Seth McKee, and an anonymous reviewer
from the Alexander and Diviya Magaro Peer Pre-Review Program at
Harvard's Institute for Quantitative Social Science.}
\\Department of Statistics\\Pennsylvania State University
\vspace{0.05in}
 \and \textbf{Christopher T. Kenny}
\\Data-Driven Social Sciences\\Princeton University
\vspace{0.05in}
 \and \textbf{Tyler Simko}
\\Department of Political Science\\University of Michigan
\vspace{0.05in}
 \and \textbf{Emma Ebowe}
\\Department of Government\\College of William and Mary
\vspace{0.05in}
 \and \textbf{Michael Y. Zhao}
\\Harvard College
\vspace{0.05in}
 \and \textbf{Kosuke Imai}
\\Department of Government\\
Department of Statistics\\Harvard University
\vspace{0.05in}
 }
\date{August 10, 2025}

\begin{document}
\allsectionsfont{\sffamily}

\maketitle

\begin{abstract}
Political actors often manipulate redistricting plans to gain electoral
advantages, a process known as gerrymandering. Several states have
implemented institutional reforms to address this problem, such as
establishing map-drawing commissions. Estimating the impact of such
reforms is challenging because each state structures its processes and
rules differently. We model redistricting as a sequential game whose
equilibrium solution summarizes multi-step institutional interactions as
a univariate score. We argue this score measures the leeway political
actors have over the partisan lean of the final plan. Using a
differences-in-differences design, we demonstrate that reforms reduce
partisan bias and increase competitiveness when they constrain partisan
actors. We perform a counterfactual policy analysis to estimate the
effects of enacting recent reforms nationwide. Though commissions
generally reduce bias, reforms that restrict partisan actors in multiple
ways, like removing veto points (Michigan), are more effective than
commissions where parties retain some control (Ohio).
\end{abstract}

\textbf{\textit{Keywords}}\quad redistricting~\textbullet~differences-in-differences~\textbullet~formal
modeling~\textbullet~continuous treatment


\section{Introduction}\label{introduction}

Democratic institutions play a crucial role in preventing political
actors from pursuing policies that prioritize their own interests over
the broader public good \citep[Federalist papers 53 and
54:][]{federalist54, federalist53}. Further, increasing political
polarization and the erosion of democratic norms in many countries have
encouraged more recent attention to placing additional institutional
constraints on policy makers
\citep{levitsky2023tyranny, little2023measuring, mccarty2019polarization}.
To safeguard institutions, advocates have suggested reforms that
insulate democratic processes from partisan control.\footnote{For
  example, see \href{https://www.peoplenotpoliticiansoregon.com/}{People
  Not Politicians Oregon} or Ohio's
  \href{https://citizensnotpoliticians.org/}{Citizens Not Politicians}.}
For example, the Electoral Count Reform and Presidential Transition
Improvement Act of 2022 represents an effort to make it more difficult
for partisan actors to manipulate the presidential electoral
certification process in the United States.

We study reform in American congressional redistricting, a political
process often exploited by partisan actors to enact districting plans
that favor their own party. This manipulation, known as \emph{partisan
gerrymandering}, has been widespread in the past two redistricting
cycles \citep{50states2010, kenny2023widespread, warshaw_2022}.
Redistricting plans that disproportionately favor a certain party can
limit how responsive a party's share of seats in the legislature is to
changes in its vote share and can reduce the electoral power of racial
minorities
\citep[e.g.][]{canon2022race, canon1999representation, grofman1991identifying, polsby1991third}.

\subsection{Methodological challenges and proposed
approach}\label{methodological-challenges-and-proposed-approach}

Reform efforts to limit gerrymandering are often designed to constrain
partisan map drawers. They include the establishment of independent
map-drawing commissions and the introduction of court oversight over
proposed plans \citep{cain2012}. Estimating the causal impact of these
institutional reforms, however, is challenging for three reasons. The
first is the problem of \emph{treatment complexity}. Redistricting
reform efforts must intervene in a complex and multidimensional process.
The specific rules governing each state's redistricting process differ
in many ways, including who proposes initial maps and whether or not
courts can intervene.

Second, states that adopt redistricting reforms may differ in both
observable and unobservable ways from states that do not. These
differences include statewide variations in institutional
characteristics and political contexts. This \emph{confounding bias}
problem is common to any observational study and must be addressed to
estimate credible causal effects.

Finally, the outcome of the institutional process---a redistricting
plan---is also complex. For example, partisan features may be confounded
by other factors such as a state's geography and demographics
\citep[e.g.,][]{cottrell2019using}. If a large number of Democratic
voters live in cities, any redistricting plan with compact districts may
end up creating a small number of heavily Democratic-leaning districts
rather than efficiently allocating Democratic votes across more
districts. We must overcome this \emph{outcome complexity} to accurately
measure the partisan bias of each enacted plan.

\begin{figure}[t]

\centering{

\pandocbounded{\includegraphics[keepaspectratio]{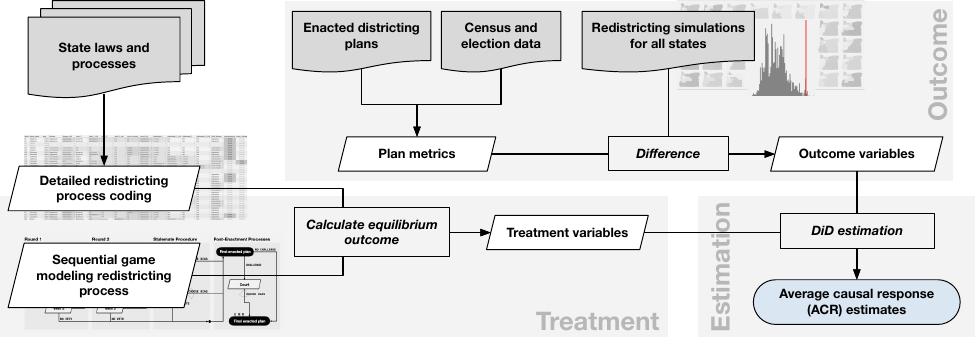}}

}

\caption{\label{fig-analysis-proc}Schematic summary of our methodology.
Our approach is designed to address three key methodological challenges
in the study of redistricting reform. First, we address \emph{treatment
complexity} by modeling the redistricting process as a zero-sum
sequential game to estimate theoretically informed parameters that serve
as our treatment. Second, we address \emph{outcome complexity} by
generating representative distributions of simulated redistricting plans
for each state, which adjust for state-specific changes in political
geography. Finally, we address \emph{confounding bias} in causal effect
estimation with a difference-in-differences design that uses simulated
alternative to strengthen the credibility of the parallel trends
assumption.}

\end{figure}%

In this paper, we propose a new methodological approach that addresses
the above challenges to study the causal impact of redistricting reform.
Figure~\ref{fig-analysis-proc} summarizes our methodology. To deal with
the \emph{treatment complexity}, we first collect and standardize
information about each state's relevant laws, starting from the initial
map drawer (i.e., the redistricting commission or legislature) and
following through varying stalemate processes and opportunities for
court intervention. We then reduce the differing districting procedures
across states to a single theoretically informed parameter by modeling
redistricting reform as a zero-sum sequential game. We use our original
dataset of institutional procedures to characterize the players and
available moves in the game. Analyzing the game allows us to measure the
ability of partisan players to maximize the partisan lean of a
redistricting plan. The Nash equilibrium of the game is a measure of the
``leeway'' that a single party has over the final redistricting plan.

We use two variations of each state's game to produce two measures of
leeway. The first is the \emph{realized} leeway, which uses the observed
parties of the players to compute the equilibrium. The second is the
\emph{maximum} leeway, which instead computes the equilibrium under
one-party control. The latter allows us to measure the strength of
institutions separately from party control, which varies across states
and over time. We emphasize that both of these leeway variables are
summary measures of our high-dimensional treatment variables of interest
(i.e., institutional features) and are not functions of the outcome
variable of interest. Therefore, our methodology places an observational
study of institutional changes under the design-based approach of causal
inference \citep{rubin2008}.

Specifically, we address the \emph{confounding bias} problem by using
our leeway measures as a continuous treatment variable in a
differences-in-differences (DiD) design applied to the 2010 and 2020
redistricting cycles. For example, states with Democratic control tend
to enact more liberal policies \citep{caughey2017incremental}. This
approach addresses potential confounding by comparing changes in states
that have enacted reforms to those in similar states that have not,
under a parallel trends assumption \citep{callaway2021cdid}. We use this
model to estimate how changes in the map drawer's leeway influence
resulting plans across a set of both partisan and nonpartisan outcomes,
such as the responsiveness to swings in partisan preferences and the
number of expected seats per party.

Our approach uses a formal model to summarize a multi-dimensional
treatment variable, and then applies a causal inference methodology to
adjust for bias due to unobserved confounding. However, we do not adopt
a structural modeling approach, which would require modeling a state's
decision to adopt particular reforms for their redistricting process.
The reason is that, while we are able to measure all the institutional
features of redistricting processes, it is impossible to observe all the
factors which affect a state's decision to adopt redistricting reforms.
Instead, we use a DiD design to address the issue of unobserved
confounding.

Lastly, we deal with the issue of \emph{outcome complexity} by
separating the causal effect of state-specific institutional reforms
from changes in political geography over time. To do this, we generate a
sample of alternative redistricting plans via a simulation algorithm
\citep{smc} following each state's rules and political geography, but
without regard to any partisan information
\citep{50stateSimulations, 50states2010}. We use these simulated plans
as a nonpartisan baseline for both 2010 and 2020 redistricting cycles,
and compute the difference in outcome variables between the enacted and
simulated plans. This adjusts for state-specific changes in political
geography, making the parallel trends assumption more plausible.

\subsection{Summary of findings}\label{summary-of-findings}

We find that more restrictive redistricting processes reduce partisan
bias by constraining map drawers. For example, changing from a
single-party legislature to an independent commission leads to a
reduction of about 0.5 excess seats. These effects are substantively
large, as congressional gerrymandering generally results in gains of
less than one or two seats per state \citep{kenny2023widespread}.
Similarly, we find smaller but positive effects of constraining reforms
on electoral responsiveness. We estimate that a similar change from
legislature to commission would increase the share of competitive seats
within a state from 25\% to 38\% on average.

A key advantage of our methodological approach is the ability to perform
a counterfactual analysis of institutional reforms. We investigate how
enacting recent procedural reforms nationwide could reduce widespread
partisan gerrymandering. We quantify how the partisan bias and
responsiveness of adopted plans could counterfactually change if all
states nationwide adopted three kinds of commission structures currently
enacted in several states: (1) an Ohio-style approach that requires
supermajorities and uses a bipartisan backup commission, (2) a New
York-style commission with a nonpartisan map drawer but several partisan
veto points later in the process, and (3) a Michigan-style commission
with a nonpartisan commission drawer, no partisan veto points, and the
potential for court review.

We find that commissions can generally reduce existing pro-Republican
bias, but the details of commission structure matter. In particular,
unlike reforms adopted in Ohio and New York, a Michigan-style
nonpartisan commission has no partisan veto points. We find that
implementing Michigan-style reforms nationwide leads to an additional
6.2 Democratic seats, on average. All three styles of reforms increase
electoral responsiveness.

\subsection{Contributions to the
literature}\label{contributions-to-the-literature}

We make two primary substantive contributions to the literature on
redistricting reform. First, we help address a debate in the literature
about whether, and how, reforms impact redistricting plans. Existing
work on the effectiveness of redistricting reforms presents conflicting
arguments and findings, with \citet{nelson2023} concluding that ``the
efficacy of redistricting reforms is contested in political science''
(p.~207). For example, some studies argue that commissions produce
fairer plans, largely by removing self-serving electoral incentives from
legislative map drawers
\citep{carson2004effect, carson2014reevaluating, edwards2017institutional, lindgren2013effect, litton2012road, mcdonald2004comparative, nelson2023, keena2021gerrymandering}.
In contrast, others reach much less optimistic conclusions about the
efficacy of redistricting commissions and find limited or no impact of
reforms on outcomes like competitiveness and partisan bias
\citep{cottrill2012effects, henderson2018gerrymandering, kousser2018reform, miller2013redistricting, seabrook2017drawing}.
We contribute to this debate by formalizing and presenting an approach
that focuses on how reforms impact \emph{leeway}---the relative control
political actors have over the resulting redistricting plan.

Some work has focused on the long-term decrease of partisan bias in
redistricting. \citet{caughey2022dynamic} highlights how the size of the
partisan bias has largely decreased since the 1940s, especially after
the reapportionment revolution, though the trend has flattened in recent
cycles. We focus on this most recent period. While we do not explain the
long-term decrease, we offer policy evaluations on how restrictive
reforms could help restart the decrease.

Second, we contribute to the literature on redistricting reform by
presenting the most comprehensive empirical evidence yet available about
how the full process of redistricting shapes political outcomes. Most
existing studies examine reform structures by comparing outcomes from
single aspects (e.g., who draws initial maps or whether courts can
intervene) of much more complex redistricting process
\citep{carson2004effect, carson2014reevaluating, edwards2017institutional, nelson2023}.
We argue that redistricting is best examined by analyzing its entire
process. Crucially, our approach allows us to demonstrate that various
reforms implemented together can be more effective than a single reform
alone.

This process-based approach requires several methodological innovations
for studying redistricting reform. We believe that the proposed new
methodological approach can also be applied to the study of other
institutional systems. Reform efforts in redistricting have produced
diverse institutional changes across states. While most scholars have
classified reforms into different categories
\citep{cain2012, edwards2017institutional, warshaw_2022, nelson2023},
such an approach may miss important nuances and potential interactions
between features of these institutional changes.

In contrast, we use formal modeling to place these complex institutional
characteristics on a continuous univariate scale and summarize how they
constrain partisan actors. This theoretically driven approach, which we
demonstrate accurately predicts empirical patterns, makes it possible to
apply a difference-in-differences strategy to our complex setting for
credible causal inference.

We also advance the literature on redistricting reform by addressing the
aforementioned three methodological challenges in our unified approach.
First, most studies of redistricting electoral reforms have ignored
\emph{treatment complexity} by focusing on only one or two aspects of
the reform process at a time. For example; some researchers use an
indicator for the existence of a redistricting commission
\citep[e.g.,][]{carson2004effect, carson2014reevaluating}; while others
account for different types of commissions
\citep[e.g.,][]{nelson2023, edwards2017institutional}. Similarly, much
of the existing causal research on redistricting has focused on a single
aspect of the process, such as being directly impacted by the
\emph{Shelby} decision \citep{komisarchik2021throwing} or being placed
in a packed district \citep{fraga2022partisan}. In contrast, we model
the entire process of redistricting using a unified formal theoretic
framework. This approach enables us to examine how various institutional
features affect redistricting outcomes rather than studying each feature
in isolation.

Second; much of prior work has been descriptive in nature and primarily
relied upon cross-sectional comparisons
\citep[e.g.,][]{carson2004effect, carson2014reevaluating, edwards2017institutional, nelson2023, keena2021gerrymandering, best2021, warshaw_2022}.
This can lead to \emph{confounding bias} if states that adopted
redistricting reforms politically differ from those that did not.
Notably, many medium-to-large Democratic states have adopted these
reforms. We address confounding bias by examining the over-time changes
in institutional rules and employing a differences-in-differences
design.

Finally, we address \emph{outcome complexity} through the use of
redistricting simulations that account for factors such as political
geography. Specifically, these simulated alternative districts allow us
to differentiate the impact of redistricting reforms from that of
changes in political geography over time. The period between
redistricting cycles in the United States is a full decade, allowing
political geography to undergo meaningful changes. Most studies, with
some exceptions \citep[e.g.,][]{best2021, warshaw_2022}, do not account
for changes in underlying political geography when studying
redistricting reform. Even studies that address the outcome complexity
do not address the above treatment complexity issue.

Beyond redistricting, our methodological approach can be seen as a
general strategy for the causal analysis of complex institutional
reforms. By combining a game-theoretic treatment model with causal
inference methods, we are able to leverage the strengths of these two
approaches. A game-theoretic approach is a substantively effective way
to model formal institutions with specific rules, making it possible to
map multidimensional policies to a univariate summary in a
theoretically-informed way. Once this summarization of various
institutional reforms is done, we can apply standard causal inference
methods to estimate the effects of counterfactual policies by mapping
them directly to this univariate treatment variable.

Our work relates to a broader methodological literature in the social
sciences. In economics, for example, \citet{chetty2009sufficient}
advocates the combined use of structural and reduced-form approaches via
a sufficient statistic in a manner similar to our approach. In
sociology, \citet{lundberg2021estimand} emphasize the importance of
theoretically informed quantities of interest in causal analysis.
Finally, in political science, \citet{canen2023quantifying} call for the
integration of rigorous theoretical and empirical approaches in causal
research. We demonstrate how such an analysis can be done in the
estimation of causal effects of redistricting reforms.

\section{Overview of Redistricting Processes and
Reforms}\label{overview-of-redistricting-processes-and-reforms}

Before introducing our methodological framework, we provide a brief
overview of redistricting processes and reforms. Every decade following
the U.S. Census, states and localities use many different procedures to
redraw their legislative district boundaries. For example, states differ
on whether legislatures or independent redistricting commissions propose
initial plans for new Congressional district maps. If these actors fail
to produce a plan, state laws further vary on how these stalemates are
handled. Some states pass responsibilities to a court (e.g., Virginia),
a backup commission (e.g., Ohio), or a group of state party leaders
(e.g., Iowa). Even plans that pass the proposal step can face a veto
from other actors, such as the governor and state legislature.

\begin{figure}[t]

\centering{

\includegraphics[width=1\linewidth,height=\textheight,keepaspectratio]{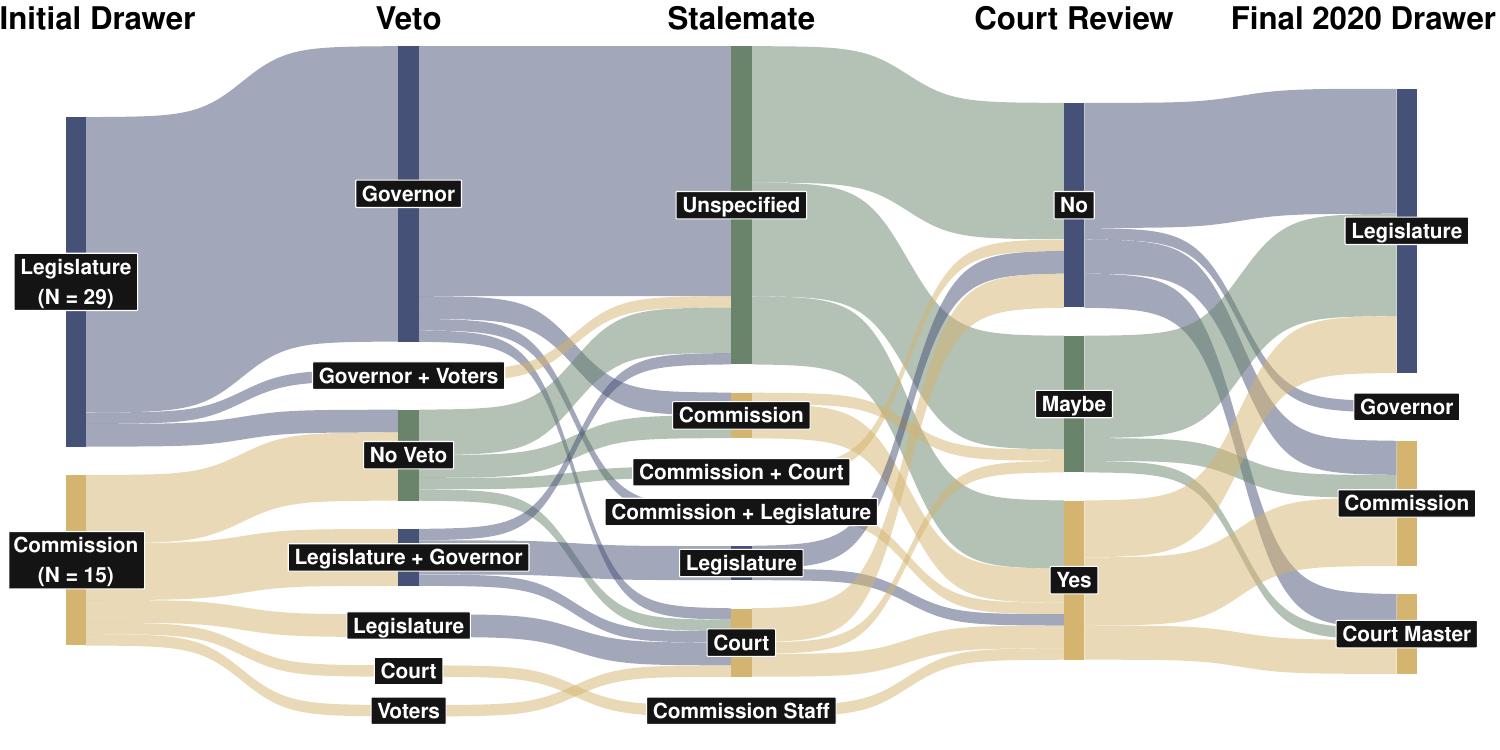}

}

\caption{\label{fig-process}Redistricting procedures for all 44 states
with more than one district in 2020. Each vertical column indicates a
separate step in the redistricting process, and nodes indicate different
procedures that each state can adopt at that step. The width of each
area connecting the nodes is proportional to the number of states with
that specific combination of procedure at both ends. Yellow nodes
indicate actors or institutions that are not explicitly partisan, while
blue nodes indicate explicitly partisan actors or choices. Green nodes
indicate cases where the procedure is not known or does not exist. Here,
we collapse multiple potential stalemate and veto procedures into one
step for visual clarity (e.g., Governor + Voters indicates the
possibility of a first veto by a governor, and a second by the voters).}

\end{figure}%

As explained in Section~\ref{sec-data-inst} below, we collect
information about the redistricting process used in each state for the
2010 and 2020 cycles. Figure~\ref{fig-process} summarizes this dataset
and illustrates the diversity of 2020 redistricting procedures across
states. For example, 29 states drew initial plans in their legislature,
while 15 states used an independent redistricting commission instead.
There is significant variation in the procedures following this initial
draw, with seven distinct veto mechanisms across all the states. Much of
this variation originates from state-specific policy campaigns that have
adopted widely different goals driven by local actors
\citep{keena2021gerrymandering}. For example, California created a
redistricting commission that has the power of drawing congressional
district boundaries through a series of ballot propositions in the early
2000s. In recent years, several states have also implemented
redistricting reforms in response to organized efforts to limit partisan
influence. They include Michigan whose voters approved a constitutional
amendment to establish an independent redistricting commission in 2018.

These procedural differences make it difficult to attribute causal
effects to particular institutional designs or reforms. Most existing
work has turned to classification schemes that simplify this variation
by assigning the control of map redrawing to a single drawer, typically
the creator of the initial or final plan
\citep{edwards2017institutional, carson2014reevaluating}. A more complex
alternative would be to create indices that count the number of times a
particular action appears in the state procedure (e.g., the number of
veto points). For example, as shown in Figure~\ref{fig-process}, even
states that have a commission propose their initial plan vary
drastically in how the process continues afterwards.

While these simplifications make the study of complex procedures
tractable, they risk oversimplifying complex procedures in two ways.
First, attributing institutional outcomes to a single actor ignores the
fact that the final redistricting plan is produced through a series of
steps. Take, for example, the New York redistricting process in 2020. An
independent bipartisan commission had the power to propose an initial
plan, but failed to agree on a single plan. State law required the
stalemated process to move to the legislature, which adopted a plan that
Republican and civil rights groups criticized as a Democratic
gerrymander. These groups challenged the adopted plan in a series of
lawsuits, and in 2022 the New York Court of Appeals struck down the plan
and tasked a court-appointed special master with drawing a remedial
plan. Though the final plan scores well on quantitative fairness metrics
\citep{kenny2023widespread}, simply classifying a ``court'' as the sole
map drawer for New York in 2020 ignores the partisan interests involved,
failing to capture the complexity of the procedures that led to the
final plan.

Second, classification schemes can overlook strategic interactions,
where the behavior of certain actors depends on the presence or
characteristics of others. For example, a commission might draw a
different map if it knew that the map could later be reviewed by a
court. Or, the potential of a governor's veto may limit the likelihood
of a partisan gerrymander by the state legislature, but not in cases
where the governor and legislative majority share a partisan
affiliation. More detailed coding of procedural schemes can account for
some of these interaction effects and increase realism, but will
necessarily decrease statistical power, making it more difficult to
estimate causal effects.

Thus, we face a methodological dilemma: while some simplification of
institutional features is necessary, common approaches to doing so
obscure critical characteristics of redistricting processes. In the next
section, we propose a theoretically grounded approach that models
redistricting as a zero-sum sequential game and uses its Nash
equilibrium as a treatment variable to summarize these complex
institutions.

\section{A Theoretical Model of Institutional
Leeway}\label{a-theoretical-model-of-institutional-leeway}

To study the impact of redistricting reforms and processes, we must
first address the problem of \emph{treatment complexity}. We develop a
standardized set of 14 institutional features that capture the most
important actors in congressional redistricting processes across states.
We collect data on these features so that it is possible to compare
redistricting processes consistently across states, despite the
differences in the laws and bodies governing redistricting in each
state. This step corresponds to the left top corner of
Figure~\ref{fig-analysis-proc}.

Using this data, we develop a sequential redistricting game to summarize
these features in a theoretically informed manner (see the ``treatment''
box of Figure~\ref{fig-analysis-proc}). Specifically, we use the Nash
equilibrium of this game as a one-dimensional measure of the
institutional leeway political actors have over the partisan lean of the
resulting redistricting plan. Finally, we empirically validate the
proposed measure by demonstrating that it is not particularly sensitive
to model specification and predicts redistricting outcomes well.

We emphasize that the proposed measure is a theoretically motivated
summary of our high-dimensional treatment variable of interest. Since
this measure does not use outcome variables, we are able to apply
standard causal inference methodology to adjust for observed and
unobserved confounding factors.

\subsection{Data on the relevant institutional
features}\label{sec-data-inst}

\begin{figure}

\centering{

\pandocbounded{\includegraphics[keepaspectratio]{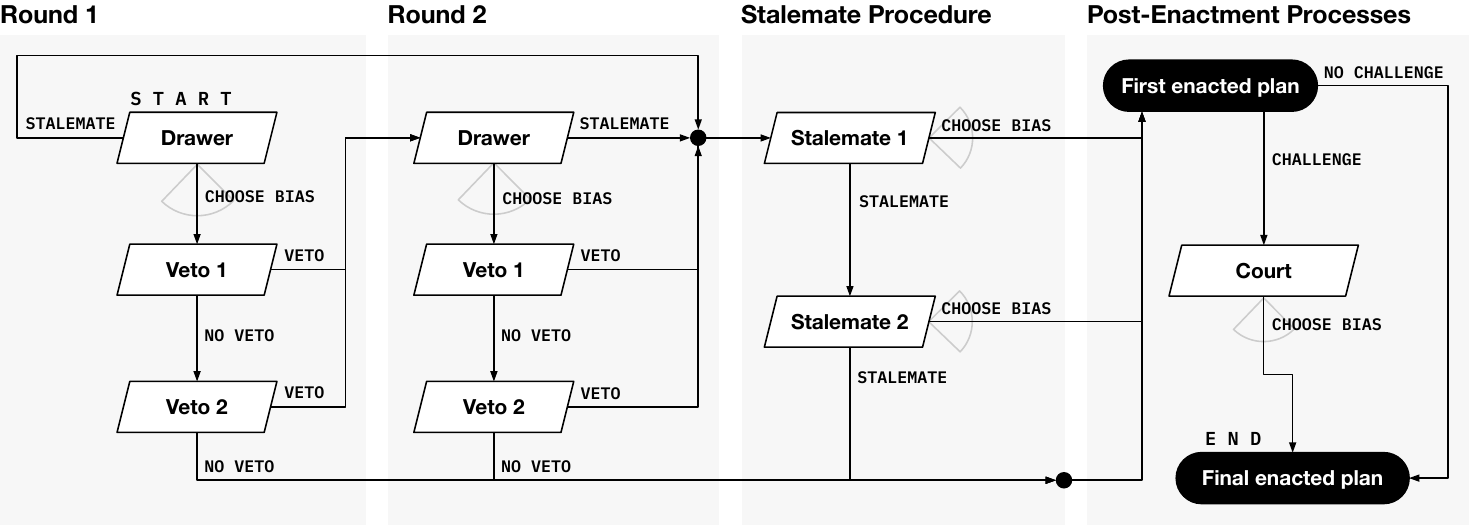}}

}

\caption{\label{fig-game}Prototypical game tree used to model
redistricting in every state. States differ in which party, if any,
controls each node, and which nodes are present in the state's process.}

\end{figure}%

Our standardized coding of relevant institutional features is based on
the prototypical redistricting process shown in Figure~\ref{fig-game}.
First, an initial map drawer proposes a plan that may be vetoed by other
actors. If the plan is not vetoed, it can be challenged in court. If it
is vetoed, there is another round of map drawing. If a plan is vetoed
twice, or if the initial map drawer cannot agree on a plan, then a
different institution (often a court) must resolve the stalemate and
adopt a plan.

Every state's process can be described as a subset of this prototype.
For example, in Michigan, a commission draws congressional districts,
resolves any stalemates, and there is an explicit mechanism for state
court review. Thus, Michigan's process would be described by the
``Drawer'' step in Round 1 and ``Stalemates'' steps only in
Figure~\ref{fig-game}.

For each state, we record which institutional body, if any, acts at each
step and which party, if any, controls that institution. We also
collected additional information relevant to modeling redistricting
processes and court review, such as whether the state's redistricting
plans were subject to DOJ preclearance before 2013,\footnote{The
  \emph{Shelby} decision was handed down in 2013, ending DOJ
  preclearance.} and which institutional body ended up drawing the plan
that was used in the first postcensal elections. Most procedural details
are straightforward, and the information for coding is readily available
in public data from each state (see e.g.,
\url{https://redistricting.lls.edu/}). Furthermore, many states have
similar redistricting procedures that are easily classified under the
categories we defined earlier.

Appendix~\ref{sec-app-coding} explains in detail how each of these
variables was coded and describes special cases.
Figure~\ref{fig-process} of the previous section graphically summarizes
this dataset for 2020 while Table~\ref{tbl-coding} of the Appendix
presents all the variables across states in both 2010 and 2020.

\subsection{The redistricting game}\label{sec-game}

Coding the details of each state's redistricting process preserves
important procedural information, compared to categorizing each state
into a small number of groups such as ``legislature-controlled'' and
``independent commission.'' However, the detailed coding presents a
challenge for causal inference, since the treatment---a state's
redistricting process---is now high-dimensional. Out of the 87
state-decade processes we code,\footnote{Only 43 states had two or more
  congressional districts in the 2010 cycle; for 2020, Montana did as
  well, bringing the total to 44.} there are 58 distinct combinations of
procedural variables, 41 of which are completely unique.

The combination of high-dimensional treatment and a limited sample size
means that there may not be enough information to estimate the causal
effect of changing from one specific configuration of procedural
variables to another, without further assumptions.

We address this methodological challenge by leveraging two basic
substantive assumptions about the redistricting process. First, each
party aims to draw a map that favors it as much as possible, and second,
the parties are constrained by statutory and constitutional rules in
doing so. Specifically, we treat the redistricting process depicted in
Figure~\ref{fig-game} as a sequential zero-sum game\footnote{The
  zero-sum assumption rules out cooperation between the parties to,
  e.g., protect incumbents. Given the role of the game here as a summary
  of institutional features rather than a structural model that predicts
  the outcome, we believe this is an acceptable limitation.} with two
players---the Democratic and Republican parties---each trying to
maximize the degree to which the drawn plan favors their party.

In each state, the nodes in the game tree can be controlled by different
parties, or by neither party (e.g., when a supermajority is required to
adopt a plan and neither party has supermajority control). Any
split-control nodes, as well as a node for state court action, are
considered moves by nature. Moves at one node do not affect the set of
actions available at other nodes. Some nodes involve discrete choices
such as whether to veto a plan or not, others are labeled ``choose
bias,'' meaning that the player at that node draws a plan with a chosen
amount of partisan bias favoring either party. This bias is exactly the
utility for the party of the chosen plan: a plan with bias \(x\) is
worth \(x\) to the Republicans and \(-x\) to the Democrats. We need not
quantify exactly what the utility measure is as a function of a specific
plan chosen; it suffices to let the parties try to maximize an abstract
univariate measure of partisan bias. We let the bias score range from
\(-4\), indicating a maximum Democratic advantage, to \(+4\), indicating
a maximum Republican advantage.

For concreteness, consider the case of redistricting in Oregon. In 2020,
the first move belonged to the Democratic controlled legislature. It has
to pick the amount of partisan bias \(-4<x<4\) in the plan that it
adopts. If this plan is ultimately the final enacted plan, Republicans
receive utility \(x\) and Democrats receive utility \(-x\). If the
legislature fails to adopt a plan, the first stalemate move in Oregon
belongs to a nonpartisan commission.

If a plan is adopted in the first round in Oregon, it proceeds to face a
potential veto by the Democratic governor or a possible court review.
The court may decide to accept a legal challenge, decide in favor of the
plaintiffs, and redraw the map; this choice is considered a move by
nature. If the court review results in a redrawn map with partisan bias
\(x'\), then the Republicans receive utility \(x'\) and the Democrats
receive utility \(-x'\). In Oregon, court review is explicitly allowed
on partisan grounds and challenges under the federal Voting Rights Act
are possible, so there is moderate probability that the
commission-adopted plan will be overturned. If no plan is enacted by the
commission in the first round, the commission is again tasked with
drawing a plan. If the commission fails to enact a plan again, courts
must step in and redraw district lines to ensure compliance with federal
constitutional ``one person, one vote'' apportionment requirements. This
is also considered a move by nature.

To complete the description of the game, we must specify the rules for
determining the expected outcomes for moves by nature. There are two
kinds of moves by nature: map drawers controlled by neither party
exclusively, and the results of court challenges. The full specification
of these moves can be found in Appendix~\ref{sec-app-game}, but we
briefly summarize them here. We make three assumptions: (1) nonpartisan
map-drawers whose choices are subject to veto will generate maps which
favor the party controlling the veto after a first veto has been made,
(2) stalemate map-drawing will produce a map that is moderately balanced
but tends to be influenced by any biases present in the most recent
redistricting proposal, and (3) split-control map-drawers will stalemate
with some probability and produce similar results to nonpartisan
map-drawers the rest of the time. Similar assumptions about partisan
map-drawers are not needed, since partisan drawers are assumed to act
strategically within the game.

Finally, we decompose the court challenge process into five components:
the probability that a legal challenge is possible, the probability that
a challenge is made when possible, the probability that a court sides
with plaintiffs, the expected remedy a court orders in those cases, and
the probability and expected effect of a challenge based on the federal
VRA. The latter applies only to states previously subject to DOJ
preclearance. Each of these five components has a parametric
specification, detailed in Appendix~\ref{sec-app-game}. The partisan
control of state courts is accounted for in the specification of these
various components, thus allowing judicial polarization to enter the
picture without assuming absolute strategic coordination between the
state party and its allies in the judiciary. Overall, the court
challenge process is not modeled strategically, since non-party actors
are more often than not the ones that initiate litigation, and have
different incentives than the party actors.

All in all, the game specification depends on 19 parameters which govern
the moves by nature, with most of these parameters relating to the court
challenge process. Rather than fix these parameters to arbitrary
constants, we place a prior distribution on each parameter over a range
of probable values. We simulate 200 different draws from this joint
prior distribution; each draw generates a slightly different game
specification. We then average our results across the random
draws.\footnote{We average, rather than propagating each draw through
  the estimation step. This has computational benefits and reduces the
  complexity of an outcome model (the outcome model would need to
  account for a different scale of treatment values across different
  draws). More importantly, however, the variation in the estimated
  equilibria, as described below, is quite small compared to other
  sources of uncertainty.}

\subsection{Equilibrium solution as a treatment
variable}\label{equilibrium-solution-as-a-treatment-variable}

To effectively summarize a high-dimensional treatment, we use the
subgame perfect Nash equilibrium of the redistricting game. The utility
of the equilibrium solution captures the expected partisan bias of plans
that arise out of a state's redistricting process, under the current
party control of the state's institutions. All of the multi-step
institutional interactions and negotiations that might happen as part of
the redistricting process are thus reduced to a univariate score. The
upshot is that the 14 procedural variables can be reduced into a single
variable that measures the leeway political actors have over the
partisan lean of the final redistricting plan.

To calculate the equilibrium itself, we numerically solve the game via
backward induction. This requires up to four levels of nested
optimization in some states. Section~\ref{sec-michigan} describes a full
example of this process for Michigan, and Appendix~\ref{sec-app-ex-al}
walks through a more complex, multi-step example for Alabama. This
solution process is automated across all of the states and is carried
out on each of the 100 different draws from the prior on the game's
parameters. This process requires the use of several tuning parameters
described at length when we present the full details of the game in
Appendix~\ref{sec-app-game}. Further, Appendix~\ref{sec-app-court} shows
that the values of our calculated equilibria are not sensitive to the
choice of initial values for these tuning parameters.

For our causal analysis, we use three treatment measures. The first
treatment measure is exactly the game's Nash equilibrium, averaged
across the prior. This measure depends on which parties control each
node in the game tree. In our analysis of nonpartisan outcomes, we also
use the absolute value of the equilibrium to capture the magnitude but
not the sign of the expected bias. Finally, we also generate a third
treatment measure that does not depend on the current party control.
This is calculated by assigning a single party (here, the
Democrats\footnote{The game is completely symmetric between parties,
  with one exception: due to racial polarization in the U.S., challenges
  to redistricting plans under the VRA are much more likely for
  Republican-leaning plans, and when these challenges prevail the remedy
  almost always results in a plan more favorable to Democrats. In
  keeping with the goal of calculating the \emph{maximum} leeway in each
  state, we therefore assign states to Democratic control under the
  counterfactual scenario here. Assigning to Republican control instead
  would slightly reduce the calculated leeway for those states
  previously subject to DOJ preclearance.}) to control each node in the
game tree that belongs to a partisan actor---legislature, governor, or
partisan commission---and then recalculating the average Nash
equilibrium. We refer to this equilibrium as the \emph{maximum leeway}
of a state's redistricting process, since it captures the expected bias
under a worst-case partisan outcome where all of the levers of state
government are controlled by one party. The first and second treatment
measures we refer to as the \emph{realized leeway} and \emph{absolute
realized leeway}, since they depend on the realized values of the party
control for each state institution.

\subsection{Realized and maximum leeway
scores}\label{realized-and-maximum-leeway-scores}

\begin{figure}[t]

\centering{

\pandocbounded{\includegraphics[keepaspectratio]{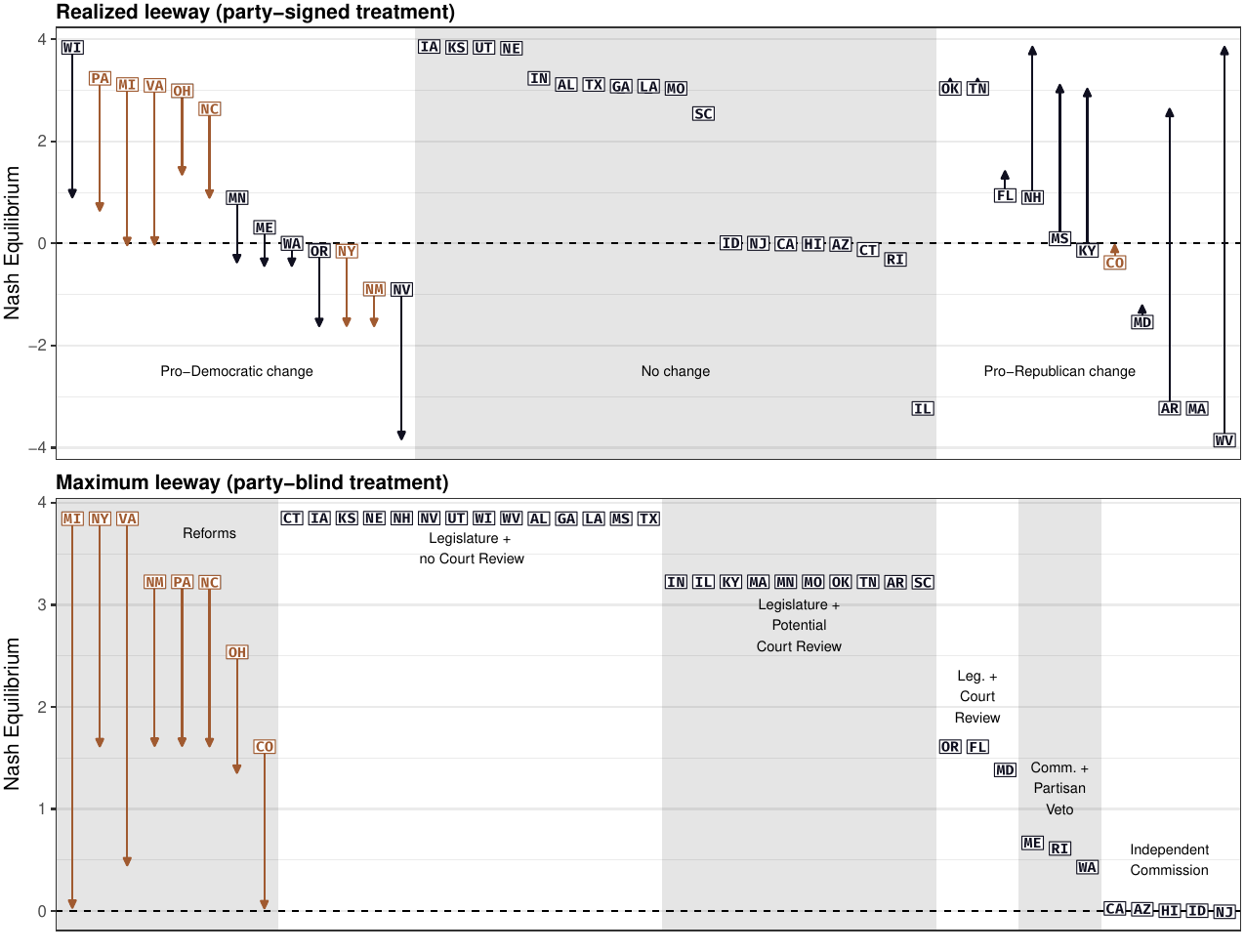}}

}

\caption{\label{fig-doses}Treatment values for each state in 2010, with
values for 2020 indicated by arrows, where different. States in orange
are those which experienced a reform to their redistricting procedures,
either by legislation, constitutional amendment, or a court ruling that
allowed for state court review of alleged partisan gerrymanders.}

\end{figure}%

Figure~\ref{fig-doses} visualizes these treatment measures for all the
states we study; an arrow indicates a change in the treatment value from
2010 to 2020. The realized leeway scores (game equilibria) cover the
entire range of possible partisan biases, from West Virginia in 2010
with complete Democratic control of state government, to Wisconsin,
Iowa, Kansas, Utah, and Nebraska, where Republican trifectas were
unconstrained by Democrats or by the VRA in both cycles. The maximum
leeway scores likewise span the range of possible bias, but take on
fewer values, since the various actual combinations of partisan control
of state institutions are no longer considered.

Intuitively, states introducing a commission (e.g., New York, Virginia,
or Colorado) drastically reduced their leeway to match states with
similar sets of rules in 2020. States that had intervening litigation to
clarify the interpretation of state redistricting rules (e.g., New
Mexico, North Carolina, or Pennsylvania) see appropriate changes in
leeway, even absent a commission. Further, in our party-signed
treatment, states that had a flip in party control of the whole system
see large changes in leeway in the correct direction (e.g.~West
Virginia). Finally, states with minor changes to the total control of
state institutions see minor changes in leeway (e.g., North Carolina).

\subsection{Full process example: Michigan}\label{sec-michigan}

Here, we build further intuition for our approach by summarizing each
step described above for Michigan, a state that experienced a large
change in leeway from 2010 to 2020 by adopting an independent
commission. Appendix~\ref{sec-app-ex-al} walks through a more complex
example for this process in Alabama.

First, we outline the institutional structure of the redistricting
process to establish the steps in the game. In Michigan, an independent
commission drew the initial map in 2020. See
Appendix~\ref{sec-app-coding} for details on all steps for each state.
Michigan's commission makes the first move, which is to pick the amount
of partisan bias \(-4<x<4\) in the plan that it adopts. As above, the
bias in this plan translates into partisan utility---if this proposal is
adopted, Republicans receive utility \(x\) and Democrats receive utility
\(-x\).

The following steps also depend on the state's specific institutional
structure. In Michigan, if this initial proposal is adopted, the plan
faces potential court review. Alternatively, if the commission's first
proposal fails, the stalemate move also belongs to the commission (see
Appendix Table~\ref{tbl-coding}). While Michigan has no partisan veto
points, plans in other states could also face potential vetos at this
point (or, in states like Iowa, could even face vetos from two sources).

With this game structure established, we solve the equilibrium as above
using backward induction. Michigan's redistricting process requires one
optimization step due to its lack of partisan veto points, though see
Appendix~\ref{sec-app-ex-al} for a more complex example with multiple
subgames. This means that the last partisan move in the game is the
initial (Round 1) commission proposal described above.

Our overall equilibrium solution for Michigan's 2020 structure is a bias
of \(-0.03\), suggesting very little bias in favor of either party. This
near-zero bias reflects the lack of partisan control in this particular
structure. Because our solution approach reflects institutional designs
at each time period, our equilibria estimates also capture intuitive
changes in leeway over time. For example, in 2010, Michigan
redistricting was entirely controlled by the legislature and governor,
who were both Republican. This game structure results in the highest
possible value of maximum leeway (party-blind) near 4, as a single party
controls all important game steps. However, our near-zero 2020
equilibria reflects the adoption of a strong independent commission. In
the new process, the commission is bound by strict criteria, the process
allows for court review, and even the commissioners themselves are
selected in part by a lottery system. With the reform, Michigan's
maximum leeway changed from the highest observed value to zero, meaning
that partisan actors have no control over redistricting outcomes, on
average.

\subsection{Empirical validation of the proposed treatment
vaiables}\label{empirical-validation-of-the-proposed-treatment-vaiables}

Building treatment variables through a game-theoretic model provides
significant advantages in dimension reduction and interpretability, but
it also comes with the risk that the model is misspecified. Although it
is difficult to completely verify the validity of each component of the
model, we take several steps towards empirically validating the model as
a whole, showing that the resulting treatment variable predicts the
observed outcomes well. We also later conduct a robustness analysis for
model misspecification in the context of causal effect estimation (see
Appendix Section~\ref{sec-app-robust}).

The model was designed to be flexible enough to capture the actual
redistricting processes in each state. Any misspecification is therefore
due to how the moves by nature are specified, or in the underlying setup
of two opposing partisan actors competing on a single zero-sum
dimension. First, we find that the treatment values are not sensitive to
specific parameter values, up to monotonic transformations. Across the
200 random draws from the prior, the average pairwise Spearman
correlation between the Nash equilibria for each state is above
0.99.\footnote{i.e., for each pair of random draws, we calculate the
  Spearman correlation between the two vectors of calculated state
  equilibria, then average these correlations across all pairings.} This
also gives us confidence that the specific choice of the prior is not
influencing the results.

\begin{figure}[t]

\centering{

\pandocbounded{\includegraphics[keepaspectratio]{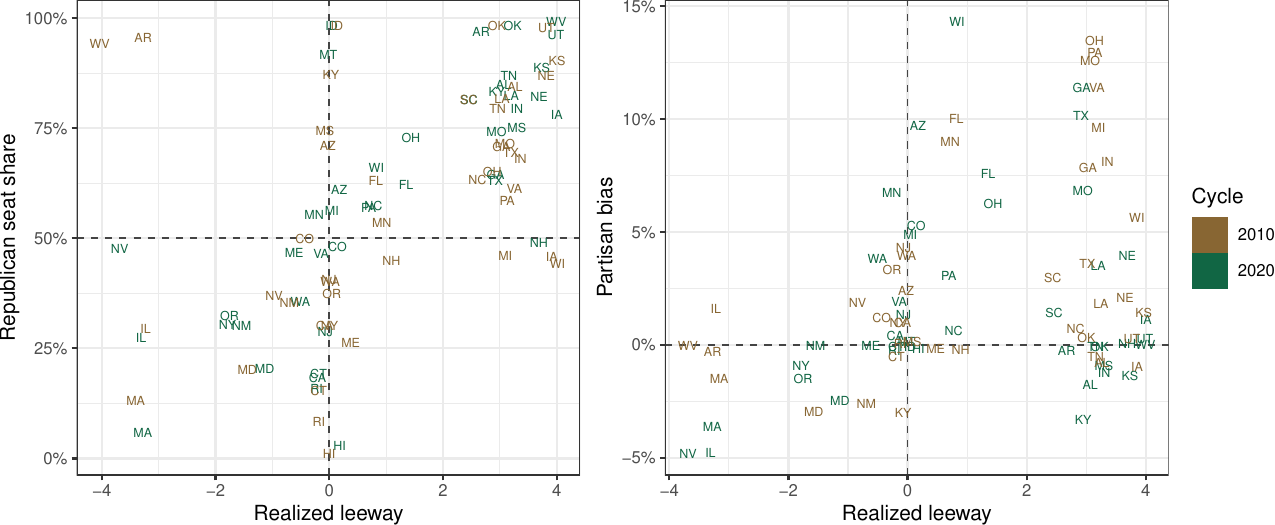}}

}

\caption{\label{fig-trt-corr}Measures of partisan advantage versus
treatment values for states' enacted plans for the 2010 and 2020
redistricting cycles. Points are slightly jittered to avoid
overplotting.}

\end{figure}%

Second, as shown in Figure~\ref{fig-trt-corr}, there is a substantial
correlation in the expected direction between the treatment values and
the measures of partisan advantage. The left panel plots the expected
share of seats won by Republicans versus the realized leeway measure;
recall that the leeway measure is positive for plans that favor
Republicans. The right panel shows the partisan bias
\citep{king1987democratic}, measured at the state's baseline vote share,
versus the realized leeway. The greater values of partisan bias
correspond to plans that systematically favor one party, with positive
values favoring Republicans.

For both outcome measures, the purely \emph{a priori} model predictions
of the expected partisan bias do in fact correlate with the actual
partisan bias of the plans that come out of each state's redistricting
process. To what extent these correlations can tell us about the causal
effects of reforms is of course another matter, and the primary question
addressed in the next section.

\begin{table}

\caption{\label{tbl-eq-path}Correspondence between the institution that
drew the final redistricting plan for each state and the most likely
outcome based on the equilibrium path of the redistricting game.}

\centering{

\begin{tabular}{l>{}r>{}r>{}rr}
\toprule
\multicolumn{1}{c}{ } & \multicolumn{3}{c}{Most likely in equilibrium} & \multicolumn{1}{c}{ } \\
\cmidrule(l{3pt}r{3pt}){2-4}
Final drawer & Legislature & Commission & Court & Total\\
\midrule
Legislature & \cellcolor[HTML]{A08CB9}{26.9} & \cellcolor[HTML]{F6F4E5}{0.0} & \cellcolor[HTML]{D69CC3}{20.1} & 47\\
Commission & \cellcolor[HTML]{F5E9D3}{1.9} & \cellcolor[HTML]{E1A5C0}{18.1} & \cellcolor[HTML]{F5E9D3}{3.0} & 23\\
Court & \cellcolor[HTML]{F4DEC9}{5.8} & \cellcolor[HTML]{F6F4E5}{0.9} & \cellcolor[HTML]{F2C6BE}{10.3} & 17\\
Total & \cellcolor{white}{34.6} & \cellcolor{white}{19.0} & \cellcolor{white}{33.4} & 87\\
\bottomrule
\end{tabular}

}

\end{table}%

We also compare the equilibrium path in the model with the actual
institution that drew the final map in each state. Because the moves by
nature in the game are random, the actual equilibrium path may be a
probabilistic mixture over multiple possible paths. We average these
probabilities over the 200 draws of the game to arrive at an overall
probability that a legislature, commission, or court draws the final map
in each state, given its procedure and party control.

Table~\ref{tbl-eq-path} compares the most likely final map-drawer,
according to these probabilities, to the actual institution that drew
the final redistricting plan. The correspondence is generally excellent,
considering that the predictions are made based on theory alone. There
is a tendency, however, for the model to predict court-drawn maps when
they are in fact drawn by legislatures. The overall likelihood of a
court intervention is itself a model parameter that is varied across
random draws, and as noted above the ranking of the states by leeway is
basically unchanged by different parameter values. Therefore, we expect
this tendency to over-predict court intervention to at most slightly
understate the equilibrium value for states with court review, since a
higher likelihood of intervention would tend to pull the equilibrium
towards zero.

Finally, it is natural to wonder whether a more accurate model could be
obtained by estimating the model parameters from the observed data
rather than specifying their prior distributions. That is, one could
find the parameter values that maximize the correlation between the
treatment and the outcome. We do not take this approach to cleanly
separate the estimation of causal effects from the construction of a
treatment variable \citep{rubin2008}. This enables us to employ a causal
identification strategy that does not assume the correct specification
of the game-theoretic model. Instead, we use the model to summarize a
high-dimensional treatment variable in a theoretically informed way
without looking at the outcome variable.\footnote{We did attempt to fit
  the game parameters to the observed data for the 2010 cycle alone via
  maximum likelihood, but we found that the observed data were only
  minimally informative about most of the parameter values. This is good
  news, insofar as it implies that a wide range of parameter values are
  all compatible with the observed data.}

\section{Estimating the Causal Effects of Institutional
Leeway}\label{sec-estimation}

We now discuss how our approach addresses the remaining challenges
regarding outcome complexity and confounding bias in causal effect
estimation (see the ``estimation'' and ``outcome'' boxes of
Figure~\ref{fig-analysis-proc}). We use a difference-in-differences
design with the continuous treatment variable of institutional leeway to
adjust for both observed and unmeasured state-specific confounding
factors. We also use a simulation approach to adjust for changes in
political geography that have happened during this time period.
Specifically, we generate a representative sample of non-partisan
redistricting plans for both 2010 and 2020 redistricting cycles to
quantify state-specific changes in political geography. By accounting
for such changes, we are able to better isolate the impact of
redistricting reforms from that of political geography.

\subsection{Difference-in-differences design}\label{sec-did}

To estimate the causal effects of changes in leeway on redistricting
outcomes, we use a differences-in-differences (DiD) design with a
continuous treatment variable. This strategy addresses potential
confounding by comparing changes in states that have enacted reforms to
changes in similar states that have not modified their redistricting
process across the 2010 and 2020 redistricting cycles. We assume that,
in the absence of reforms, the states with institutional changes would
have experienced the same trend in outcome variables as those states
without reforms.

Formally, let \(\vb Z_{it}\) be the 14-dimensional vector of
institutional features for state \(i\) at time \(t\) discussed in
Section~\ref{sec-data-inst}. We use \(t=0\) and \(t=1\) to denote 2010
and 2020 redistricting cycles, respectively. Let \(u^*(\vb z)\)
represent the equilibrium outcome (game utility) for the average Nash
equilibrium of the redistricting game described by the process
\(\vb z\). Then, we can define a univariate treatment variable
\(D_{it}=u^*(\vb Z_{it})\) for each state-decade, which represents the
``dose'' of institutional leeway given to political actors.

The key assumption made by our use of the Nash equilibria as the
treatment variable is that the institutional features affect the outcome
only through this treatment variable, i.e., \[
Y_{it}(\vb z)=Y_{it}(\vb z') \qq{for any} \vb z, \vb z' \text{ with } u^*(\vb z) = u^*(\vb z'),
\] where \(Y_{it}(\vb z)\) denotes a generic potential outcome variable
for state \(i\) at time \(t\) with institutional features
\(\vb z\).\footnote{For identification, it in fact suffices to assume
  only that
  \(\E[Y_{it}(\vb z)\mid\vb X=\vb x]=\E[Y_{it}(\vb z')\mid\vb X=\vb x]\)
  for any \(\vb z, \vb z'\) with \(u^*(\vb z) = u^*(\vb z')\); however,
  the stronger individual-level assumption is invariant under
  transformations and ensures that no additional assumptions are
  required for estimation.} Under this \emph{sufficiency} assumption, we
can simply write the potential outcomes as
\(Y_{it}(d) = Y_{it}(u^*(\vb z))\) for the treatment dose
\(d=u^*(\vb z)\). While the assumption is not directly testable with the
data we have, we evaluate robustness to its violations in
Appendix~\ref{sec-app-robust} by additionally controlling for \(\vb z\)
in effect estimation; the resulting estimates are not statistically
distinguishable from the main specification's estimates.\footnote{It is
  also possible that we may have not measured all relevant features of
  redistricting institutions. The existence of such unmeasured features
  may alter the interpretation of causal effects of interest
  \citep{vand:hern:13}. For example, the causal effect of a change in
  redistricting institutions would be interpreted as the effect of such
  a change averaging over other unmeasured institutional changes that
  are associated with it.}

Our target estimand is the conditional average treatment effect (CATE)
for a change from the treatment level \(d\) to \(d'\) given a set of
covariates \(\vb X_i\) that are not affected by the treatment. The CATE
is defined as, \[
\mathrm{CATE}_{\vb x}(d', d) = \E[Y_{i1}(d') - Y_{i1}(d) \mid \vb X_i = \vb x].
\] Estimating \(\mathrm{CATE}_{\vb x}(d',d)\) for any pair of values
\(d', d\) and covariate values \(\vb x\) requires the identification of
a full dose-response curve. Our identification approach relies on a
strong conditional parallel trends assumption introduced by
\citet{callaway2021cdid}. Specifically, we assume that the average
change in outcomes for states experiencing a change in treatment dosage
from \(d\) to \(d'\) depends only on the observed covariates, and not
the observed dosage value.

Our covariates include a range of variables that existing work in the
redistricting literature identifies as predicting redistricting outcomes
and are not affected by redistricting reforms. We include
(pre-treatment) 2008 Democratic presidential vote share, an indicator
for being in the South,\footnote{Southern states are defined as Alabama,
  Arkansas, Florida, Georgia, Kentucky, Louisiana, Mississippi, North
  Carolina, South Carolina, Tennessee, Texas, Virginia, and West
  Virginia. This set corresponds to Confederate states, plus Kentucky
  and West Virginia, which also had a history of Democratic dominance in
  the 20th century and notable differences between party preferences at
  the state and national levels.} the logarithm of the number of
districts in 2020, the change in the number of districts between 2010
and 2020, the logarithm of the average number of state corruption
convictions by year between 2000 and 2010, and an indicator for whether
states allow ballot initiatives. For the maximum leeway treatment, which
uses no information about partisan control of state institutions, we
also control for changes in partisan control of the map-drawing body and
of the state supreme court.

Formally, we require the following conditional parallel trends
assumption for all \(d,d'\): \[
\E[Y_{i1}(d') - Y_{i0}(d) \mid \vb X_{i} = \vb x]
= \E[Y_{i1}(d') - Y_{i0}(d) \mid \vb X_{i} = \vb x, D_{i0}=d, D_{i1}=d'].
\] This assumption is analogous to the parallel trends assumption in the
traditional binary difference-in-differences design, but differs in that
it refers to changes in dosage across a continuous measure, rather than
a single level change from zero. Compared to a traditional
selection-on-observables assumption, the conditional parallel trends
does not rule out confounding factors that are constant over time. With
this assumption, we can identify our estimand as \[
  \mathrm{CATE}_x(d, d')
  = \E[Y_{i1} - Y_{i0} \mid \vb X_{i} = \vb x, D_{i0}=d, D_{i1}=d'] - \E[Y_{i1} - Y_{i0} \mid \vb X_{i} = \vb x, D_{i0}=d, D_{i1}=d],
\] where \(Y_{it} = Y_{it}(D_{it})\) is the observed outcome. Averaging
over the marginal distribution of \(\vb X_{i}\), we can also estimate
the average treatment effect (ATE).

Evaluating the plausibility of the parallel trends assumption here is
complicated by a lack of public data on precinct-level election returns
for the 2000 redistricting cycle or earlier cycles. Without this data,
we cannot generate simulated districts for decades before 2010, which
prevents us from conducting placebo checks using the pre-trends, a
common practice for differences-in-differences studies. We are able,
however, to conduct a placebo check using an outcome that should be
unaffected by redistricting reform, as we discuss in
Section~\ref{sec-placebo} below.

Thus, at least partially, we are left to justify the assumption based on
substantive grounds. To violate the conditional parallel trends
assumption, the change in outcomes (such as Democratic seats) that would
have been observed in Michigan, had it not adopted reforms, would have
to systematically differ from the corresponding change in other
non-reform states like Wisconsin, after controlling for covariates. The
covariates we include are some of the strongest predictors of treatment
adoption and redistricting outcomes, which substantially increases the
plausibility of this assumption.

Second, aspects of the treatment and outcome themselves weigh against
the possibility of a violation. Changes in redistricting procedure are
usually advanced outside of legislatures, often via citizen referenda
funded by nonpartisan good-governance groups who are not responsible for
map-drawing. The political parties also coordinate redistricting
strategy at a national level. Further, state legislatures and other
map-drawing bodies experience turnover between decades.

Together, these factors suggest that the change in Michigan's outcomes
would have looked much like that of a non-reform state's, had it not
adopted reforms. In the specific example of Wisconsin, political
developments since 2023 further justify this conclusion. A narrow state
supreme court victory by a justice who campaigned, in part, in
opposition to gerrymandering, led to a successful court challenge to
state legislative maps that resulted in a significant decrease in the
partisan bias of those maps (on common bias metrics). Challenges to the
congressional maps are currently pending.

\subsection{Adjusting for political
geography}\label{adjusting-for-political-geography}

To further increase the credibility of the conditional parallel trends
assumption above, we adjust for the changes in political geography
between the two redistricting cycles. Partisan preferences and the
geographic distribution of voters can change in different ways within
each state over time. For example, Michigan, which experienced reform,
became more Republican during our sample period, while Georgia, which
did not, became more Democratic. Without this adjustment, an estimated
effect of the Michigan reform would be biased by the differential change
in the states' political geographies.

To make the adjustment, we use representative sets of nonpartisan
redistricting plans within each state that respect each state's specific
redistricting rules. These simulation samples were separately generated
for 2020 \citep{50stateSimulations} and 2010 \citep{50states2010} using
the algorithm of \citet{smc}. A detailed discussion of these simulations
and their limitations may be found in Appendix~\ref{sec-app-sims}. We
subtract the mean outcome in each state's simulated sample, denoted by
\(\widetilde{Y}_{it}\), from the observed outcome \(Y_{it}\) before
estimating the causal effects under the DiD design. We argue that this
subtraction of the outcomes based on simulated baseline plans from the
observed outcomes accounts for the change in political geography in each
state.

One limitation of basing simulations off of each state's rules is that
any changes in outcome variables from 2010 to 2020 that are due to
changes in rules would be subtracted away. Since redistricting rules are
sometimes changed as part of larger reform efforts, part of the overall
effect of redistricting reform would not be included in the measured
effect, attenuating the overall estimated effects. However, these rules
changes occurred in only a few states, and their effect on partisan
outcomes was measured by \citet{kenny2023widespread} and found to be
minimal.

Thus, we assume that once we adjust for the state-specific change in
political geography, states with different changes in institutional
features have parallel trends in the potential outcomes. Formally, if we
denote the difference between the potential outcome and the simulated
outcome as \(\Delta Y_{it}(d)=Y_{it}(d)-\widetilde{Y}_{it}\), our
conditional parallel trends assumption becomes \[
\E[\Delta Y_{i1}(d') - \Delta Y_{i0}(d) \mid \vb X_{i} = \vb x]
= \E[\Delta Y_{i1}(d') - \Delta Y_{i0}(d) \mid \vb X_{i} = \vb x, D_{i0}=d, D_{i1}=d']
\] for all \(d, d'\). Then, as above, the CATE is identified as \[
  \mathrm{CATE}_{\vb x}(d, d')
  = \E[\Delta Y_{i1} - \Delta Y_{i0} \mid \vb X_{i} = \vb x, D_{i0}=d, D_{i1}=d'] - \E[\Delta Y_{i1} - \Delta Y_{i0} \mid \vb X_{i} = \vb x, D_{i0}=d, D_{i1}=d],
\] where \(\Delta Y_{it} = Y_{it} - \widetilde{Y}_{it}\).

\subsection{Estimation of causal effects}\label{sec-est-model}

We estimate this causal estimand with a Bayesian linear regression
model, where the response is the change in each simulation-adjusted
outcome between 2010 and 2020, \(\Delta Y_{i1}-\Delta Y_{i0}\) (Appendix
Section~\ref{sec-app-raw} presents the descriptive analysis of raw
changes). The predictors are the change in treatment level
\(D_{i1}-D_{i0}\), the baseline treatment level \(D_{i0}\), the
covariates \(\vb X_i\), and the interaction of the treatment change with
baseline treatment and with each of the covariates. This is a
regression-adjusted differences-in-differences estimator
\citep{heckman1997did}. In contrast to the common two-way fixed effects
estimator, this approach allows for heterogeneous treatment effects. The
difference from most regression-adjusted estimators is the continuous
treatment, which necessitates a modeling choice on the form of the
dose-response curve \citep{callaway2021cdid}.

Given the small sample size (\(n=87\)), the moderate number of
covariates (\(p=15\) with interactions for the modal specification), and
the high noise level, we believe the linear specification and Bayesian
estimation are appropriate. The coefficient priors help avoid
overfitting to a few samples, and uncertainty in the ATE is
automatically quantified; details on these priors are included in
Appendix~\ref{sec-app-prior}. A more flexible regression model beyond
linear would be unlikely to increase predictive power, given the small
sample size, and would further risk overfitting. However, as a
robustness check, we also fit a nonparametric Bayesian Additive
Regression Trees (BART) model \citep{chipman2010bart} and include those
results in Appendix~\ref{sec-app-bart}. The results are qualitatively
the same, but the estimated effect magnitudes are attenuated.

\section{Estimated Causal Effects of Redistricting
Reforms}\label{estimated-causal-effects-of-redistricting-reforms}

The outcome of the redistricting process is a complete congressional
districting plan. A plan can be evaluated in a number of ways---how many
seats each party is expected to win, how many of the seats are
competitive, how well the partisan composition of the delegation matches
the voters' preferences, and so on. We begin this section by introducing
quantitative measures of several partisan and nonpartisan aspects of
districting plans. These constitute the outcome variables for the causal
estimates we present in the second half of the section. Appendix
Section~\ref{sec-app-raw} presents the results of descriptive analyses
that are largely consistent with the main results shown below.

\subsection{Outcome measures}\label{outcome-measures}

Our measures can be divided into two buckets: nonpartisan outcomes that
quantify how much gerrymandering is present or how responsive a
districting plan is to shifts in public opinion, and partisan outcomes
that capture which party is advantaged by a districting plan. For the
nonpartisan outcomes, we also use the absolute value of the realized
leeway as an additional treatment variable so that neither treatment nor
outcome considers the partisan direction.

All of these measures are calculated from district-level election
results. Since there is significant exogenous variation in election
results due to swings in the national political environment, rather than
using actual 2012 and 2022 House election results to evaluate
districting plans, we use a statistical model to estimate the
distribution of election results across future hypothetical elections. A
statistical election model also allows us to apply these same measures
to the simulated districting plans, under which no elections have taken
place.

We adopt the model from \citet{kenny2023widespread}, which assumes
district election outcomes can be decomposed into a baseline
district-level vote share plus district-specific and national
swings.\footnote{The full details of this approach can be found in the
  Methods section and SI Appendix, Section B of
  \citet{kenny2023widespread}} The model is closely related to the
stochastic uniform partisan swing model of \citet{gelman1994unified} and
the congressional model of \citet{ebanks2023american}. As the baseline
district-level vote, we use the 2008 presidential results for the 2010
redistricting cycle and an average of the 2016 and 2020 presidential
results for the 2020 cycle, each logit-shifted so that the national
partisan vote is exactly 50/50 \citep{vest2016, vest2020}. As in
\citet{kenny2023widespread}, we use historical congressional election
data since 1976 to estimate the variance of district and national
swings. We calculate exact expectations of all of our outcome measures
against the predictive distribution of the model via numerical
integration.

Our primary partisan measure of districting plans is the expected number
of seats won by the Republican party. As discussed in
Section~\ref{sec-estimation}, we subtract the average outcomes in the
simulated baseline redistricting plans
\citep{50stateSimulations, 50states2010} from the observed outcome in
the enacted plan to adjust for the state-specific change in political
geography, which is a potential confounding factor
\citep{cottrell2019using}. After this adjustment, the seat outcome
ranges from \(-1.57\) in Illinois in 2020 (favoring Democrats) to
\(1.78\) in Texas in 2020 (favoring Republicans). These
simulation-differenced seats can be directly interpreted as a
measurement of bias due to partisan gerrymandering, with positive values
indicating a Republican bias beyond what would be expected based on the
state's political geography alone.

The expected number of Republican seats is an interpretable measure, but
may not be completely comparable across states. Depending on political
geography and especially the total number of districts in each state,
the natural variation in the number of seats outcome may vary
significantly between states. To address this, we also include as an
outcome measure the simulation \(z\)-score of Republican seats, which is
calculated by taking the simulation-differenced seats outcome and
dividing it by the standard deviation of Republican seats in the
simulation set \citep{kenny2023widespread}. This puts all the states'
outcomes on a common scale, increasing the plausibility of the parallel
trends assumption and the homoskedasticity assumption of the estimation
model.

For a nonpartisan outcome measure, we take the absolute value of these
two partisan measures. Both the absolute difference and the absolute
\(z\)-score of Republican seats measure how far the enacted plan
deviates from the nonpartisan simulation baseline in terms of partisan
composition.\footnote{For the transformed outcomes based on \(z\)-score
  and absolute value, we do not interpret the estimated causal effects
  in terms of the original outcome variable (i.e., the expected number
  of Republican seats). Instead, the estimates should be interpreted as
  the effects on the transformed outcomes. This also means that the
  parallel trends assumption is assumed to hold on the transformed
  outcome variable.}

Finally, we also measure the responsiveness of districting plans to
changes in the national electoral environment. Responsiveness is
measured as the rate of change in the share of Republican seats given an
infinitesimal change in Republican vote share nationwide. Responsiveness
is closely linked to the presence of competitive seats; the more
competitive seats there are, the larger the change in seat share will be
for a given shift in vote share. Indeed, electoral responsiveness is the
primary motivation for the creation of competitive congressional
districts.

The responsiveness measure ranges from 0.044 in Wyoming in 2020 to 7.91
in New Hampshire in 2020, with most plans' values lying between 0.5 and
3. The interpretation of these values is as follows; in New Hampshire, a
1pp increase in a party's vote share leads to a 7.91pp change in the
party's expected seat share. This relatively large increase makes sense
in context---both of the state's congressional districts have Republican
vote share within a few points of 50\%. As with the other measures, we
subtract the mean responsiveness of the simulated plans from the enacted
plan's responsiveness when estimating causal effects.

In Appendix~\ref{sec-app-addl}, we extend the analysis to a series of
alternative measures of partisan bias, such as the efficiency gap
\citep{stephanopoulos2015}, and find qualitatively similar results.

\subsection{Empirical findings}\label{empirical-findings}

\begin{figure}[t]

\centering{

\pandocbounded{\includegraphics[keepaspectratio]{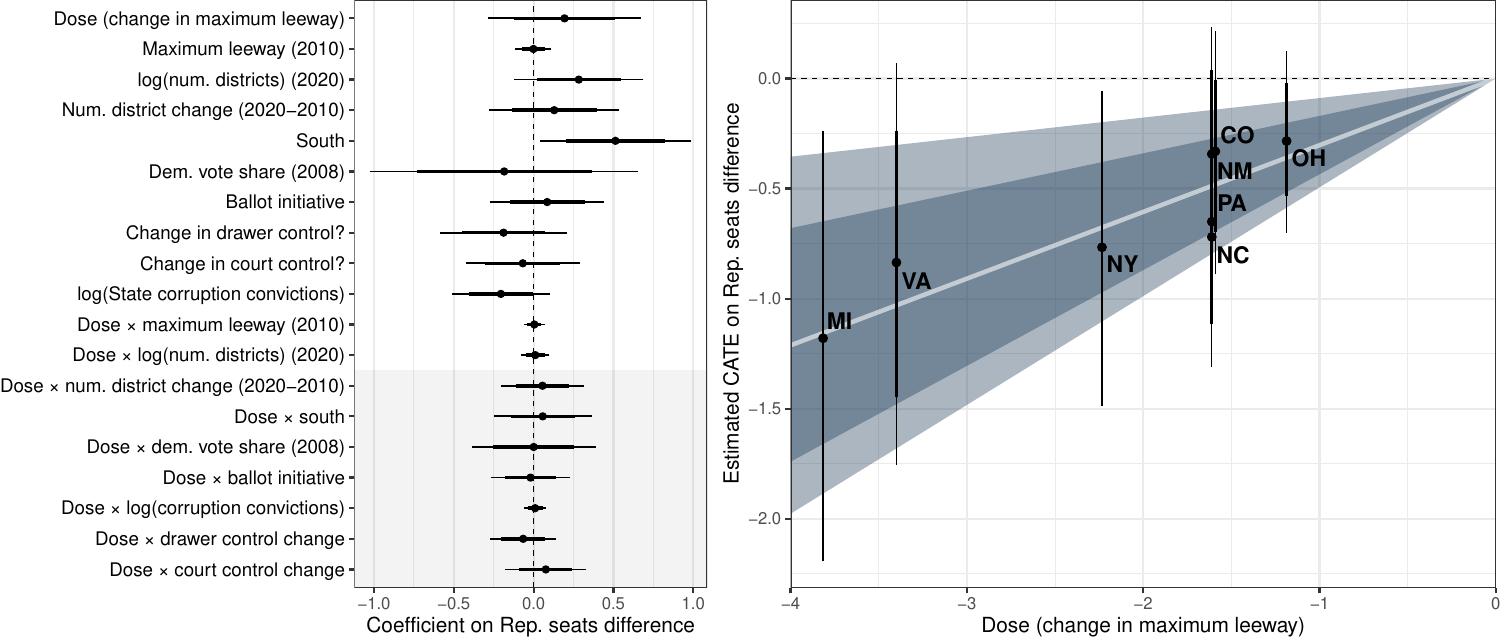}}

}

\caption{\label{fig-model-ex}Fitted model coefficient estimates for the
Republican seat outcome measure (left) using the maximum leeway
treatment, and estimated conditional average treatment effects for each
reformed state's covariate combination plotted against the state's dose
(right). The model-based dose-response curve is underlaid in blue. 80\%
and 95\% credible intervals are shown throughout.}

\end{figure}

We first study the effects of changes in maximum leeway on Republican
seats. Recall that maximum leeway represents the worst expected bias
under a single-party control. The left panel of
Figure~\ref{fig-model-ex} shows the estimated coefficients of the
estimation model discussed in Section~\ref{sec-estimation} fit to the
simulation-adjusted seats variable. All estimated coefficients for this
model and all other models in the paper are contained in
Appendix~\ref{sec-app-coef}. The coefficients in
Figure~\ref{fig-model-ex} are shown on the scale of the outcome, where a
positive coefficient indicates a relationship with a positive change in
Republican seats from 2010 to 2020. The model \(R^2\) is around 0.4,
highlighting the potential for the control variables to confound any
observed correlation between reform and the outcome measure.

Given the fitted outcome model, we can estimate the conditional average
treatment effect (CATE) for any given starting and final value of
maximum leeway and any combination of covariates. The right panel of
Figure~\ref{fig-model-ex} plots these CATEs for each state that
experienced a change in maximum leeway between 2010 and 2020, using the
specific covariates of that state. There is a clear dose-response
relationship between the decrease in maximum leeway and the decrease in
the expected number of Republican seats.\footnote{This pattern was also
  observed in robustness checks that used a flexible Bayesian Additive
  Regression Trees (BART) model for estimation.} A state with Michigan's
dose (nearly 4) has an estimated CATE of around \(-1.25\), with 95\%
credible interval (--2.0, --0.4), and so would be expected to gain just
over 1 Democratic seat, on average. Similarly, a state with Ohio's dose
(just over 1) and covariates would be expected to gain about half of a
Democratic seat (--0.6, --0.1). Motivated by this pattern, we also
include in the right figure a dose-response curve calculated by
calculating a CATE for dose level and averaging over the observed
covariate distribution.\footnote{Because of the interaction between the
  dose variable and the covariates (which do not have mean zero), the
  overall slope estimate and its variance differ from the estimate and
  variance of the coefficient on Dose in the linear model.}

Since the dose-response pattern is well approximated by a linear
relationship, we can summarize it by its slope, a quantity also known as
the average causal response (ACR). The product of the ACR and a given
dose corresponds to the estimated causal effect of that dose, averaged
across states.

\begin{figure}[t]

\centering{

\pandocbounded{\includegraphics[keepaspectratio]{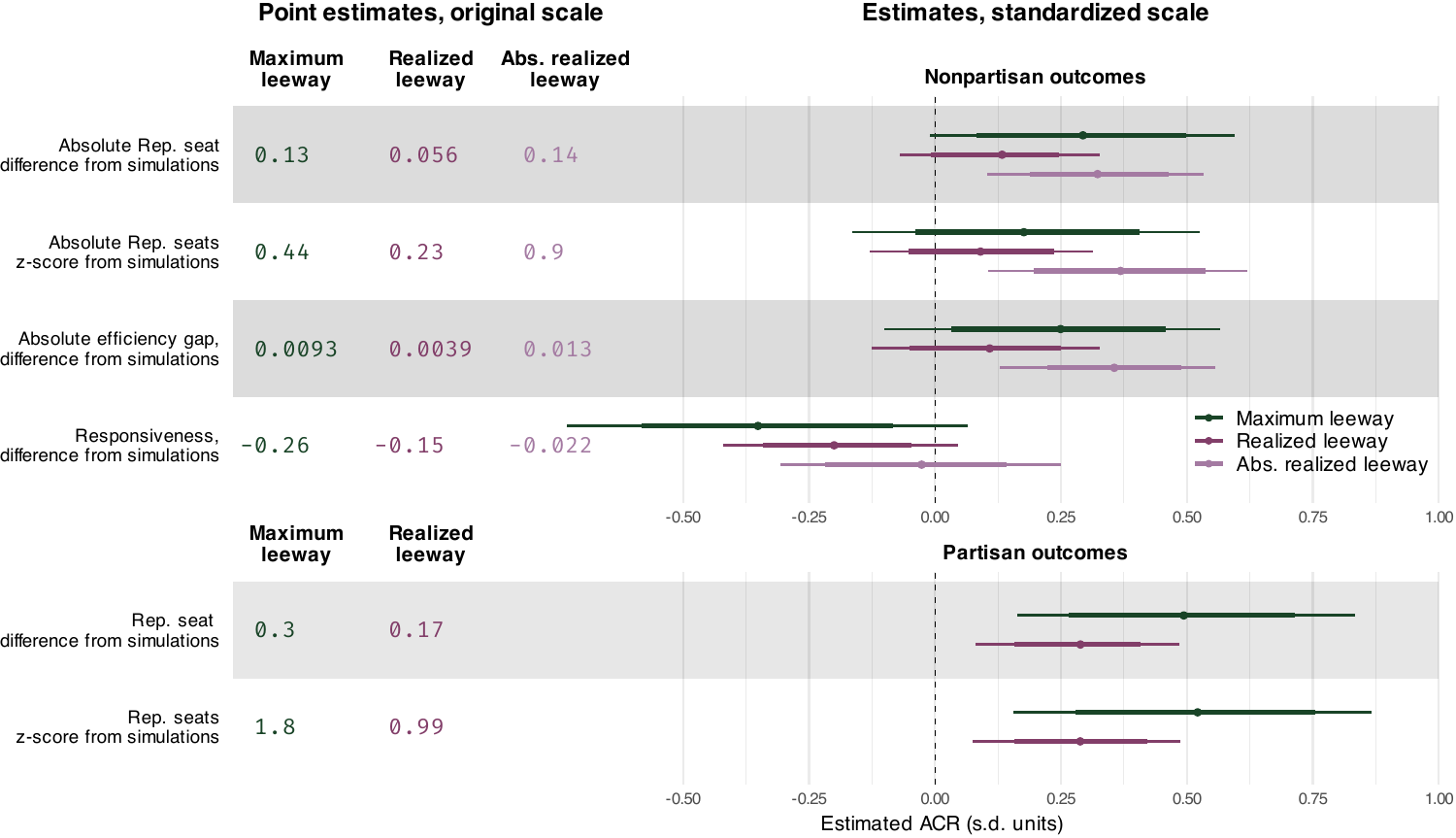}}

}

\caption{\label{fig-acr}Average causal response (ACR) of leeway on
redistricting outcomes. The points correspond to the mean estimated ACR,
while the lines represent 80\% and 95\% credible intervals. Intervals
are colored by the treatment variable used. The numbers in the columns
display the mean ACR on each outcome's response scale. The estimates and
intervals on the right are displayed in units of outcome standard
deviations, to allow for comparability between outcomes. For partisan
outcomes, a positive number indicates a pro-Republican effect and a
negative number indicates a pro-Democratic effect for a positive dose.}

\end{figure}

Figure~\ref{fig-acr} presents the ACR for Republican seats and the other
outcome variables, using both the maximum and realized leeway treatment
values. As mentioned above, for the nonpartisan outcomes, we also use
the absolute value of the realized leeway as a treatment variable. The
right side of the figure plots the ACR estimates in terms of the
standard deviation of the respective outcome variable per unit change in
treatment. This means that a point estimate of 0.5 is interpreted as an
increase of 0.5 standard deviations of the outcome variable for a 1-unit
change, or an increase of 2 standard deviations for a 4-unit change.

Across the three nonpartisan outcomes measuring the amount of
gerrymandering, we find a consistent effect of leeway. The estimated
ACRs are all positive, meaning that an increase in leeway leads to a
greater gap between the enacted plan and the simulated baseline. That
is, states that reduce leeway through reform efforts are expected to
rein in partisan gerrymandering. The magnitude of the effect is
substantial: a reform effort like Michigan's, with a decrease in the
maximum leeway of 4 units, would be expected to reduce the difference in
Republican seats from the baseline by about 0.5 seats (--0.006, 1.040),
or 1.7 simulation standard deviations (--2, 5)---enough to move an
extreme partisan gerrymander to a fair plan. The ACRs for the absolute
realized leeway are very similar to those for the maximum leeway; the
ACRs for the fully signed realized leeway are positive but smaller in
magnitude, and have enough uncertainty that the 80\% credible intervals
cross zero. Thus, there is some weak evidence that giving Democrats more
control of the redistricting process by increasing their leeway or
reducing Republican leeway would also reduce gerrymandering.

We also find that reforms increase electoral responsiveness. The point
estimates of the ACRs for both the maximum leeway and the realized
leeway agree; the posterior probability that the effect is negative is
97\% and 96\%, respectively. The estimated effect for the absolute
realized leeway is near zero. We estimate that a reduction in the
maximum or realized leeway of 4 units would increase responsiveness by
1.1 (--0.042, 2.142), meaning that a 1pp increase in a party's vote
share would translate to an additional 0.6pp in seat share on top of the
average 2pp increase in seat share across all states.

Another way to interpret the effect on responsiveness is in terms of the
share of competitive seats, where a seat that is counted as competitive
in proportion to how close its baseline partisanship is to 50\%. This is
equivalent to a linear rescaling of the responsiveness
outcome.\footnote{Specifically, we measure the competitiveness of a
  single seat as the derivative of the election model's win probability
  for that seat at its baseline vote, divided by the maximum derivative,
  which is obtained at a 50/50 baseline vote. Based on our election
  model, a district with 50/50 baseline vote is counted as 1.0
  competitive seats, and a district with 60/40 baseline vote is counted
  as just 0.14 competitive seats. This approach avoids arbitrary cutoffs
  in what counts as a competitive district.} In terms of competitive
seats, we estimate that a reduction in leeway of 4 units would increase
the share of competitive seats in an average state from 25\% to 38\%
(24\%, 51\%).

In terms of partisan outcomes, the estimated ACRs on Republican seats
are all positive, for both realized and maximum leeway. This is in line
with an interpretation that \emph{increasing} leeway benefits the party
in control, as one might expect. Recall that the doses from
Figure~\ref{fig-doses}, for the realized leeway estimates, a dose of 4
represents approximately going from a Democratic legislature to an
independent commission or an independent commission to a Republican
legislature. For such a dose of 4, we would expect Republicans to gain
around 0.7 seats (0.21, 1.18), or 7.2 simulation standard deviations
(2.4, 12.2)---again, a substantial effect.

\subsection{Placebo check}\label{sec-placebo}

As a further validation of our overall approach to estimating causal
effects, we conduct a placebo check by applying the same estimation
procedure to a political outcome variable that should not be affected by
redistricting reform but is closely related to the outcome variable of
interest. For this placebo outcome variable, we use the Democratic
two-party vote share in the first presidential election following each
census (2012 and 2024). As national contests, presidential elections are
insulated from state-specific reform efforts. Moreover, issues like the
economy, immigration, and health care dominate presidential campaigns,
not state-specific governance issues such as redistricting reform. Thus,
we expect there to be no effect of redistricting reforms on presidential
vote share.

\begin{figure}

\centering{

\pandocbounded{\includegraphics[keepaspectratio]{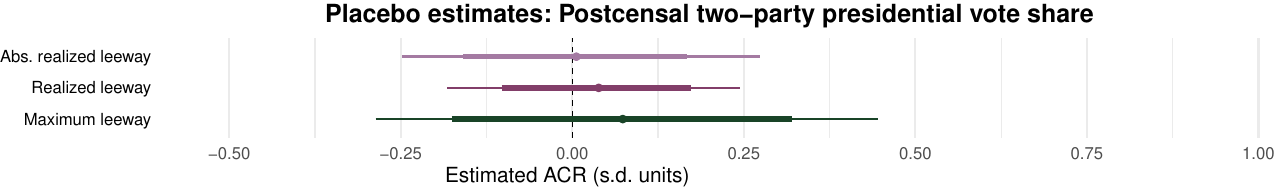}}

}

\caption{\label{fig-acr-placebo}Average causal response (ACR) of leeway
on placebo outcome. The points correspond to the mean estimated ACR,
while the lines represent 80\% and 95\% credible intervals. Positive
numbers correspond to an increase in Democratic vote share.}

\end{figure}%

Figure~\ref{fig-acr-placebo} displays the estimated ACRs for the placebo
outcome using each of the three treatment variables. All three estimates
are nearly zero, with credible intervals that easily cover zero. If the
estimates were significantly different from zero, we would suspect the
validity of the sufficiency assumption, the parallel trends assumption,
or the estimation method. That all placebo estimates are close to zero
is therefore strong evidence of the plausibility of these assumptions
and our estimation strategy overall.

\FloatBarrier

\section{Redistricting Commissions: Counterfactual Policy
Analysis}\label{redistricting-commissions-counterfactual-policy-analysis}

What if redistricting reforms were adopted nationwide? Redistricting
commissions are the most commonly adopted map-drawing reform. A total of
15 states, including Michigan, New Jersey, and Arizona, currently use
some form of commission. Advocates often argue that commissions can lead
to fairer redistricting plans by further removing legislators from the
process \citep[e.g.,][]{ocp}.\footnote{See also
  \href{https://www.peoplenotpoliticiansoregon.com/}{People Not
  Politicians Oregon} or Ohio's
  \href{https://citizensnotpoliticians.org/}{Citizens Not Politicians}.}

In this section, we conduct a counterfactual policy analysis of the
nationwide adoption of redistricting commissions. Though hypothetical,
counterfactual policy analyses are useful tools for evaluating
institutional reform proposals \citep{cervas2019presidential}. Our
approach is well suited for this analysis, in part because commissions
can vary drastically in the extent to which they limit the influence of
partisan actors.

\subsection{Analysis procedure}\label{analysis-procedure}

To conduct the analysis, we apply fitted causal regression models based
on the party-aware treatment to predict electoral outcomes under a
series of hypothetical scenarios, in which every state adopts a
redistricting commission with particular institutional powers, but
partisan control of state institutions remains unchanged. Although each
of the 15 states uses a redistricting commission with different
structures and rules, our model is able to characterize these different
commissions in terms of their partisan leeway and use these treatment
values to predict electoral outcomes under a given commission structure.

We caution that any counterfactual policy analysis of this type involves
substantial extrapolation from the observed data, and its results should
therefore be interpreted as exploratory rather than confirmatory. While
the regression models do allow for effect heterogeneity, the critical
sufficiency assumption rules out unmodeled heterogeneity that may in
practice lead to different reform effects.

We study three types of commission structures currently enacted in
several states: (1) a New York-style commission with a nonpartisan map
drawer and several partisan veto points; (2) an Ohio-style reform with
legislature-drawn map and several partisan and bipartisan veto points
and stalemate procedures, including a bipartisan commission stalemate
procedure; (3) and a Michigan-style reform, with a nonpartisan
commission, no partisan veto points, and the potential for court
review.\footnote{Although states could change their redistricting
  processes in the future, we use shorthands like ``Michigan-style'' to
  refer to the current (as of 2024) institutional design of each state's
  commission structure.} Though each of these states has a commission of
some kind, they differ in important ways---for example, Michigan's
elimination of partisan veto points removes the ability of partisan
actors to veto unfavorable plans.

\begin{figure}[t]

\centering{

\pandocbounded{\includegraphics[keepaspectratio]{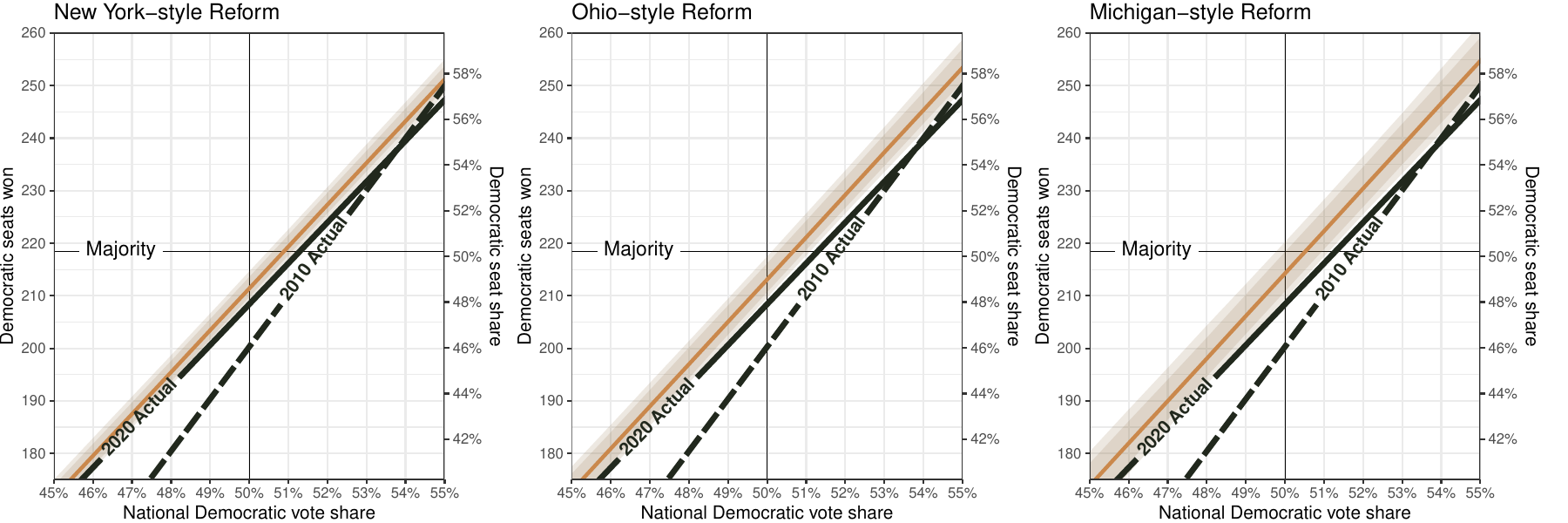}}

}

\caption{\label{fig-reform}\textbf{Commissions Can Reduce Partisan
Bias}. The figures show three predicted seats-votes curves if all US
states adopted new redistricting institutions with: (1) a New York-style
commission with a nonpartisan map drawer and several partisan veto
points; (2) an Ohio-style legislature-drawn map and several partisan and
bipartisan veto points; and (3) a Michigan-style reform, with a
nonpartisan commission, no partisan veto points, and the potential for
court review. Hypothetical commission structures are plotted as orange
lines (with 80\% and 95\% credible intervals), with reference lines for
actual plans for both 2020 and 2010 in black.}

\end{figure}%

We visualize the impact of each reform structure using a seats-votes
curve. Seats-votes curves are often used to study a measure of partisan
bias, operationalized as the difference in the share of seats and votes
at a given point on the curve
\citep{tufte1973pb, king1987democratic, katz2020}. A seats-votes curve
for which a 50\% vote share translates exactly to a 50\% seat share
represents the baseline for partisan bias. We approximate the
seats-votes curve around 50\% vote share as a line, and estimate the
effect of reforms on its slope and intercept separately. Up to a linear
transformation, the slope corresponds to responsiveness and the
intercept to seat share. These two outcome variables are differenced
using the simulation sets, and a regression model is fit to these
differenced outcomes, as discussed in Section~\ref{sec-estimation}.

\subsection{Findings}\label{findings}

We find that nationwide adoption of commissions would generally reduce
partisan bias, and that reforms that place additional restrictions on
partisan actors (as in Michigan) are generally more effective.
Figure~\ref{fig-reform} shows a seats-votes curve for each
counterfactual commission scenario (the orange lines), which display the
predicted number of total Democratic seats in the US House (y-axis) for
a given national Democratic vote share (x-axis).
Figure~\ref{fig-reform-by-state} in the appendix presents the estimated
state-level effects, which show that the reform effects in states with
opposite partisan control (e.g., Texas versus Illinois) may, in part,
cancel out. The model predicts that all commission structures would
reduce the existing Republican advantage in the House driven by both the
inefficient geographic distribution of Democratic voters
\citep{chen_cottrell_2016, chen_rodden_2015} and net partisan
manipulation favoring Republicans \citep{kenny2023widespread}. We also
compare these estimated relationships to the actual seats-votes curves
for 2010 and 2020 (the black lines). We note that the lines for 2010 and
2020 intersect the 50\% vote line below the 50\% seat line, since a
combination of the underlying geographic distribution of Democratic
votes and gerrymandering efforts nationwide disadvantages the Democratic
party.

We find that New York- and Ohio-style reforms produce relatively
moderate improvements in partisan bias and responsiveness. The point
estimates for the number of Democratic seats gained under these reforms
is 3.3 for New York-style reforms, with 95\% credible interval (0.75,
6.16), and 5.1 for Ohio-style reforms (1.3, 9.3). Responsiveness for New
York- and Ohio-style reforms is also similar as for every 1pp increase
in vote share, Democrats gain an estimated 8 seats with New York-style
reforms, and 8.1 seats with Ohio-style reforms. These responsiveness
values are only slightly higher than the actual responsiveness following
2020 redistricting. Ohio-style reforms appear more effective overall,
though the substantive differences are small; we are more than 99\%
confident that Ohio-style reforms produce a larger effect on partisan
bias than New York-style reforms and 88\% confident that Ohio-style
reforms increase responsiveness more than New York-style reforms do.

Further, we find that Michigan-style reforms have the greatest effects
in both responsiveness and Democratic seats, since partisan actors are
the most constrained by the presence of a nonpartisan commission, no
partisan veto points, and the potential for court review. We estimate
that the number of Democratic seats gained under Michigan-style reforms
is 6.2 with 95\% credible interval (1.1, 11.9), and that for every 1pp
increase in vote share, Democrats gain an estimated 8.1 seats (7.5,
8.7), an increase of 0.28 over the baseline. We are 83\% confident that
Michigan-style reforms have a larger effect than Ohio reforms, and 98\%
confident in a larger effect than New York-style reforms. We are also
82\% confident that Michigan-style reforms increase responsiveness.

All three proposed reforms reduce the deviation from partisan symmetry
by increasing net Democratic seats. However, greater effects are
produced when, as in Michigan-style reforms, multiple nodes of the
redistricting game constrain leeway by ensuring partisan nodes are
nonpartisan, and do not precede partisan vetoes. Meanwhile, examples of
New York-style and Ohio-style reforms demonstrate how constraining
partisan actors at different nodes, through different reforms, may
produce substantively similar effects.

\section{Conclusion}\label{conclusion}

In this paper, we propose a methodology for estimating causal effects of
complex institutional reforms and apply it to study the impacts of
redistricting reforms in the United States. Although redistricting
reforms differ across states in important procedural details, we show
how to obtain a theoretically informed univariate summary measure of
such reforms through a game-theoretic model. We then combine the results
of this formal model with a standard causal inference research design
strategy to obtain credible causal effect estimates even with a limited
sample size. We find that redistricting reforms reduce the partisan bias
of enacted plans by constraining the leeway of partisan actors.

Our methodological approach enables us to go beyond the estimation of
causal effects. Specifically, we conduct a counterfactual policy
analysis to infer the consequences of adopting a series of different
redistricting reforms nationwide. We find that adopting redistricting
commissions generally reduces the current Republican advantage, and this
reduction is substantially larger when the reforms place greater
restrictions on partisan actors. For example, a Michigan-style reform
(which combines a nonpartisan commission with no partisan veto points)
would yield a greater pro-Democratic effect than commissions adopted in
Ohio and New York (which maintain the potential for some partisan
control). While we apply this approach to three commission reforms, the
same model can be applied to other reforms of interest. Future research
should also apply our methodology to other redistricting problems, such
as the one for state legislatures.

The key idea behind our methodology is also applicable to studies of
other complex institutions. For example, most scholars analyze
democratic institutions by using the Polity score, which is a 21-point
scale index based on six ``component variables'' concerning executive
recruitment, executive constraints, and political participation. Instead
of summing these scores to obtain the overall Polity score, our proposed
methodology suggests that researchers consider developing a more
theoretically informed measure of democracy, based directly on their
substantive questions of interest. We believe that such an approach
preserves critical aspects of institutional complexity while improving
the credibility of causal analysis.

\hypertarget{refs}{}

\begin{CSLReferences}{0}{0}\end{CSLReferences}

\appendix

\renewcommand\thefigure{\thesection.\arabic{figure}}

\renewcommand\thetable{\thesection.\arabic{table}}

\setcounter{figure}{0}

\section{Details of institutional coding}\label{sec-app-coding}

This appendix details the coding of each state's redistricting
processes. The actual coded variables for each state in 2010 and 2020
are shown in Table~\ref{tbl-coding}.

\subsection{Summary of encoding}\label{summary-of-encoding}

Below, we provide a simplified summary of the coding, prior to the full
details of the coding scheme.

\subsection{Variable coding}\label{variable-coding}

\subparagraph{Drawer.}\label{drawer.}

What kind of body draws the maps?

\begin{itemize}
\tightlist
\item
  \textbf{Legislature}: The state legislature directly draws the maps.
  \emph{Example:} Florida 2020.
\item
  \textbf{Commission}: Embodies any type of commission, including
  independent commission, bipartisan commission, or a legislative
  commission. \emph{Example:} California 2020 had an independent
  commission.
\item
  \textbf{N/A}: There is no redistricting. \emph{Example:} Wyoming 2020
  only has one congressional district and so does not redistrict.
\end{itemize}

\subparagraph{Drawer Control.}\label{drawer-control.}

Who effectively controls the map-drawing body?

\begin{itemize}
\tightlist
\item
  \textbf{Democrats}. \emph{Example:} Nevada 2020 had Democratic
  majorities in both chambers of the state legislature.
\item
  \textbf{Republicans}. \emph{Example:} North Carolina 2020 had
  Republican majorities in both chambers of the state legislature.
\item
  \textbf{Split}: The map is drawn by the legislature, and partisan
  control of the legislature is split across both chambers.
  \emph{Example:} Minnesota 2020 had a Democratic majority in the state
  house and a Republican majority in the state senate, so the
  legislature was under split partisan control.
\item
  \textbf{Nonpartisans}: The map-drawing body is composed of people
  chosen through nonpartisan or explicitly bipartisan means.
  \emph{Example:} California 2020 had an independent commission.
\item
  \textbf{N/A}
\end{itemize}

\subparagraph{Veto 1.}\label{veto-1.}

Who has veto power over the proposed maps?

\begin{itemize}
\tightlist
\item
  \textbf{Legislature}: The legislature can choose to adopt or
  substitute their own maps for those presented to it by the initial map
  drawers. \emph{Example:} Iowa 2020. The legislature can veto the
  commission's maps.
\item
  \textbf{Governor}. \emph{Example:} Alabama 2020.
\item
  \textbf{Voters}: Maps can be subject to public referendum.
  \emph{Example:} California 2020.
\item
  \textbf{N/A}: No body has veto authority, or the map drawer has the
  votes/ability to override vetoes. \emph{Example:} Kentucky 2020.
  Though the governor has statutory authority to veto maps from the
  legislature, the legislature had a Republican supermajority that
  overrode the Democratic governor's veto.
\end{itemize}

\subparagraph{Veto 1 Control.}\label{veto-1-control.}

Who controls the body with veto 2 power?

\begin{itemize}
\tightlist
\item
  \textbf{Democrats}. \emph{Example:} Illinois 2020. The Democratic
  governor can veto the legislature's plan.
\item
  \textbf{Republicans}. \emph{Example:} Georgia 2020. The Republican
  governor can veto the legislature's plan.
\item
  \textbf{Split}: The veto power is held by the legislature, which is
  split in its partisan control across both of its chambers.
  \emph{Example:} Washington 2010. The legislature has veto power over
  the commission's maps, but the legislature was under split partisan
  control.
\item
  \textbf{N/A}: No body has veto authority, or the map drawer has the
  votes/ability to override vetoes. \emph{Example:} Tennessee 2020. The
  legislature has a Republican supermajority, and though the Republican
  governor has the statutory authority to veto maps, a supermajority
  renders this power effectively moot.
\end{itemize}

\subparagraph{Veto 2.}\label{veto-2.}

Who else has veto power over the proposed maps?

\begin{itemize}
\tightlist
\item
  \textbf{Governor}. \emph{Example:} Iowa 2020. After the legislature
  receives maps from the legislature and approves a set of maps, the
  governor can veto the redistricting bill.
\item
  \textbf{N/A}
\end{itemize}

\subparagraph{Veto 2 Control.}\label{veto-2-control.}

Who controls the body with veto 2 power?

\begin{itemize}
\tightlist
\item
  \textbf{Democrats}. \emph{Example:} Maine 2020. The Democratic
  governor can veto the plan from the commission which was approved by
  the legislator.
\item
  \textbf{Republicans}. \emph{Example:} Iowa 2020. The Republican
  governor can veto the plan from the commission which was approved by
  the legislator.
\item
  \textbf{N/A}: No body has veto 2 authority, or the map drawer has the
  votes/ability to override vetoes. \emph{Example:} Utah 2020. The
  legislature (which has veto power over the commission) has a
  Republican supermajority, and though the Republican governor has the
  statutory authority to veto maps, a supermajority renders this power
  effectively moot.
\end{itemize}

\subparagraph{Court Review.}\label{court-review.}

Is there a legal avenue for map challenges to partisan gerrymandering in
state court?
\citep{rudensky2023status, wang2019laboratories, douglas2014right}

\begin{itemize}
\tightlist
\item
  \textbf{Yes}: Clear enabling provisions in state law or constitution
  OR a legal challenge brought at any point in the past which was not
  dismissed. \emph{Example:} New Mexico 2020. A state court accepted
  (but ultimately denied) a lawsuit from the New Mexico GOP challenging
  Congressional lines as a partisan gerrymander.
\item
  \textbf{Maybe}: Possible enabling provisions in state law or
  constitution AND no legal challenge brought at any point.
  \emph{Example:} South Carolina 2020.
\item
  \textbf{No}: No enabling provisions in state law or constitution; or
  legal challenge clearly dismissed. \emph{Example:} Kansas 2020. The
  State Supreme Court ruled that the use of partisan factors in
  redistricting is constitutionally permissible.
\item
  \textbf{N/A}
\end{itemize}

Note that court review is a hidden variable that is not clear until a
legal challenge is actually brought forth and either succeeds or is
rejected. Indeed, even the specter of litigation may influence how other
actors in redistricting approach the drawing of maps. When legal
challenges were brought forth for the first time in a state in the 2020
redistricting cycle and the relevant laws under which the court ruled
were in place in the 2010 redistricting cycle, we retroactively apply
the permissibility of court review to the 2010 cycle as well under the
assumption that had a challenge been brought in 2010, it would have
proceeded the same way.

\subparagraph{Court Control.}\label{court-control.}

Which party controls the state supreme court (or equivalent)?
\citep{parker2023polarization}

\begin{itemize}
\tightlist
\item
  \textbf{Democrats}. \emph{Example:} Illinois 2020.
\item
  \textbf{Republicans}. \emph{Example:} Indiana 2020.
\item
  \textbf{Split}: Equally divided court during the window for which
  challenges could have been brought for 2012 and 2022 maps.
  \emph{Example:} Nevada 2020.
\item
  \textbf{Nonpartisans}: Supreme court is truly nonpartisan, not just
  having nonpartisan elections. \emph{Example:} South Carolina 2020.
\item
  \textbf{N/A}
\end{itemize}

\subparagraph{Stalemate 1.}\label{stalemate-1.}

Who decides the outcome if the initial map drawer has a stalemate?

\begin{itemize}
\tightlist
\item
  \textbf{Court}: Court has statutory power to intervene in the event of
  a stalemate. \emph{Example:} New Jersey 2020.
\item
  \textbf{Commission}. \emph{Example:} Oregon 2020.
\item
  Commission Staff. \emph{Example:} Colorado 2020.
\item
  \textbf{Unclear}: no clear statutory or constitutional provision.
  \emph{Example:} Louisiana 2020.
\item
  \textbf{N/A}
\end{itemize}

\subparagraph{Stalemate 1 Control.}\label{stalemate-1-control.}

Who controls the initial stalemate outcome?

\begin{itemize}
\tightlist
\item
  \textbf{Democrats}. \emph{Example:} Maine 2020. The Maine Supreme
  Court, which can intervene in stalemates, is controlled by Democrats.
\item
  \textbf{Republicans}. \emph{Example:} Indiana 2020. The Indiana
  Supreme Court, which can intervene in stalemates, is controlled by
  Republicans.
\item
  \textbf{Split}: Stalemates are broken by a body that must have
  bipartisan support. \emph{Example:} Ohio 2020. Stalemates are broken
  by a backup commission with the support of at least two members from
  each party (see exceptions below.)
\item
  \textbf{Nonpartisans}: includes any type of commission selected in a
  nonpartisan or explicitly bipartisan way. \emph{Example:} Oregon 2020.
\item
  \textbf{N/A}
\end{itemize}

\subparagraph{Stalemate 2.}\label{stalemate-2.}

Who decides the outcome if the first stalemate-breaking body itself has
a stalemate?

\begin{itemize}
\tightlist
\item
  \textbf{Court}: Court has statutory power to intervene in the event of
  a stalemate. \emph{Example:} Connecticut 2020.
\item
  \textbf{Legislature}. \emph{Example:} Ohio 2020.
\item
  \textbf{Unclear}. \emph{Example:} Minnesota 2020.
\item
  \textbf{N/A}
\end{itemize}

\subparagraph{Stalemate 2 Control.}\label{stalemate-2-control.}

Who controls the second stalemate outcome?

\begin{itemize}
\tightlist
\item
  \textbf{Democrats}. \emph{Example:} Connecticut 2020.
\item
  \textbf{Republicans}. \emph{Example:} Ohio 2020.
\item
  \textbf{N/A}
\end{itemize}

\subparagraph{Final Drawer.}\label{final-drawer.}

Who drew the map in place for the 2012 or 2022 election cycle?

\begin{itemize}
\tightlist
\item
  \textbf{Legislature}: Includes re-drawing a remedial map after court
  challenges. \emph{Example:} South Carolina 2020.
\item
  \textbf{Commission}: Includes re-drawing a remedial map after court
  challenges. \emph{Example:} Washington 2020.
\item
  \textbf{Governor}. No example available.
\item
  Court master: Court ordered a special master to draw the final plan.
  \emph{Example:} Pennsylvania 2020.
\item
  Court D remedy: Court picked a remedial map from a
  Democratic-affiliated plaintiff or intervenor. \emph{Example:}
  Colorado 2020.
\item
  Court R remedy: Court picked a remedial map from a
  Republican-affiliated plaintiff or intervenor. No example available.
\item
  \textbf{N/A}
\end{itemize}

\subparagraph{Preclearance.}\label{preclearance.}

Was any part of the state subject to DOJ preclearance in 2010 (prior to
Shelby County v. Holder)? \citep{doj2013}

\begin{itemize}
\tightlist
\item
  \textbf{Yes}. \emph{Example:} Georgia 2010.
\item
  \textbf{No}. \emph{Example:} Massachusetts 2010.
\end{itemize}

\subsection{Notes on coding}\label{notes-on-coding}

Most criteria are straightforward, and the information for coding can
easily be found in public data for each state. Furthermore, most states
have similar redistricting policies that are easily classified under the
categories we defined earlier. However, there are unique exceptions that
we will treat below. In particular, we coded the below issues as
follows.

\subparagraph{Veto overrides.}\label{veto-overrides.}

\begin{itemize}
\tightlist
\item
  In Kansas 2020, Kentucky 2020, Louisiana 2020, Maryland 2010, Maryland
  2020, Massachusetts 2010, Massachusetts 2020, Montana 2020, New York
  2020, Rhode Island 2010, and Rhode Island 2020, the legislature had
  the necessary supermajority to override any vetoes, so we coded the
  governor's veto as N/A.
\end{itemize}

\subparagraph{Unique commission
compositions.}\label{unique-commission-compositions.}

\begin{itemize}
\tightlist
\item
  In Idaho in 2010 and 2020, commissions are re-formed upon stalemate.
  Members are picked by both parties in a way that forms a 3-3 even
  split from both parties, except that the State Supreme Court picks the
  members of the commission if the parties don't decide in time. We code
  the control of this commission as having a partisan split.
\item
  In Indiana in 2010 and 2020, stalemates are broken by a five-member
  backup commission composed of the majority leader from each house, the
  chair of the redistricting committee from each house, and a state
  legislator appointed by the governor. Since both houses and the
  governorship were controlled by Republicans, we coded this commission
  as being controlled by the Republican party.
\item
  In Michigan 2020, if the commission does not adopt a final plan, a
  plan is randomly chosen. We code this as a stalemate 1 breaker by the
  nonpartisan commission.
\item
  In Ohio 2020, if the legislature fails to redistrict by bipartisan
  majorities, the work falls to a commission that must approve a map by
  a bipartisan vote. The commission is composed of appointees by state
  executive officers and legislative parties. If the commission cannot
  do so, it can pass a map by a majority vote without bipartisan
  support, but the map only lasts two election cycles instead. This is
  encoded as a draw by a split state legislature with the first
  stalemate broken by a nonpartisan commission and the second stalemate
  broken by Republicans, since Republicans control all state executive
  offices and have a functional majority on the commission.
\end{itemize}

\subparagraph{Courts.}\label{courts.}

\begin{itemize}
\tightlist
\item
  In Kansas in 2010, the legislature failed to redistrict, and a federal
  court stepped in to draw the maps instead. But since this was not done
  under any statutory power and only done once, we consider this an
  exception to the normal redistricting process in Kansas rather than a
  procedurally institutionalized stalemate breaker. Similarly in
  Wisconsin in 2020, the State Supreme Court approved ``least change''
  maps drawn by the governor after a stalemate, but we coded the
  stalemate as unclear for the same reasons.
\item
  On the other hand, in Minnesota in 2010 and 2020 we have coded the
  court as a stalemate breaker because it has consistently intervened in
  previous stalemates.
\item
  In Minnesota in 2020, a court-appointed panel drew the final map. We
  code this as equivalent to that of a court master.
\end{itemize}

\subparagraph{Relevant history that did not affect
encoding.}\label{relevant-history-that-did-not-affect-encoding.}

\begin{itemize}
\tightlist
\item
  In North Carolina, the 2012 map was ultimately struck down but only
  after the election. Furthermore, the 2022 map was an interim plan only
  used for the 2022 election.
\item
  In Virginia in 2010, the state legislature initially deadlocked, but
  drew a map in time after state legislative elections in 2011. Thus,
  the stalemate procedure for 2010 was coded as unclear.
\end{itemize}

\begin{table}

\caption{\label{tbl-coding}Institutional coding for all states.}

\centering{

\pandocbounded{\includegraphics[keepaspectratio]{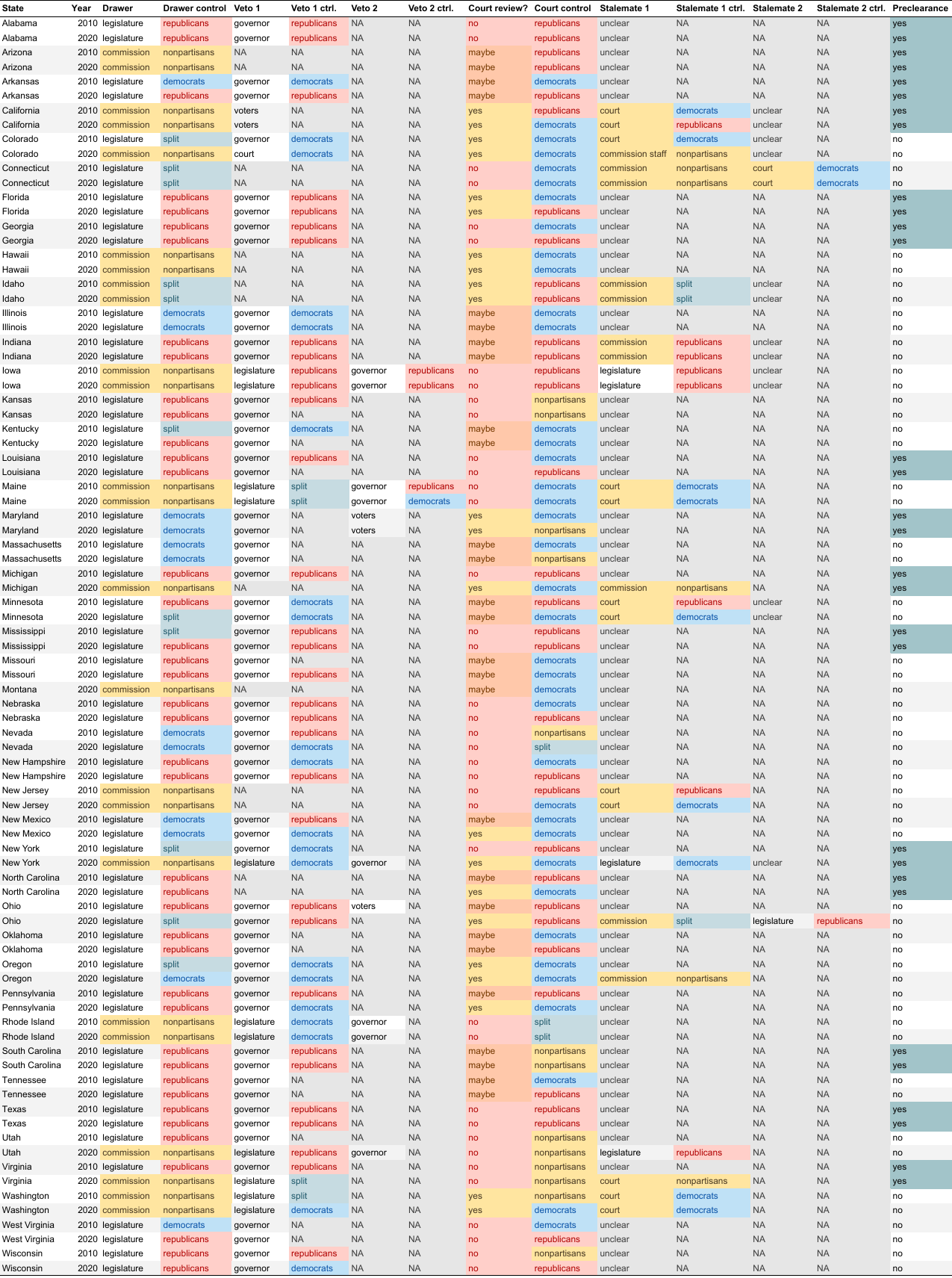}}

}

\end{table}%

\FloatBarrier

\section{Redistricting Game Specification}\label{sec-app-game}

The game tree is specified in Figure~\ref{fig-game} and is described in
Section~\ref{sec-game}. Here we detail the parametric specification of
the moves by nature, starting with the post-enactment court review
process. The parameters in the model all receive probabilistic priors,
rather than being fixed to specific values. These priors place most of
their mass across a broad range of plausible values, based on our
experience and observation of past redistricting processes.

\subsection{Post-enactment processes}\label{post-enactment-processes}

We decompose the court challenge process into five components, as
follows:

\begin{enumerate}
\def\labelenumi{\arabic{enumi}.}
\item
  The probability a legal challenge is possible, which depends on the
  data coding, is specified as \[
  \texttt{pr\_chal\_poss} = \begin{cases*}
   \texttt{chal\_poss\_conf} & if Court Review is coded as ``Yes'' \\
   \texttt{chal\_poss\_maybe} & if Court Review is coded as ``Maybe'' \\
   1-\texttt{chal\_poss\_conf} & if Court Review is coded as ``No''
  \end{cases*}
  \] The prior on \texttt{chal\_poss\_conf} is \(\Beta(19, 1)\) and the
  prior on \(\texttt{chal\_poss\_maybe}\) is \(\Beta(6, 14)\). The
  effect of this setup is to follow the treatment coding closely but
  allow for some uncertainty in the possibility of court review, rather
  than, e.g., precluding any possibility of court challenge in a state
  because it was coded as ``No.'' State courts may change their
  interpretations of state laws and constitutions in any given case.
\item
  The probability a challenge is made when one is possible, which
  depends on the extremity of the partisan bias of the first enacted
  plan, \(x\). Let \(F(x)=\pi^{-1}\arctan(x)+\half\) be the CDF of a
  standard Cauchy distribution. The specification of this probability is
  then \begin{align*}
  \texttt{pr\_chal\_if\_poss}(x) &= F(a+bx^2) \\
  a &= F^{-1}(\texttt{chal\_prob\_bias}_0) \\
  b &= (F^{-1}(\texttt{chal\_prob\_bias}_2) - F^{-1}(\texttt{chal\_prob\_bias}_0)) / 2^2.
  \end{align*} The general form of \(F(a+bx^2)\) means that the
  probability of a challenge will be U-shaped, asymptotically
  approaching 1 as the absolute partisan bias increases. For
  interpretability, the scale and shift coefficients are parameterized
  in terms of the probability of a challenge when the first enacted plan
  has bias 0 and the probability of a challenge when the first enacted
  plan has bias 2. These parameters receive \(\Beta(4, 16)\) and
  \(\Beta(17, 3)\) priors, respectively. Figure~\ref{fig-chal-if-poss}
  shows 200 draws of the induced prior on
  \(\texttt{pr\_chal\_if\_poss}\).

  \begin{figure}

  \centering{

  \pandocbounded{\includegraphics[keepaspectratio]{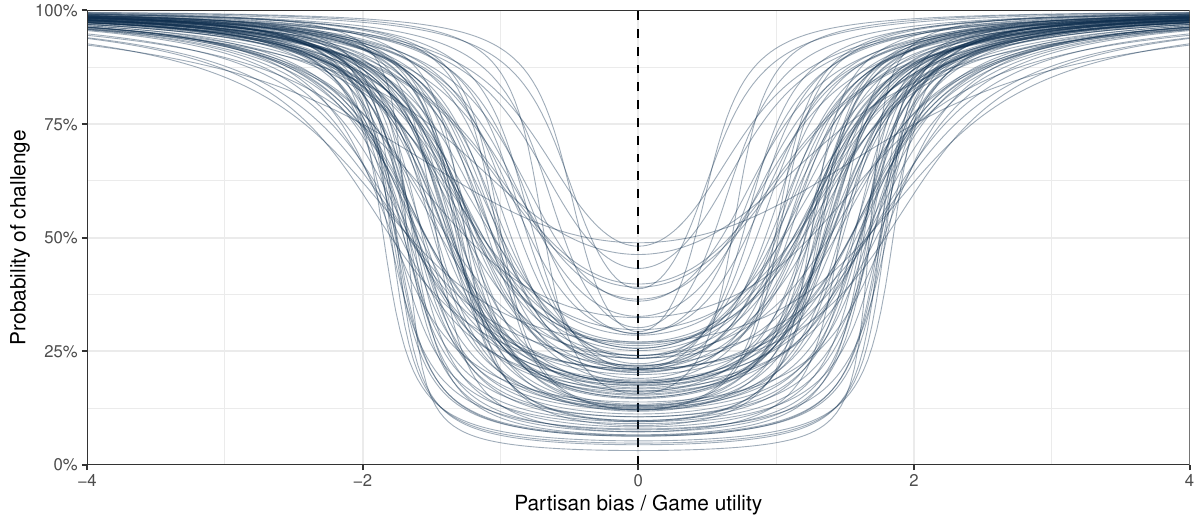}}

  }

  \caption{\label{fig-chal-if-poss}200 draws from the prior on the
  \(\texttt{pr\_chal\_if\_poss}(x)\) curve.}

  \end{figure}%
\item
  The probability a court sides with plaintiffs challenging the first
  enacted plan, i.e., that the court intervenes, which depends on the
  partisan bias of the first enacted plan as well as the partisan makeup
  of the state supreme court. We assume that courts with a partisan
  majority are more likely to intervene and find in favor of challengers
  when the partisan bias of the first enacted plan is opposite in sign
  to the court's partisan lean. Let
  \(g(x; a) = x^2(0.7 + ac\cdot x + (0.2x)^2)\), with
  \(c=\sqrt{4(2\cdot 0.7)(12\cdot 0.2^2)}/6\), which is a convex quartic
  with asymmetry controlled by a parameter \(a\); the value \(c\) is set
  so that the discriminant of the second derivative is positive, i.e.,
  that the quartic is indeed convex. Then the intervention probability
  is \[
  \texttt{pr\_intervene}(x) = \begin{cases*}
   bc\cdot g(-x;\texttt{interv\_asym}) + ac & if Court Control is coded as ``Democrats'' \\
   bc\cdot g(-x;\texttt{interv\_asym}) + ac & if Court Control is coded as ``Republicans'' \\
   ac + bcx^2 & otherwise
  \end{cases*},
  \] with \begin{align*}
  a &= F^{-1}(\texttt{interv\_prob\_bias}_0) \\
  b &= (F^{-1}(\texttt{interv\_prob\_bias}_2) - F^{-1}(\texttt{interv\_prob\_bias}_0)) / 2^2 \\
  c &= \texttt{interv\_prob\_max}
  \end{align*} The coefficients on the quartic \(g\) were adjusted so
  that the average value of \(g\) across the interval \([-4, 4]\) was
  close to the average value of \(x^2\) across the same interval. Like
  the challenge probability, the general form of \(F(ac+bcx^2)\) means
  that the probability of intervention will be U-shaped, asymptotically
  approaching \(c=\texttt{inverv\_prob\_max}\) as the absolute partisan
  bias increases. The use of \(g\) means that the probability of
  intervention will be asymmetric around 0 for partisan courts. The
  maximum \texttt{inverv\_prob\_max} has a \(\Beta(18, 2)\) prior; the
  asymmetry parameter has a \(\Beta(4, 1.5)\) prior; the relative
  probability of intervention for a neutral plan,
  \(\texttt{interv\_prob\_bias}_0\), receives a \(\Beta(4, 16)\) prior;
  and the relative probability of intervention for a plan with bias 2,
  \(\texttt{interv\_prob\_bias}_2\), receives a \(\Beta(18, 2)\) prior.
  200 draws of the induced prior on \texttt{pr\_intervene} for each type
  of court control are shown in Figure~\ref{fig-intervene}.

  \begin{figure}

  \centering{

  \pandocbounded{\includegraphics[keepaspectratio]{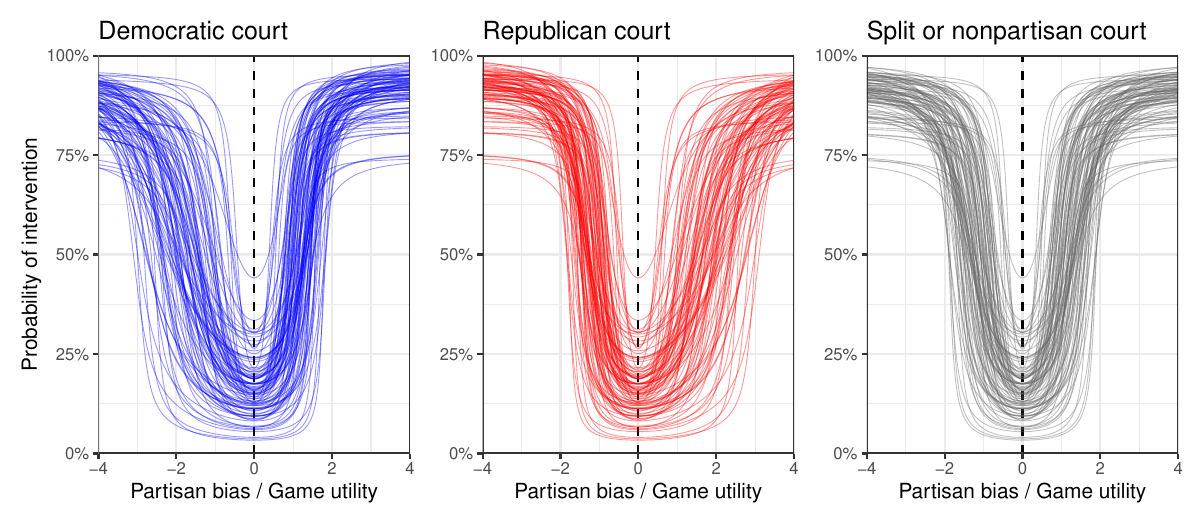}}

  }

  \caption{\label{fig-intervene}200 draws from the prior on the
  \(\texttt{pr\_intervene}(x)\) curve for each type of partisan court
  control.}

  \end{figure}%
\item
  The expected remedy a court orders when siding with plaintiffs, which
  depends on the partisan bias of the first enacted plan as well as the
  partisan makeup of the state supreme court, and is specified as
  \begin{align*}
  \texttt{court\_outcome}(x) &= \frac{\texttt{out\_nonp\_bias}_2}{\arctan(2/2)}\cdot \arctan(x/2) +
    a\cdot \texttt{out\_nonp\_part\_adv} \\
  a &= \begin{cases*}
   -1 & if Court Control is coded as ``Democrats'' \\
   1 & if Court Control is coded as ``Republicans'' \\
   0 & otherwise
  \end{cases*}.
  \end{align*} The \texttt{out\_nonp\_bias} parameter governs the
  partisan bias of a nonpartisan court remedy when the partisan bias of
  the first enacted plan is 2 (hence the subscript); it receives a
  folded Normal prior with mean 0 and standard deviation \(\half\),
  i.e., we simulate from \(\Norm(0, \half)\) and then take absolute
  values. The \texttt{out\_nonp\_part\_adv} parameter controls the
  additional partisan lean of the remedy towards the party that controls
  the court, where applicable; it receives a folded Normal prior with
  mean 0 and standard deviation \(0.4\). Figure~\ref{fig-court-outcome}
  shows 200 draws from the induced prior on \texttt{court\_outcome} for
  each type of court control.

  \begin{figure}

  \centering{

  \pandocbounded{\includegraphics[keepaspectratio]{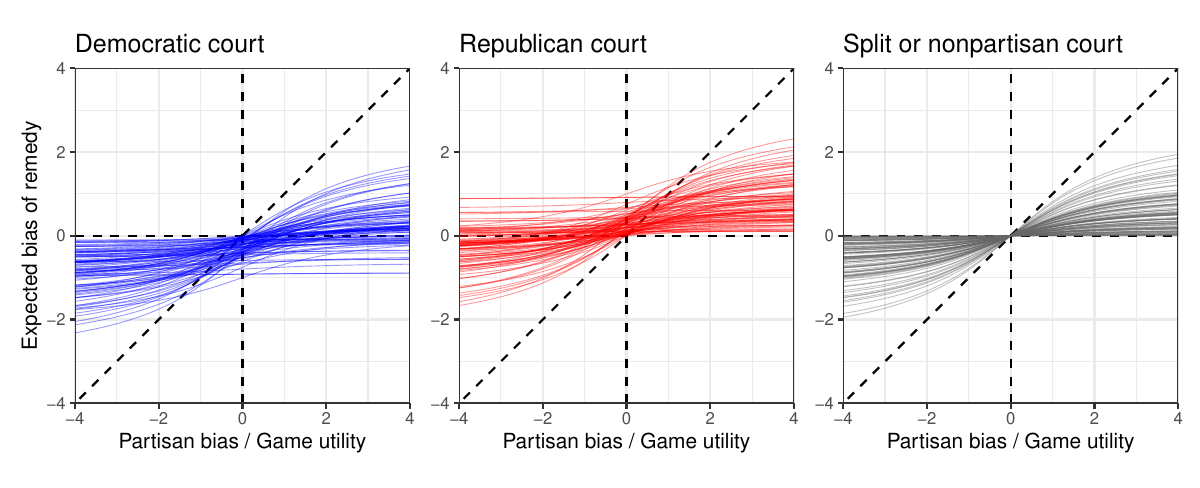}}

  }

  \caption{\label{fig-court-outcome}200 draws from the prior on the
  \(\texttt{court\_outcome}(x)\) curve for each type of partisan court
  control.}

  \end{figure}%

  Notice that, in expectation, \(\texttt{court\_outcome}(x)\) is
  increasing in \(x\). This reflects our experience in following
  partisan gerrymandering litigation in state courts. For example, the
  Ohio state legislative plan was challenged multiple times in court,
  the outcome of each case inched towards a neutral map (according to
  simulations), rather than immediately jumping to a neutral map. In
  this case (as in many), this is because the legislature was given a
  new opportunity to draw a compliant plan; in practice, map-drawers
  tend to revise the existing plan as little as possible to stave off
  further litigation without abandoning the underlying partisan goals.
\item
  The probability and expected effect of a challenge based on the
  federal Voting Rights Act, which depends on whether a state was
  subject to DOJ preclearance pre-\emph{Shelby}, as well as the partisan
  bias of the first enacted plan. The probability of a VRA challenge is
  zero for non-preclearance states; for preclearance states it is
  \begin{align*}
  \texttt{pr\_vra\_chal}(x) &= F(a+bx) \\
  a &= F^{-1}(\texttt{vra\_chal\_prob\_bias}_0) \\
  b &= (F^{-1}(\texttt{vra\_chal\_prob\_bias}_2) - F^{-1}(\texttt{chal\_prob\_bias}_0)) / 2.
  \end{align*} The general form of \(F(a+bx)\) means the challenge
  probability is asymmetrical: challenges are more likely for plans that
  favor Republicans. This follows from the high levels of racially
  polarized voting in the U.S. and specifically the strong preference
  for Democratic candidates by minority groups. The
  \(\texttt{vra\_chal\_prob\_bias}_0\) parameter is the probability of a
  VRA challenge against a plan with no partisan bias; its prior is
  \(\Beta(2, 18)\). Similarly, \(\texttt{vra\_chal\_prob\_bias}_2\), the
  challenge probability for a Republican-favoring plan, has a
  \(\Beta(9, 1)\) prior. The left panel of Figure~\ref{fig-game-vra}
  shows 200 draws from the induced prior on \texttt{pr\_vra\_chal}.

  \begin{figure}

  \centering{

  \pandocbounded{\includegraphics[keepaspectratio]{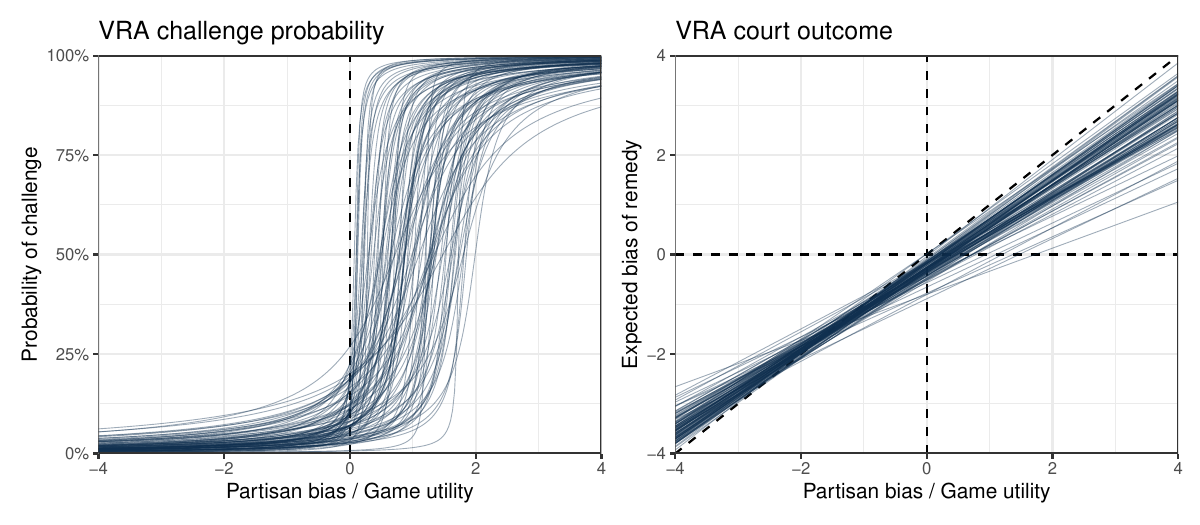}}

  }

  \caption{\label{fig-game-vra}200 draws from the prior on the
  \(\texttt{pr\_vra\_chal}(x)\) curve (left) and the
  \(\texttt{court\_vra\_outcome}(x)\) curve (right).}

  \end{figure}%

  We assume the probability of a successful VRA claim does not depend on
  the partisan bias of the first enacted plan; this parameter is
  \texttt{vra\_interv\_prob} and receives a prior of \(\Beta(4, 1.5)\).
  The court-ordered remedy for a successful VRA challenge is specified
  as \[
   \texttt{court\_vra\_outcome}(x) = \texttt{vra\_out\_slope}(x - \texttt{vra\_out\_breakeven}) +
   \texttt{vra\_out\_breakeven};
   \] the intercept \texttt{vra\_out\_breakeven} has a
  \(\Norm(-1.5, 0.5)\) prior and the slope \texttt{vra\_out\_slope} has
  a \(\Beta(16, 4)\) prior. So in expectation, VRA remedies make
  Republican-favoring plans slightly more Democratic. The right panel of
  Figure~\ref{fig-game-vra} shows 200 draws from the induced prior on
  \texttt{court\_vra\_outcome}.
\end{enumerate}

Putting these five pieces together, we can write the expected outcome of
the post-enactment processes as \begin{align*}
\texttt{exp\_court}(x) &= \texttt{pr\_int\_net}(x)\cdot\texttt{court\_outcome}(x) \\
  &\qquad+ (1-\texttt{pr\_int\_net}(x))\cdot\texttt{pr\_vra\_net}(x)\cdot\texttt{court\_vra\_outcome}(x) \\
  &\qquad+ (1-\texttt{pr\_int\_net}(x))\cdot(1-\texttt{pr\_vra\_net}(x))\cdot x \\
\texttt{pr\_int\_net}(x) &= \texttt{pr\_chal\_poss}\cdot \texttt{pr\_chal\_if\_poss}(x)\cdot \texttt{pr\_intervene}(x) \\
\texttt{pr\_vra\_net}(x) &= \texttt{pr\_vra\_chal}(x)\cdot \texttt{vra\_interv\_prob}
\end{align*}

\begin{figure}

\centering{

\pandocbounded{\includegraphics[keepaspectratio]{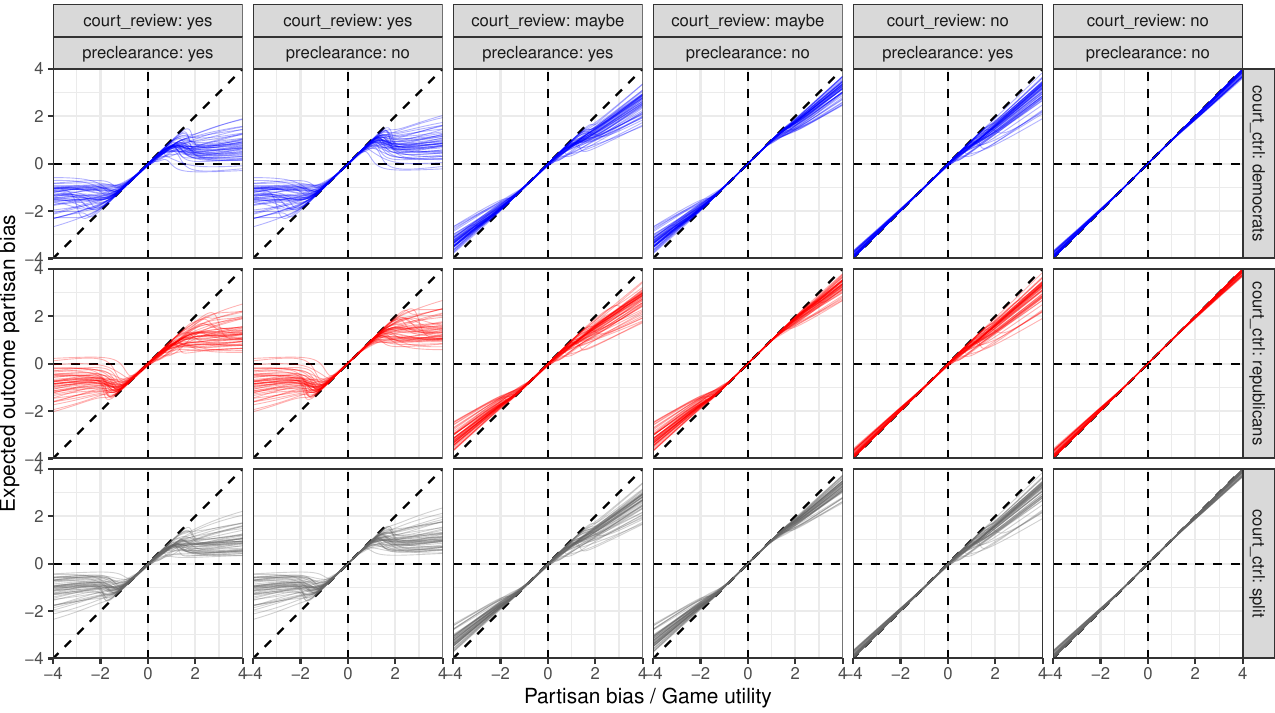}}

}

\caption{\label{fig-exp-court}50 draws from the prior on
\(\texttt{exp\_court}(x)\) for each combination of relevant procedural
variables.}

\end{figure}

Figure~\ref{fig-exp-court} shows 50 draws from the induced prior on
\(\texttt{exp\_court}(x)\) for each combination of court control, court
review, and DOJ preclearance. We can see that the most important factor
in shaping post-enactment outcomes is the presence of court review, as
might be expected. When court review is uncertain or not present, court
processes are expected to only moderately constrain outcomes.

\subsection{Stalemate process}\label{stalemate-process}

When there is no stalemate process specified, when stalemate processes
are exhausted, or when the stalemate process is controlled by neither
party, the stalemate outcome is assumed to be a linear rescaling of the
bias of the plan drawn at the previous stage, \(x\), which acts as a
sort of anchoring point, and the partisan control of the initial map
drawer. The specification is \[
\texttt{stalemate\_default}(x)
= \begin{cases*}
\texttt{stale\_slope}\cdot x - \texttt{out\_nonp\_part\_adv} & if the initial drawer is Democratic \\
\texttt{stale\_slope}\cdot x + \texttt{out\_nonp\_part\_adv} & if the initial drawer is Republican \\
\texttt{stale\_slope}\cdot x & otherwise \\
\end{cases*},
\] with \texttt{out\_nonp\_part\_adv} defined as above, and
\texttt{stale\_slope} receiving a \(\Beta(3, 17)\) prior. This reflects
a status quo bias towards the last-drawn plan.

When stalemates are specified as being resolved by state courts, and the
state court is partisan, the stalemate outcome is \[
\begin{cases*}
\texttt{stale\_slope}\cdot x - \texttt{out\_nonp\_part\_adv} & if Court Control is coded as ``Democrats'' \\
\texttt{stale\_slope}\cdot x + \texttt{out\_nonp\_part\_adv} & if Court Control is coded as ``Republicans''
\end{cases*}
\] given an input partisan bias \(x\).

\subsection{Nonpartisan veto points}\label{nonpartisan-veto-points}

When veto control is coded as NA (other than in the case of a governor
with overridable veto) or a court has a veto, the veto is exercised with
probability \[
\texttt{pr\_veto\_chal}(x) = \texttt{veto\_nonp\_prob\_max}\cdot \texttt{pr\_chal\_if\_poss}(x),
\] with the parameter \texttt{veto\_nonp\_prob\_max} receiving a
\(\Beta(3, 7)\) prior. So these veto players are unlikely to intervene,
but relatively more likely if the proposed plan is more extreme

Split-control veto players are assumed to never exercise their veto, as
are governors that can be overridden (i.e., governors facing a
veto-proof majority in both chambers; see the coding of this variable in
Appendix~\ref{sec-app-coding}).

\subsection{Split-control and nonpartisan map
drawers}\label{split-control-and-nonpartisan-map-drawers}

Map drawers with split partisan control are assumed to stalemate with
probability controlled by parameter \texttt{stale\_split\_prob}, which
has a \(\Beta(3, 5)\) prior. When they don't stalemate, split-control
map drawers are assumed to propose a map with the same partisan bias as
one drawn by a nonpartisan map drawer. While we suspect this probability
could be set to 1 without changing the results, a probability of
compromise reflects a history of compromise plans in state legislatures.

In Round 1, nonpartisan map drawers are assumed to propose a plan with
partisan bias of 0. In Round 2---after one veto has already
occurred---nonpartisan map drawers are assumed to shift the bias of
their proposed plan in the direction of the average partisanship of the
veto players. Specifically, if \(x\) is the proposal from the previous
stage, then the proposal in Round 2 is \begin{align*}
\texttt{exp\_drawer\_r2\_nonp}(x)
&= x + \texttt{veto\_party}\cdot\texttt{veto\_nonp\_shift} \\
\texttt{veto\_party} &= \begin{cases*}
-1 & if both veto players are Democratic \\
1 & if both veto players are Republican \\
-\half & if one veto player is Democratic and the other is nonpartisan or absent \\
\half & if one veto player is Republican and the other is nonpartisan or absent \\
0 & otherwise
\end{cases*}.
\end{align*} The \texttt{veto\_nonp\_shift} parameter controls the
amount of the shift and receives a \(\Norm(0.65, 0.3)\) prior. This
mechanism reflects real-world patterns of repeated map-drawing, such as
when the (Republican-controlled) Iowa legislature rejected the proposed
plan from the nonpartisan advisory commission, then adopted a second
proposal from the same commission that favored Republicans. more.

\subsection{Walk-through of equilibrium calculation for
Alabama}\label{sec-app-ex-al}

In Alabama, the legislature draws congressional districts, subject to a
governor's veto, and there is no mechanism for state court review. Thus,
Alabama's process would be described by the ``Drawer'' and ``Veto 1''
steps only in Round 1 of the Figure~\ref{fig-game} process.

The first move belongs to the Republican party, which controls the state
legislature. It has to pick the amount of partisan bias \(-4<x<4\) in
the plan that it adopts. If this plan is ultimately the final enacted
plan, the Republicans receive utility \(x\) and the Democrats receive
utility \(-x\). The second move also belongs to the Republican party,
which controls the governorship. The party must decide whether or not to
veto the adopted plan; if it does, there is a second round of drawing
and vetoing by the legislature and governor (controlled by Republicans).

If a plan is adopted at either the first or second round, it proceeds to
possible court review. The court may decide to accept a legal challenge,
decide in favor of the plaintiffs, and redraw the map; this choice is
considered a move by nature. If the court review results in a redrawn
map with partisan bias \(x'\), then the Republicans receive utility
\(x'\) and the Democrats receive utility \(-x'\). In Alabama, court
review is generally not allowed on partisan grounds, but challenges
under the federal Voting Rights Act are possible, so there is moderate
probability of the legislature-adopted plan being overturned. If the
legislature's plan is vetoed both times, the process results in a
stalemate. Since there is no enumerated stalemate procedure in Alabama,
courts must step in and redraw district lines to ensure compliance with
federal constitutional ``one person, one vote'' apportionment
requirements. This is also considered a move by nature.

To calculate the equilibrium itself, we numerically solve the game via
backwards induction. This requires up to four levels of nested
optimization. For Alabama, this means that we start with the last
partisan move in the game, which is the Round 2, Veto 1 player, the
Republican governor. She must decide, given a proposed partisan bias
\(x\) from the Round 2 map drawer, whether to veto the plan, in which
case the enacted plan is determined via the court stalemate process, or
to accept the plan and let it become law. In either case, the resulting
plan may be challenged in court.

\begin{figure}[!b]

\centering{

\pandocbounded{\includegraphics[keepaspectratio]{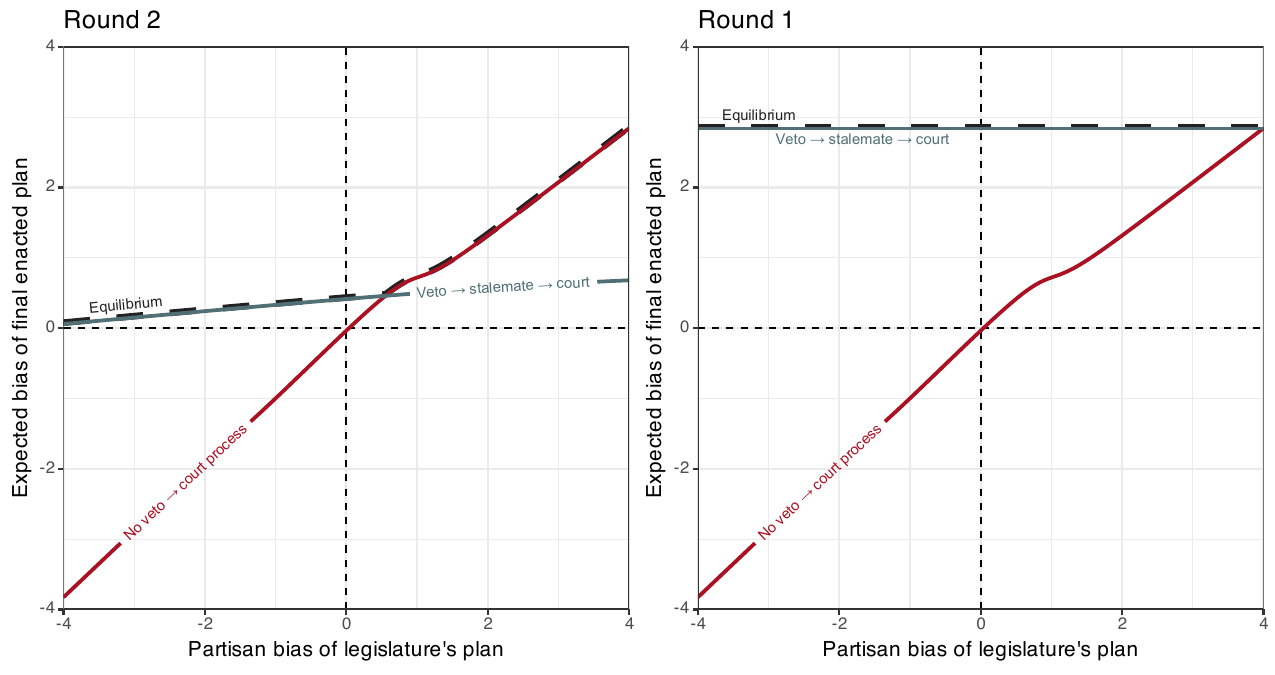}}

}

\caption{\label{fig-al-ex}Expected outcomes for ``veto'' and ``no veto''
moves by Alabama's Republican governor in Round 2 (left) and Round 1
(right), as a function of the partisan bias of the legislature's
proposed plan. The equilibrium outcome at each node is indicated by the
dotted line. The range of biases correspond to the scale introduced in
Section~\ref{sec-game}.}

\end{figure}

The expected outcome of both the stalemate-then-court and court-alone
choices are determined by the parameters governing the moves by nature.
They are shown as a function of the bias of the legislature's proposed
plan in Figure~\ref{fig-al-ex}(a), for a typical set of game parameters.
In this case, because the only realistic avenue for court review in
Alabama is a VRA claim, the expected outcome of a court challenge given
an enacted plan \(x\) is roughly \(x\). The expected bias of a plan
drawn through stalemate procedures under a Republican court is roughly
0.5, though it also increases monotonically in \(x\). Since the governor
wishes to maximize the signed partisan bias, she will veto the plan if
its bias is less than about 0.5, and not veto it otherwise.

Having solved the Round 2, Veto 1 subgame, we move backwards to the
Round 2 drawing subgame. The Republican legislature must decide on the
partisan bias of the plan it proposes, or can choose to stalemate by
failing to adopt a plan. If it stalemates, then the expected outcome
will be given by the green ``Veto'' line in Figure~\ref{fig-al-ex}(a),
and will depend on the partisan bias of the plan originally adopted by
the legislature in Round 1 before being vetoed. If it does not
stalemate, then the expected outcome is determined by the dashed
equilibrium line in Figure~\ref{fig-al-ex}(a). To maximize the partisan
bias, the legislature at Round 2 will draw a map with bias of \(+4\),
leading to expected bias (utility) of 2.84 for this set of game
parameters.

We can then move backwards again and consider the Round 1, Veto 1
subgame. Figure~\ref{fig-al-ex}(b) shows the expected outcomes under
each choice, just as before. In equilibrium, the governor will veto any
plan with bias less than 4; regardless of the governor's decision, the
expected bias is 2.84. Finally, the legislature can make any initial
proposal, since the expected outcome is the same regardless. Thus, the
overall equilibrium is a bias of 2.84. This solution process is
automated across all of the states and is carried out on each of the 100
different draws from the prior on the game's parameters.

\section{Redistricting Simulation Details}\label{sec-app-sims}

This section provides additional details on the redistricting
simulations used in the analyses here.

\subsection{Generating the
simulations}\label{generating-the-simulations}

The 2020 simulations were taken from \citet{50stateSimulations}. As that
paper documents, these simulations were generated through a consistent
workflow based on shared template code. Each set of sampled plans
underwent manual review by two human analysts, which included checks for
statistical quality based on convergence diagnostic, as well as
empirical checks on aspects of the generated plans, such as
demographics, county splits, compactness, etc. These are explained at
greater length below.

The 2010 simulations were generated in the exact same fashion, by making
modifications where appropriate to the 2020 code, and using different
district apportionments and Census data. All of the code for generating
the simulations for both 2010 and 2020 is available at
\url{https://github.com/alarm-redist/fifty-states}. That repository also
includes documentation of the specific constraints and choices made in
sampling plans for each state.

\subsubsection{Statistical quality control
checks}\label{statistical-quality-control-checks}

\begin{itemize}
\item
  Distribution of the final SMC weights. Highly variable or heavy-tailed
  weights may indicate difficulty approximating the target distribution
  and high variance for downstream estimates.
\item
  Plan diversity as measured by the variation of information metric.
  Identifies cases where many of the sampled plans are identical or
  nearly identical, which leads to discreteness artifacts and may
  indicate high variance and a possible lack of representativeness.
\item
  Gelman-Rubin \(\hat R\) statistics for each calculated summary
  statistic (demographics, vote totals, compactness, county splits,
  etc.) These check that fully independent runs of the algorithm produce
  samples from the same distribution. High \(\hat R\) values indicate
  that the algorithm has not converged and that the samples are not
  representative of their target distribution.
\end{itemize}

\subsubsection{Substantive quality control
checks}\label{substantive-quality-control-checks}

The values of the statistics below were also compared to those for the
enacted plan. Significant discrepancies, as judged by the reviewers,
could indicate a mismatch between the

\begin{itemize}
\item
  Population deviation from the ideal population. Most states had a
  limit of 0.5\% deviation; some, like Iowa, had a lower threshold.
\item
  Geographic compactness, as measured by the Polsby--Popper score or a
  graph-theoretic measure counting the number of removed edges. Some
  states used alternative measures of compactness as specified by law.
\item
  County and municipality splits.
\item
  Approximate Voting Rights Act (VRA) compliance. The number of
  districts with minority voting-age population fractions in certain
  ranges (e.g., majority-minority districts) and certain partisan
  ranges. If simulated maps produced fewer majority-minority districts
  with the same partisan characteristics as the enacted plan, then it is
  likely they would not have satisfied the VRA. We deferred judgement as
  to the required number and type of majority-minority districts to the
  enacting bodies.
\end{itemize}

\subsection{Limitations of simulation
algorithms}\label{limitations-of-simulation-algorithms}

Redistricting simulation algorithms generate samples from a particular
target probability distribution. There are two main limitations of
simulation approaches that stem from this fact.

First, sampling from complex, constrained spaces is a difficult
statistical problem and can be done only approximately. Moreover, even
if the sampling were perfect (i.e., we could generate i.i.d. samples
from the target), the use of a finite set of samples to approximate the
target distribution also leads to Monte Carlo errors. The potential risk
of unacceptably high approximation error motivates the statistical
quality control checks described above.

The second limitation is that the target distribution is designed to be
both computationally feasible to sample from and to approximate the
distribution of legal plans that would be drawn by a map drawer without
any partisan information or bias. The default distribution that our
samples are drawn from is described formally in \citet{smc}. It is
uniform on the space of all contiguous districting plans with a certain
population constraint, \emph{after reweighting} according to a certain
graph-theoretic measure of compactness. This target distribution can be
further modified, within limits, based on other features of the
simulation problem. A different choice of target distribution, which
interprets the legal constraints differently, or which makes a different
tradeoff with computational feasibility, would lead to different
samples.

Thus, there is an inherent sensitivity of the results of any simulation
analysis to the choice of target distribution. This is in many ways a
key advantage of simulation algorithms---the target distribution can be
tuned to ask different questions about districting plans in a state. But
for an analysis such as ours, it also raises questions about the
sensitivity of downstream analyses to these choices.

\subsection{Sensitivity of samples and downstream analyses to simulation
choices}\label{sensitivity-of-samples-and-downstream-analyses-to-simulation-choices}

As mentioned above, the default distribution targeted by the simulation
algorithm used here is modified by the inclusion of other constraints.
While some of these additional constraints can be easily described
mathematically and are documented alongside the simulation code, others
have effects on the target that are more difficult to express in closed
form.

The first is the incorporation of county and municipality boundaries, a
constraint that is applied in the vast majority of states. This changes
both the support of the target (it limits the number of splits of
counties or municipalities) as well as the graph-theoretic compactness
measure that appears as a weighting term in the target distribution.
Moreover, in many states, we used a hybrid approach where counties were
protected in most of the state, but inside metropolitan areas (where
many districts might appear in the same county), we used municipality
boundaries instead. This allowed us to trade off county and municipality
splits to better approximate the balance seen in the enacted plans. A
different treatment of these administrative boundaries would lead to a
different target distribution.

The second is a multi-step simulation approach used in several large
states with many districts, such as Texas, Florida, and California. In
these states, the simulation algorithm was first run inside a designated
set of counties, such as those comprising a major metropolitan area.
Rarely would the population of this area exactly divide the ideal
district population, so some population in these areas was left
unassigned. This first-stage simulation was repeated possibly in other
areas of the state as well. Then the algorithm was run a final time to
assign the remaining population to districts.

Such a multi-stage approach is supported by the design of sequential
Monte Carlo algorithms, but it does introduce an implicit constraint on
the target distribution, altering it from the default described in
\citet{smc}. Specifically, this simulation approach allowed for at most
one district to span the border of the first-stage metropolitan areas.
That is, there would be at most one district that took in the unassigned
population from the first stage and connected it to population outside
the first-stage area. Since the first-stage areas were defined by county
borders, which are already protected by the algorithm in a similar way
(i.e., only one district can span a given county border), in some states
the effect of this setup may have been absent or minimal. In other
states, this implicit constraint may have led to larger changes in the
target distribution.

These changes and other decisions were made in an attempt to produce a
sample representative of legal plans that would be drawn by a map drawer
without any partisan information or bias. How might they affect the
analyses conducted in this paper? A key point is that only the mean and
standard deviation of the simulated distribution (rather than the entire
distribution) were used to define the outcome variables in the analyses
here (except the partisan harm outcomes used in the appendices only). In
our experience, and as supported by the supplementary analysis conducted
in Appendix J of \citet{kenny2023widespread}, the first two moments are
less sensitive to alternative simulation choices than other summary
statistics, such as p-values from the tails of the distribution.

Even more importantly, the moments of the simulation distributions were
not themselves the estimand of interest here. Rather, they were used to
define an outcome variable, which was then time-differenced and fed into
a regression model. So any biases or sensitivity in the simulation
distribution that were constant over time would not affect the results.
Moreover, any such sensitivity that is uncorrelated with the explanatory
variables, after differencing, would likewise have no effect. If
sensitivity to simulation setup is considered as a form of measurement
error, then the fact that simulations are used only in the regression
outcome means that there is little risk of attenuation bias as there
might be for predictor variables that use simulations.

\section{Additional Outcomes}\label{sec-app-addl}

\begin{figure}

\centering{

\pandocbounded{\includegraphics[keepaspectratio]{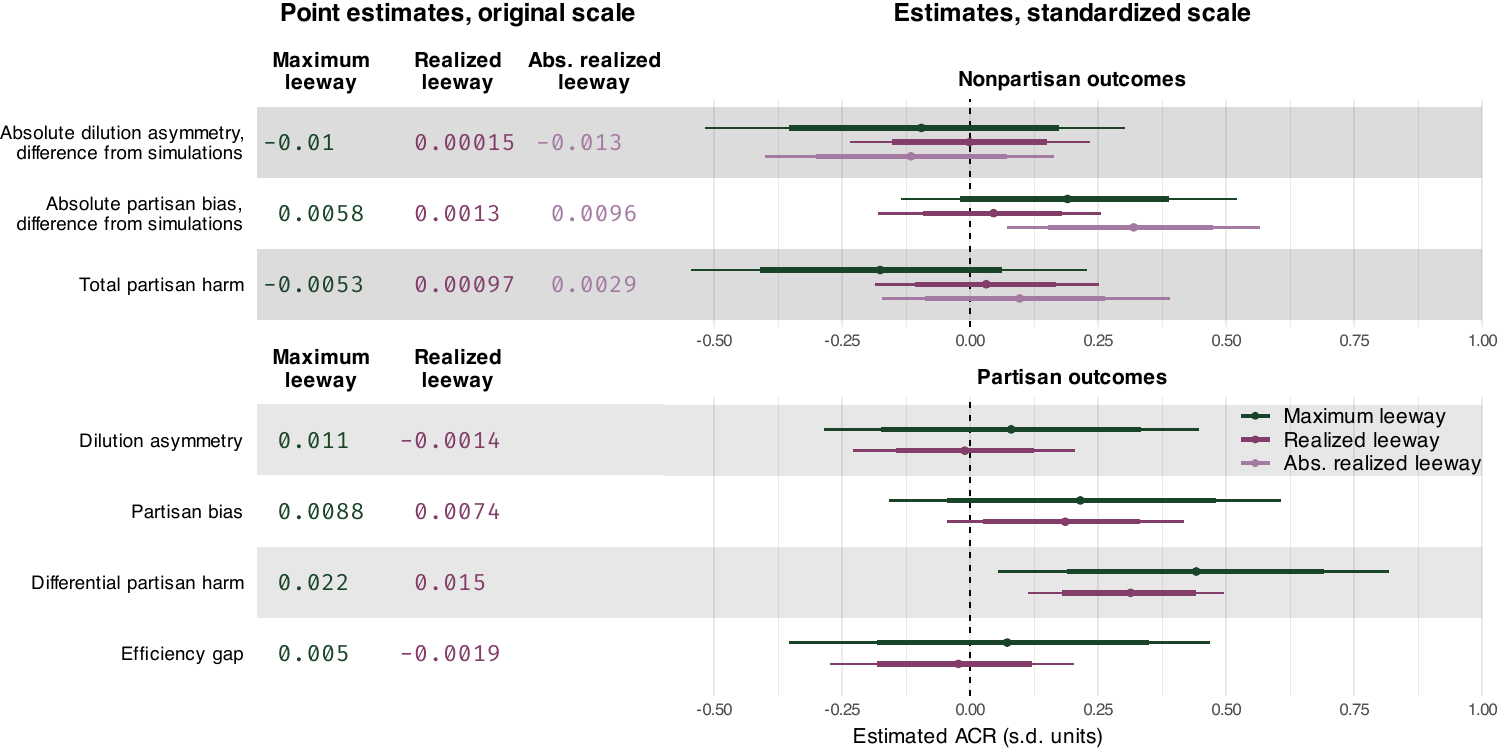}}

}

\caption{\label{fig-acr-app}Average causal response (ACR) of leeway on a
set of additional redistricting outcomes not considered in the main
text. The points correspond to the mean estimated ACR, while the lines
represent 95\% confidence intervals. Green corresponds to the
party-blind (worst case) treatment, while purple corresponds to the
party-signed treatment. The numbers in the columns display the mean ACR
on each outcome's response scale. The points and lines are displayed in
standard deviation units to allow for comparability between outcomes.
For partisan outcomes, a positive number indicates a pro-Republican
effect and a negative number indicates a pro-Democratic effect for a
positive dose.}

\end{figure}%

For completeness, we consider a series of alternative outcomes which
measure the partisan bias of redistricting plans. We start with the
efficiency gap, calculated as the difference between the wasted votes of
the two parties \citep{stephanopoulos2015}. Wasted votes are votes cast
for losing candidates and votes cast for winning candidates beyond the
threshold needed to win. The idea behind the efficiency gap is that a
fair plan should have similar numbers of wasted votes for both parties.
The partisan-signed version of the efficiency gap indicates which party
benefits from the plan in terms of vote efficiency; we use positive
values to indicate pro-Republican bias. Unlike with Republican seat
share, we do not subtract the average simulated value from the
partisan-signed efficiency gap measure. This is because the efficiency
gap is already designed to have a universal scale, where 0 indicates
partisan fairness; subtracting the simulated average would change the
definition of a fair baseline for this measure. To create a nonpartisan
measure, as with Republican seats, we take the absolute value of the
difference between the enacted and average simulated plan's efficiency
gap. Smaller values indicate a fairer plan by this measure (closer to
the normatively preferred value of 0).

Despite its popularity in the redistricting literature and litigation,
the efficiency gap faces a number of important limitations. For example,
the definition of the efficiency gap implies an ideal allocation of
evenly wasted votes between parties, which can be achieved through a
series of 75-to-25 districts. This ``three-to-one'' promotion has been
challenged as an unreasonable standard for fairness
\citep{bernstein2022measuring}. Further, because the efficiency gap
calculation is based in part on the number of seats, states with small
numbers of seats can have near discrete, ``lumpy'' distributions of the
efficiency gap \citep{nagle2017competitive, cho2017measuring}. Finally,
this seat-based measure means that relatively small changes in vote
share can cause large changes in the efficiency gap, when those changes
in vote share cause seat control to flip
\citep{stephanopoulos2018measure}.

We account for these limitations by considering a series of alternative
measures, including the dilution asymmetry \citep{gordon2024}. This
measures the difference in the percentage of each party's votes that are
wasted (cf.~the number of wasted votes used in the calculation of the
efficiency gap). Similar to the efficiency gap, we consider both the
absolute value of differences from simulations and signed raw values for
the dilution asymmetry.

A series of partisan bias measures are based on the seats-votes curve.
Under a fair plan, the seats-votes curve should be symmetric
\citep{katz2020}. We consider two groups of measures based on the
seats-votes curve. First, we consider the ``partisan bias'' which
measures the deviation from symmetry in the seats-votes curve. We use
both the raw signed value and absolute value of the difference from
simulations of the partisan bias.

Finally, we consider total and differential individual harm, which
measures how a redistricting plan negatively affects the ability for a
Democratic or Republican voter to elect a candidate of their choice,
relative to simulated redistricting plans
\citep{mccartan2022individual}. Differential partisan harm is a signed
measure where positive scores indicate Democrats are harmed at a higher
rate than Republican voters. The total partisan harm of a plan is a
nonpartisan measure, which represents the total harm to both parties.

\section{Estimation model details}\label{sec-app-prior}

The setup of the linear model used for the primary effect estimates is
described in the estimation section in the main text. In terms of the
prior specification, the intercept receives a weakly informative prior
of \(\Norm(0, (2.5\cdot\sigma_y)^2)\), where \(\sigma_y^2\) is the
variance of \(\Delta Y_t\). Each main effect coefficient receives a
weakly informative prior of
\(\Norm(0, (0.75\cdot \sigma_y/\sigma_x)^2)\), where \(\sigma_x^2\) is
the variance of each covariate \(x\). The prior standard deviation is
further scaled by 0.25 for interaction terms, following the discussion
in \citet{gelman2008weakly}. Finally, the residual standard deviation
receives an \(\Expo(1 / \sigma_y)\) prior. Computation is carried out
via NUTS-HMC as implemented in the Stan probabilistic programming
language \citep{stan}.

\section{Robustness Checks}\label{robustness-checks}

\subsection{Nonparametric outcome model}\label{sec-app-bart}

This appendix shows analogous results to Figure~\ref{fig-acr} and
Figure~\ref{fig-acr-app} but using a Bayesian Additive Regression Trees
(BART) model \citep{chipman2010bart}, rather than a linear model, for
outcome modeling. The results are qualitatively the same but the
estimated effect magnitudes are attenuated. There is no equivalent of a
coefficient estimate with BART as with linear models, so coefficient
estimates are not presented here.

\begin{figure}[t]

\centering{

\pandocbounded{\includegraphics[keepaspectratio]{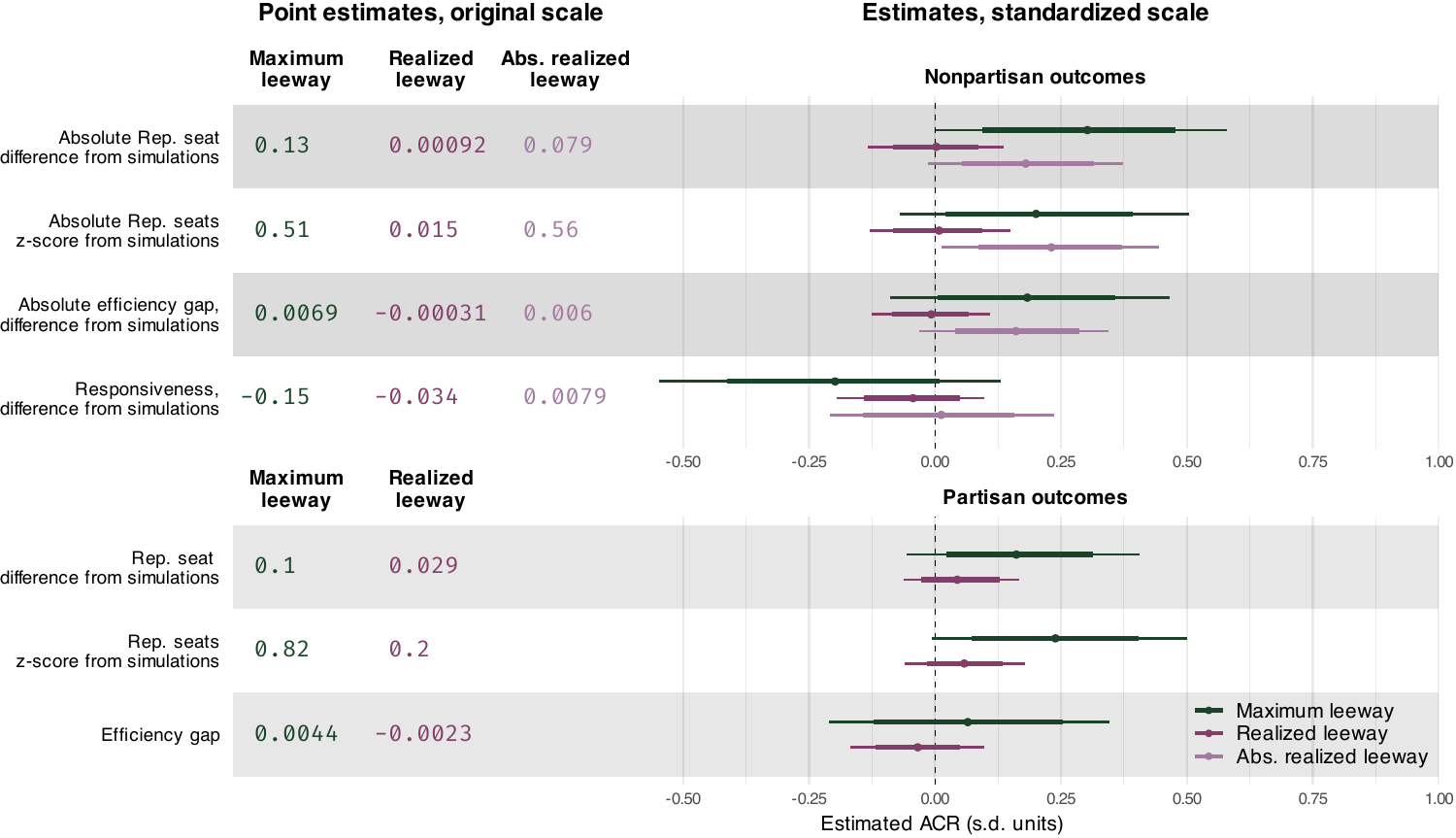}}

}

\caption{\label{fig-acr-bart}Version of Figure~\ref{fig-acr} using a
BART outcome model.}

\end{figure}%

\begin{figure}

\centering{

\pandocbounded{\includegraphics[keepaspectratio]{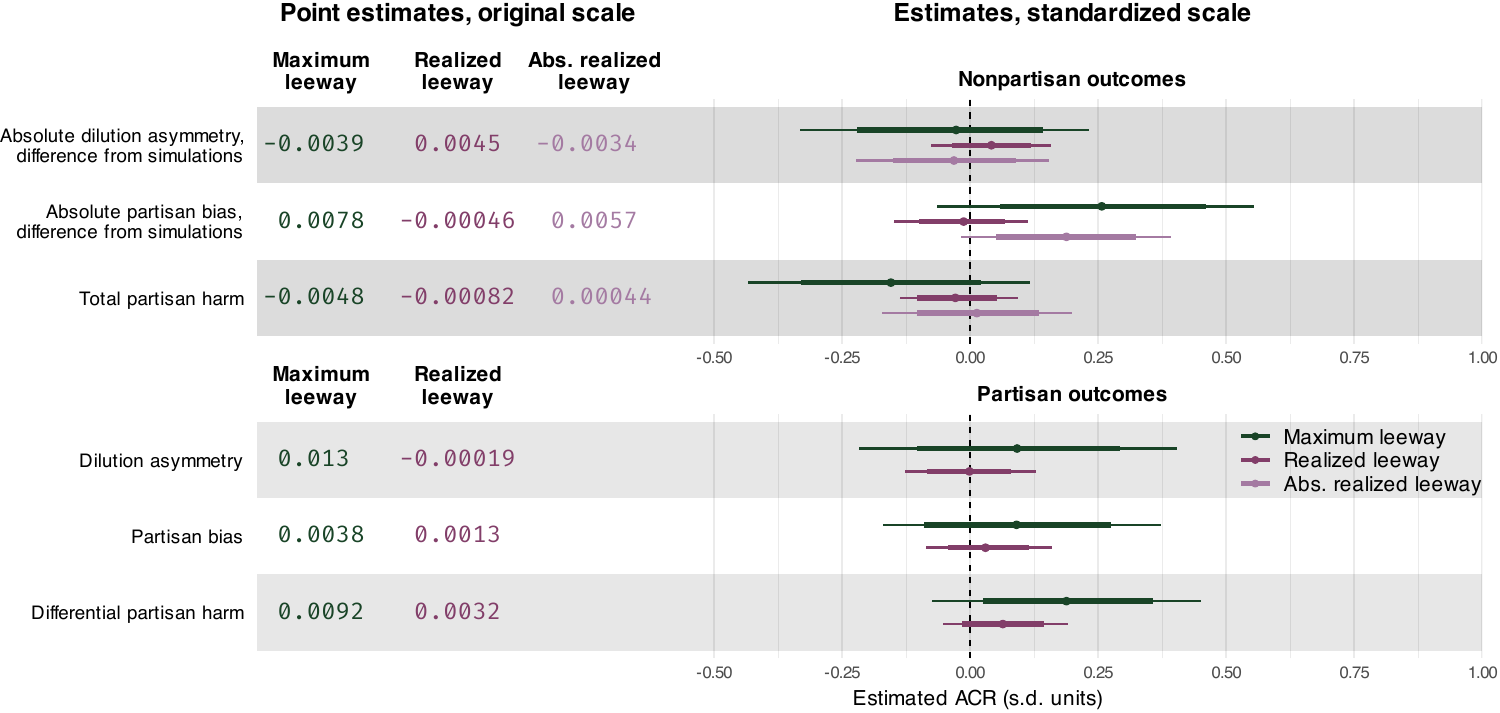}}

}

\caption{\label{fig-acr-bart-app}Version of Figure~\ref{fig-acr-app}
using a BART outcome model.}

\end{figure}%

\subsection{Including treatment components as
controls}\label{sec-app-robust}

To evaluate robustness to violations of the sufficiency assumption,
i.e., that \[
Y_{it}(\vb z)=Y_{it}(\vb z') \qq{for any} \vb z, \vb z' \text{ with } u^*(\vb z) = u^*(\vb z'),
\] we re-run the ACR estimates including \(\vb z\) as a covariate. If
the assumption were to hold, \(\vb z\) should have no effect on the
outcome, since \(u\) is controlled for. Of course, the data are limited
(\(n=43\)), and the dimension of \(\vb z\) is relatively large---the six
elements we use here must be included as levels and as differences, and
are each interacted with the change in \(u\). Thus we would likely
expect some changes due solely to random noise.

Some of the elements of \(\vb z\) do not change from 2010-2020 or
exhibit variation in only one or two states. We remove these elements,
and re-code several ``control'' variables as Democratic/Republican/Other
to further avoid exact collinearity. The final specification for
additional covariates included in ACR estimation is as follows:

\begin{itemize}
\tightlist
\item
  \texttt{drawer} in 2010
\item
  Indicator for \texttt{drawer\_ctrl} being \texttt{democrats} in 2010
\item
  Indicator for \texttt{drawer\_ctrl} being \texttt{republicans} in 2010
\item
  Indicator for \texttt{veto\_1} being \texttt{governor} in 2010
\item
  Indicator for \texttt{veto\_1\_ctrl} being \texttt{democrats} in 2010
\item
  Indicator for \texttt{veto\_1\_ctrl} being \texttt{republicans} in
  2010
\item
  \texttt{court\_review} in 2010
\item
  Indicator for \texttt{court\_ctrl} being \texttt{democrats} in 2010
\item
  Indicator for \texttt{court\_ctrl} being \texttt{republicans} in 2010
\item
  Indicator for any change in \texttt{drawer} from 2010 to 2020
\item
  Indicator for any change in \texttt{drawer\_ctrl} from 2010 to 2020
\item
  Indicator for any change in \texttt{veto\_1} from 2010 to 2020
\item
  Indicator for any change in \texttt{veto\_1\_ctrl} from 2010 to 2020
\item
  Indicator for any change in \texttt{court\_ctrl} from 2010 to 2020
\end{itemize}

\begin{figure}

\centering{

\pandocbounded{\includegraphics[keepaspectratio]{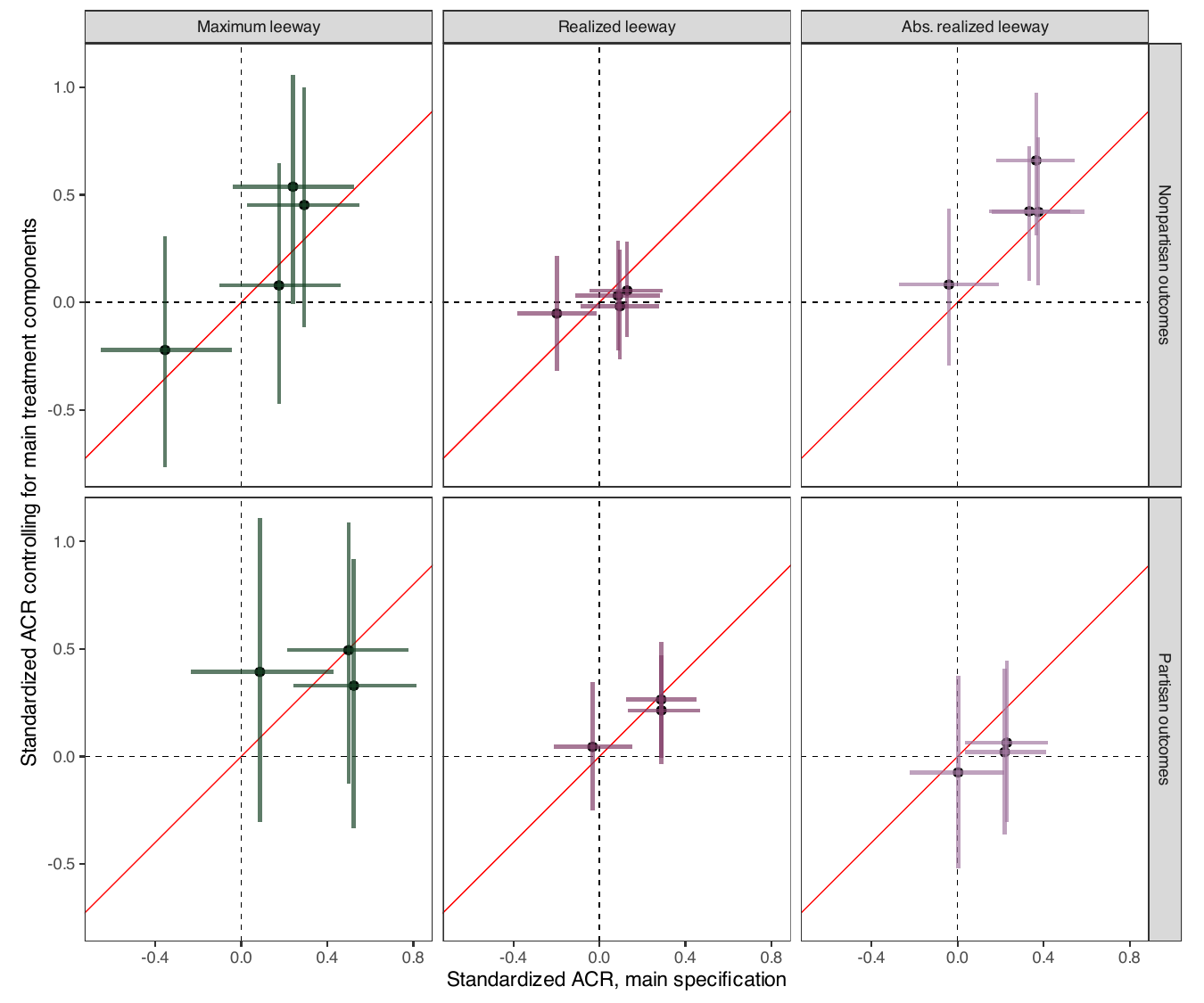}}

}

\caption{\label{fig-acr-robust-app}Comparison of estimated ACRs with
(y-axis) and without (x-axis) treatment components as controls. Point
estimates and 90\% credible intervals are shown for both specifications.
ACRs are faceted by type of outcome and the treatment specification
used.}

\end{figure}%

Figure~\ref{fig-acr-robust-app} displays the results of this robustness
check. ACRs estimated with the additional controls are plotted against
the original estimates along with 90\% credible intervals. While there
are some differences, there are no sign errors---estimates whose
credible intervals exclude 0 in one direction originally and in another
direction under the robustness check. In fact, none of the estimated
ACRs is statistically distinguishable across specifications. The
credible intervals for the alternative specification are wider, as would
be expected for adding a large number of additional controls. Altogether
these results give us confidence that, in addition to substantive
reasons for believing the sufficiency assumption holds, violations of
the assumption may not lead to statistically different conclusions.

\section{Descriptive analysis of changes}\label{sec-app-raw}

\begin{figure}

\centering{

\pandocbounded{\includegraphics[keepaspectratio]{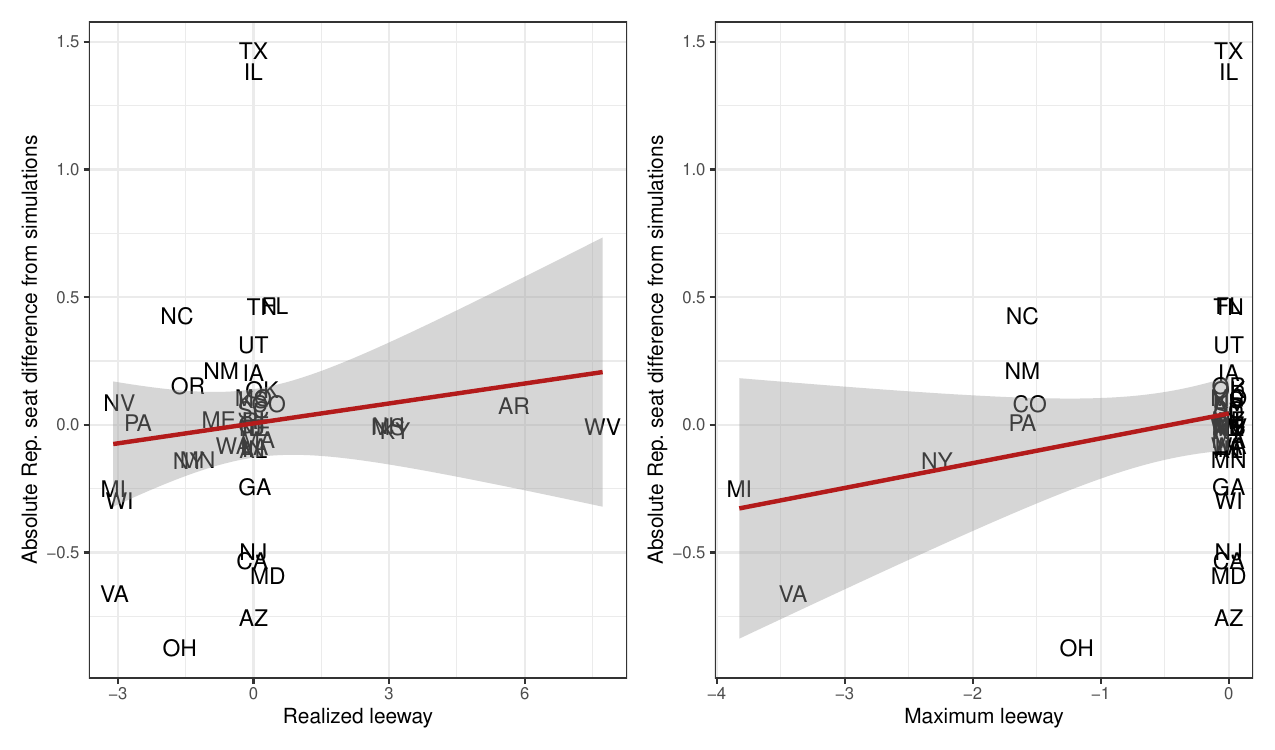}}

}

\caption{\label{fig-app-raw-abs-seat-diff}Changes in realized leeway
(left) and maximum leeway (right) against the change in the magnitude of
the Republican seats difference from simulations.}

\end{figure}

\begin{figure}

\centering{

\pandocbounded{\includegraphics[keepaspectratio]{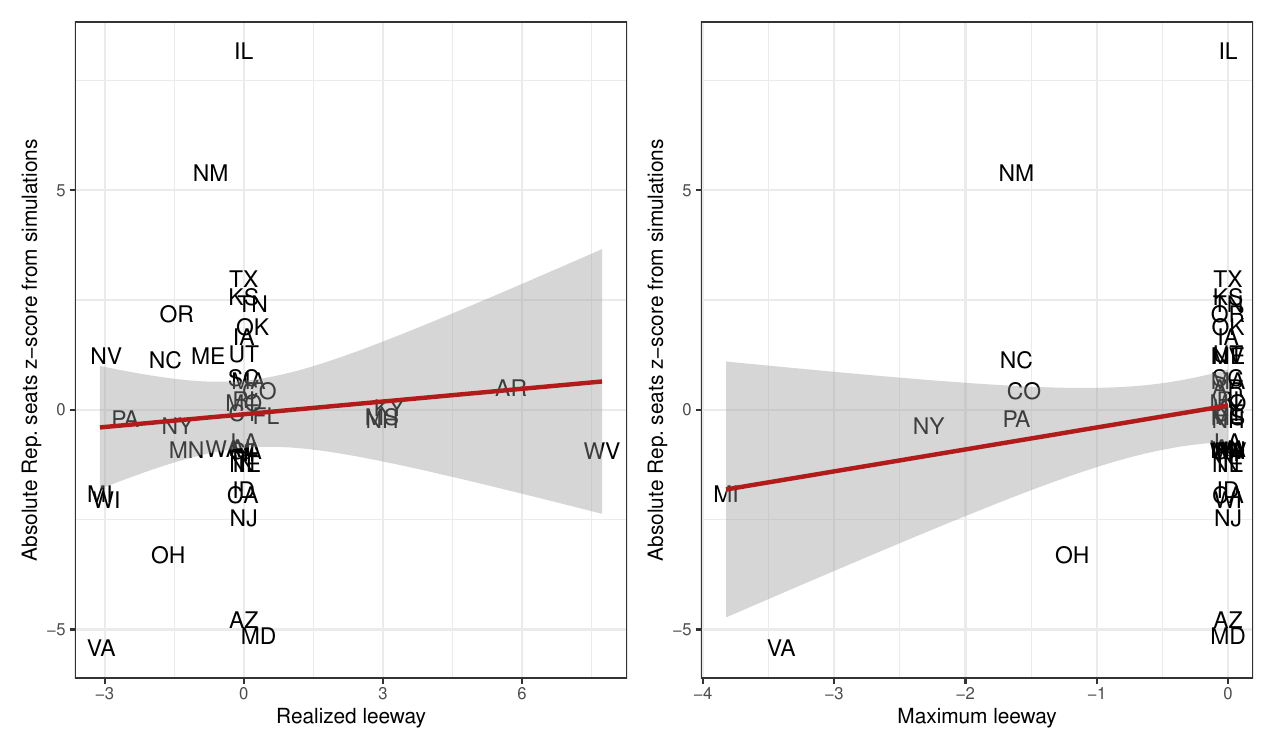}}

}

\caption{\label{fig-app-raw-abs-seat-zscore}Changes in realized leeway
(left) and maximum leeway (right) against the change in the magnitude of
the z-score of the Republican seats difference from simulations.}

\end{figure}

\begin{figure}

\centering{

\pandocbounded{\includegraphics[keepaspectratio]{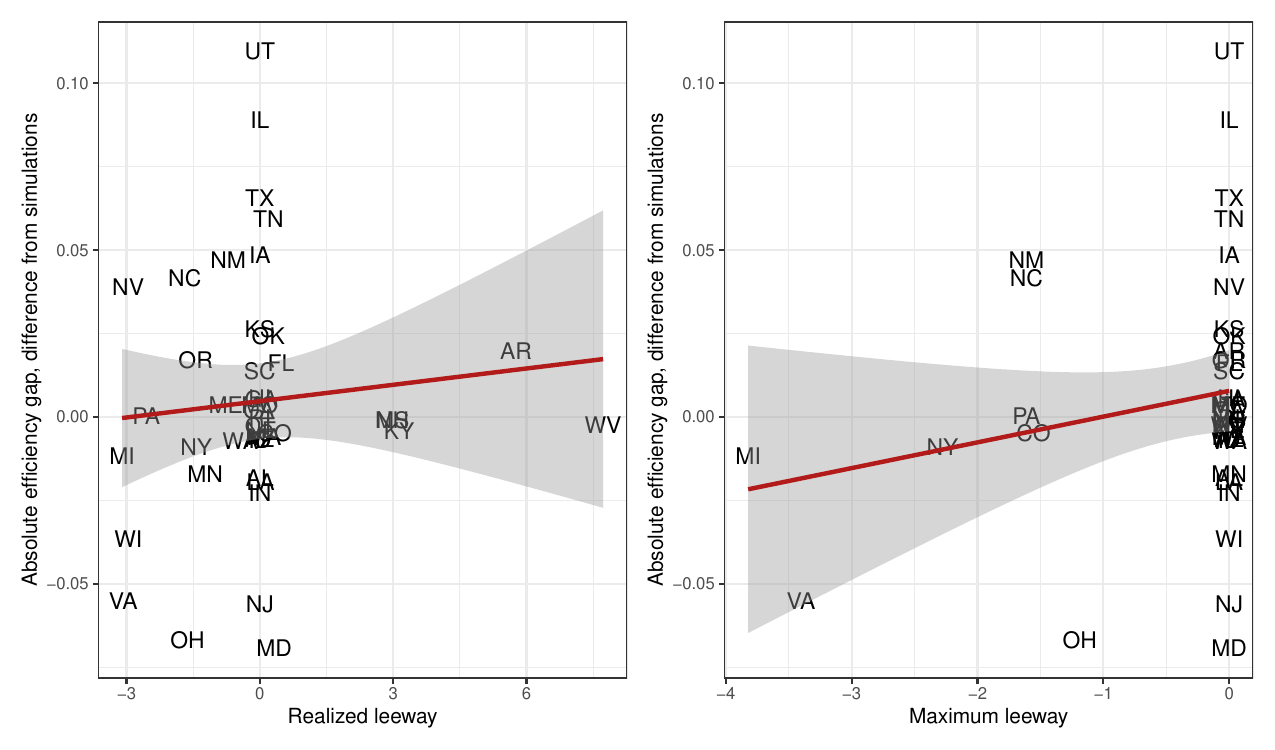}}

}

\caption{\label{fig-app-raw-abs-egap-diff}Changes in realized leeway
(left) and maximum leeway (right) against the change in the magnitude of
the efficiency gap difference from simulations.}

\end{figure}

\begin{figure}

\centering{

\pandocbounded{\includegraphics[keepaspectratio]{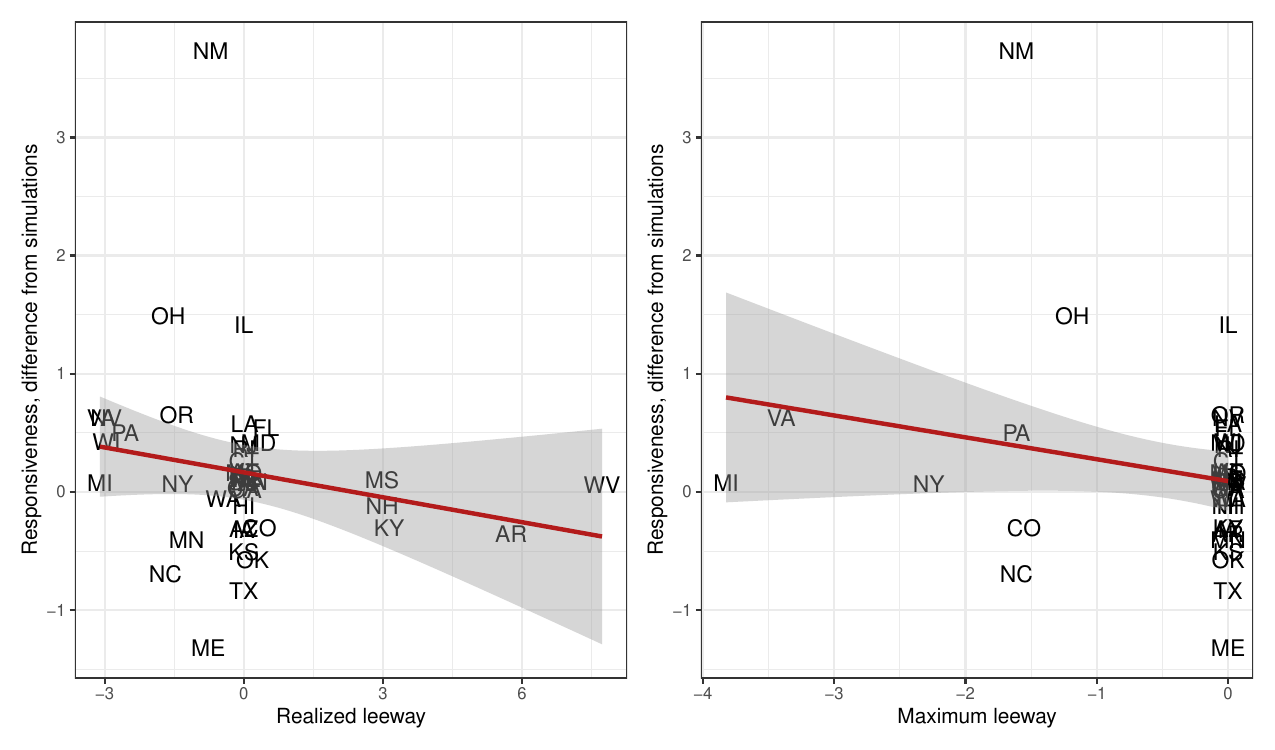}}

}

\caption{\label{fig-app-raw-resp-diff}Changes in realized leeway (left)
and maximum leeway (right) against the change in the magnitude of the
responsiveness difference from simulations.}

\end{figure}

\begin{figure}

\centering{

\pandocbounded{\includegraphics[keepaspectratio]{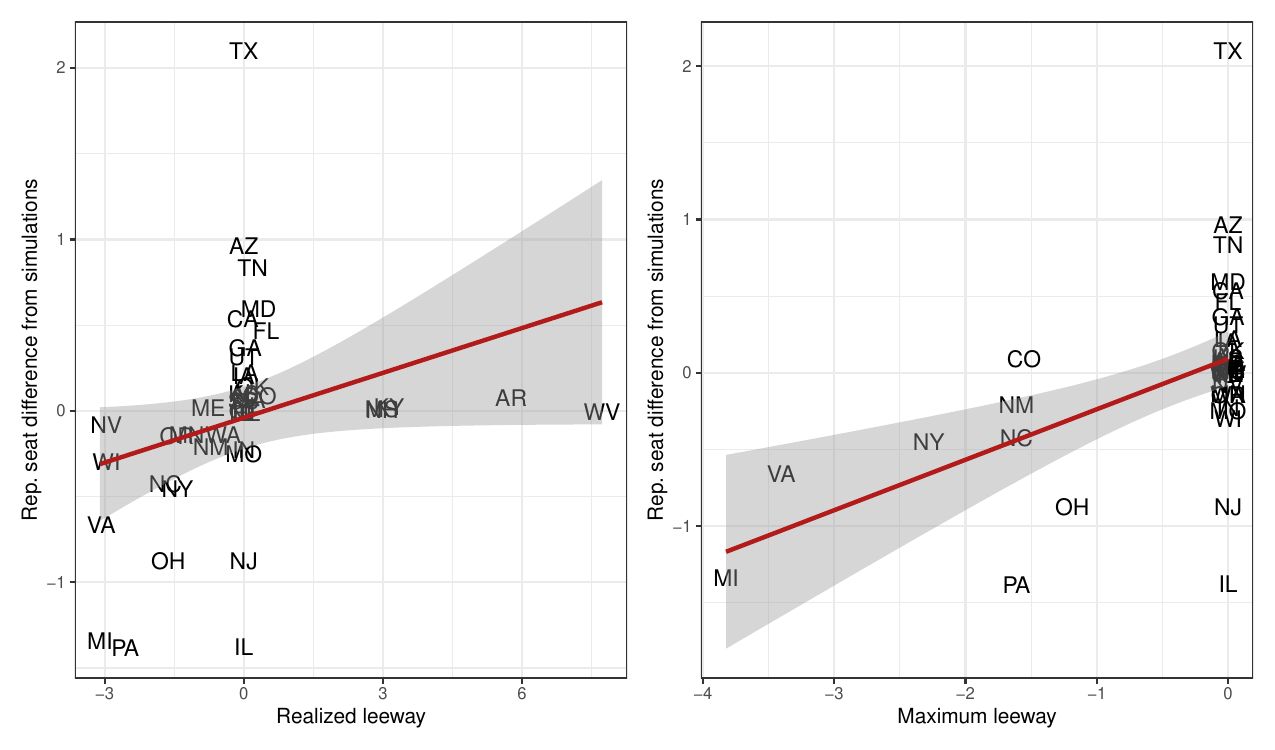}}

}

\caption{\label{fig-app-raw-seat-diff}Changes in realized leeway (left)
and maximum leeway (right) against the change in the the Republican
seats difference from simulations.}

\end{figure}

\begin{figure}

\centering{

\pandocbounded{\includegraphics[keepaspectratio]{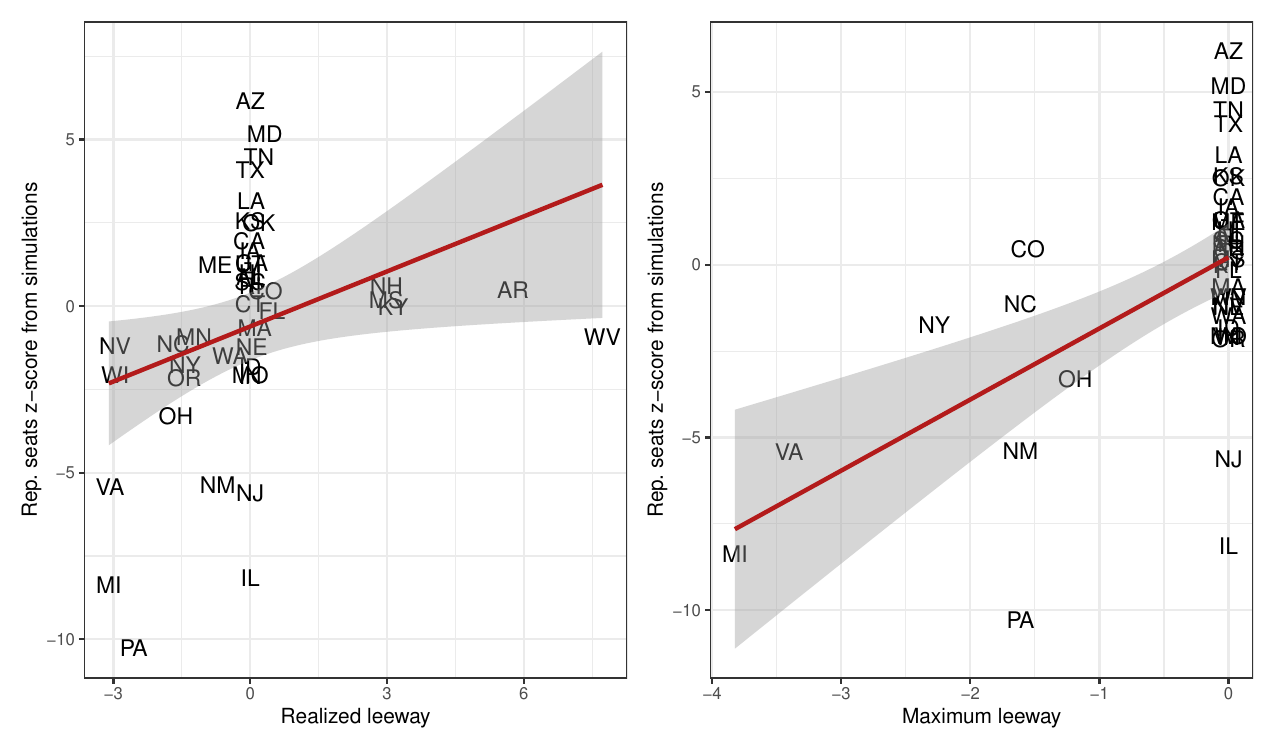}}

}

\caption{\label{fig-app-raw-seat-zscore}Changes in realized leeway
(left) and maximum leeway (right) against the change in the z-score of
the Republican seats difference from simulations.}

\end{figure}

For completeness, we conduct a descriptive analysis that examines how
the raw dosages in our leeway variables correlate with the raw changes
in the outcomes used in our main analysis.
Figure~\ref{fig-app-raw-abs-seat-diff} to
Figure~\ref{fig-app-raw-seat-zscore} plot the change in the realized
leeway (left plot) and maximum leeway (right plot) against the change in
a different outcome variable. The solid line indicates the linear model
fit while the shaded area represents the 95\% pointwise confidence
interval. Many states had no changes in their redistricting processes,
and thus no change in leeway, which is indicated by the stacking of
states at 0 for both leeway measures. While this analysis does not
adjust for observed confounding variables as we do in the main text, the
linear trends observed in these figures are qualitatively consistent
with our main results based on ACRs in Figure~\ref{fig-acr}.

\section{Robustness to reduced court intervention}\label{sec-app-court}

This paper focuses on the plans used in the first election after
redistricting, whereas litigation can change the enacted plan well after
this point. As such, plans which are labelled as having been drawn by
legislatures and, less frequently, commissions, may eventually be
redrawn by courts. Peeking at such eventualities in designing our game
theoretic model can introduce post-treatment bias, so we avoid this.
Notably, states in the South are susceptible to VRA litigation that can
take more than the 14 months between Census data release and November
general elections. They may see more restriction in the model than is
empirically true in that time period.

Below, we demonstrate that tuning parameters for solving the game do not
meaningfully impact the calculated Nash equilibria. To produce this, we
lower the model parameters across the board. We largely decrease the
likelihood of VRA litigation changing the final plan and modestly
decrease the probability that other litigation changes the final plan.

\begin{figure}

\centering{

\pandocbounded{\includegraphics[keepaspectratio]{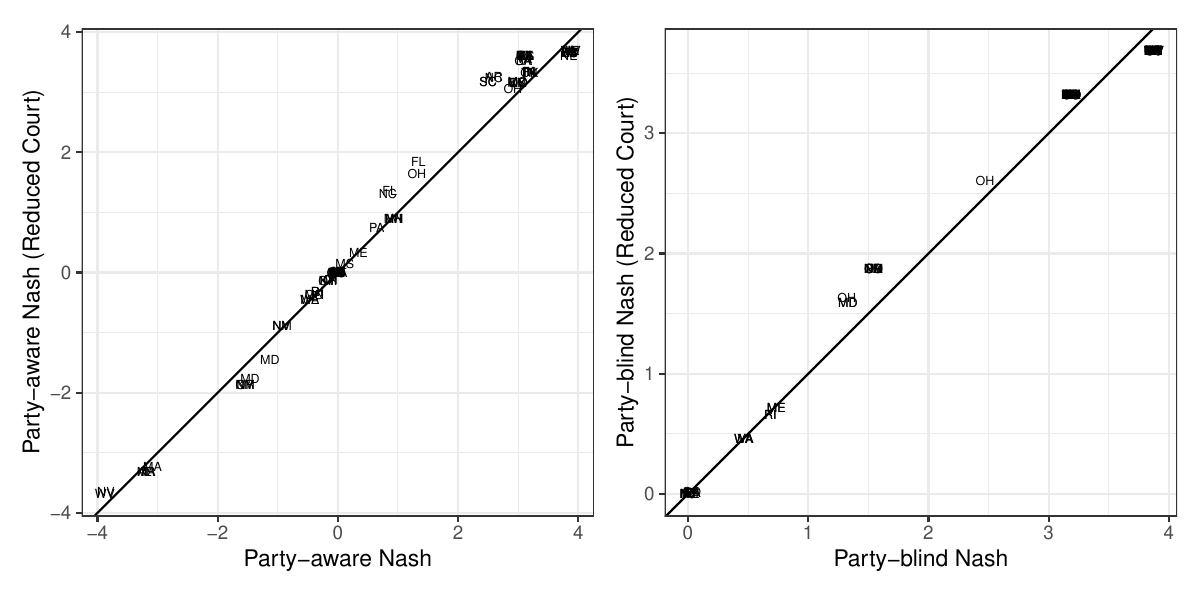}}

}

\caption{\label{fig-app-reduce-court}Changes in Nash equilibria after
decreasing the probability of court intervention}

\end{figure}%

As Figure~\ref{fig-app-reduce-court} demonstrates, greatly reducing the
probability of court intervention does not meaningfully impact the Nash
equilibria. While values do shift by small amounts, these are very small
(on the order of 0.1) compared to the total range of 0--4 or -4--4.

\section{State-level estimates of counterfactual policy
analysis}\label{state-level-estimates-of-counterfactual-policy-analysis}

\begin{figure}

\centering{

\pandocbounded{\includegraphics[keepaspectratio]{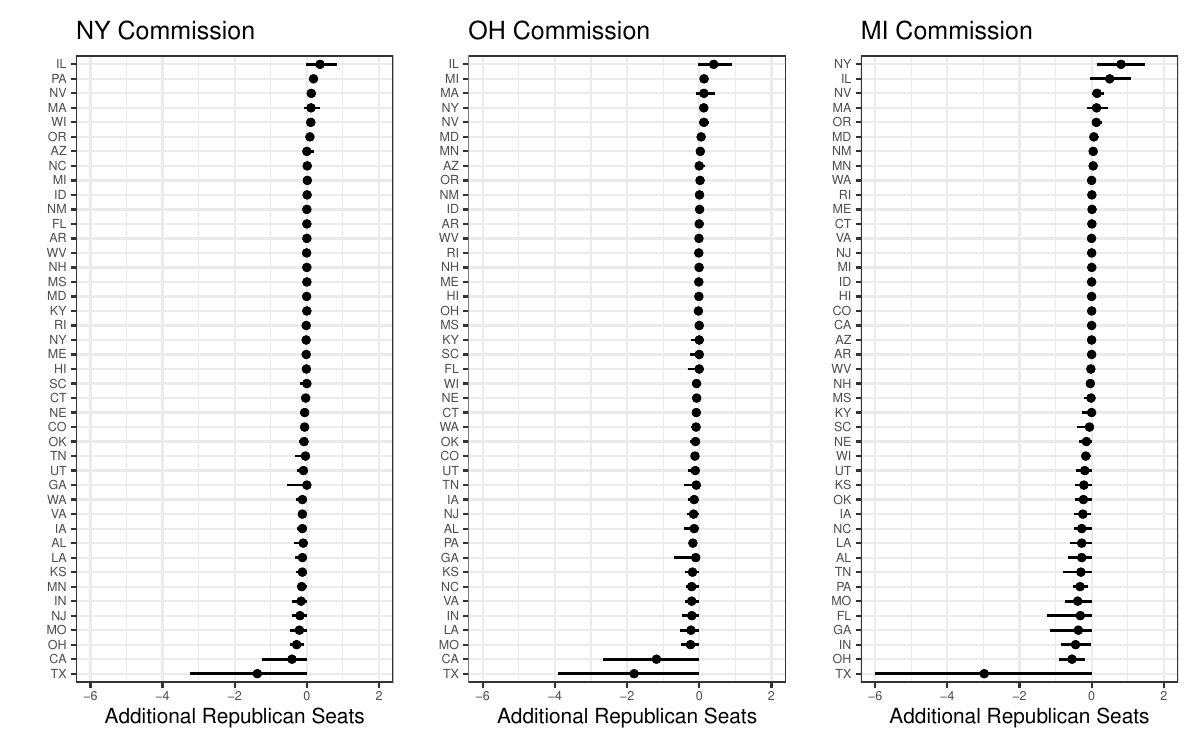}}

}

\caption{\label{fig-reform-by-state}\textbf{State-level Estimates of
Counterfactual Policy Analysis} The figures show the predicted number of
seats that would change under three redistricting institution reforms:
(1) a New York-style commission with a nonpartisan map drawer and
several partisan veto points; (2) an Ohio-style legislature-drawn map
and several partisan and bipartisan veto points; and (3) a
Michigan-style reform, with a nonpartisan commission, no partisan veto
points, and the potential for court review.}

\end{figure}

Figure~\ref{fig-reform-by-state} present the state-specific estimates of
the counterfactual policy analysis whereas Figure~\ref{fig-reform} shows
the national-level estimate that aggregates these state-level estimates.
Each bar represents the point estimate for the increase in the number of
Republican seats. Like the aggregate estimates, state-specific estimates
should also be interpreted with caution, as there may be unmodelled
heterogeneity.

\FloatBarrier

\section{Coefficient estimates for all models}\label{sec-app-coef}

Table~\ref{tbl-coefs} below contains point estimates and standard errors
for all estimated model coefficients. Each set of rows marked by a
single value in the ``Model'' column represents a separately estimated
model. ``D.2'' in the ``Spec.'' (specification) column indicates the
robustness check specification used in the corresponding appendix.

\tiny

\begin{longtable}{lllll>{}l>{}r>{}r}

\caption{\label{tbl-coefs}All model coefficients.}

\tabularnewline

\toprule
Model & Spec. & Leeway & Outcome type & Outcome & Coefficient & Estimate & Std. err.\\
\midrule
\endfirsthead
\multicolumn{8}{@{}l}{\textit{(continued)}}\\
\toprule
Model & Spec. & Leeway & Outcome type & Outcome & Coefficient & Estimate & Std. err.\\
\midrule
\endhead

\endfoot
\bottomrule
\endlastfoot
1a & Main & Maximum & Nonpartisan & Abs. Rep. seat diff. from sims. & \ttfamily{dose\_worst} & \ttfamily{0.09221} & \ttfamily{0.16980}\\
1a & Main & Maximum & Nonpartisan & Abs. Rep. seat diff. from sims. & \ttfamily{nash\_worst\_10} & \ttfamily{0.12040} & \ttfamily{0.03761}\\
1a & Main & Maximum & Nonpartisan & Abs. Rep. seat diff. from sims. & \ttfamily{dvs\_08} & \ttfamily{-0.05920} & \ttfamily{0.30000}\\
1a & Main & Maximum & Nonpartisan & Abs. Rep. seat diff. from sims. & \ttfamily{is\_south} & \ttfamily{0.19200} & \ttfamily{0.14240}\\
1a & Main & Maximum & Nonpartisan & Abs. Rep. seat diff. from sims. & \ttfamily{log(n\_distr\_20)} & \ttfamily{0.01992} & \ttfamily{0.13900}\\
\addlinespace
1a & Main & Maximum & Nonpartisan & Abs. Rep. seat diff. from sims. & \ttfamily{distr\_chg} & \ttfamily{0.57970} & \ttfamily{0.13120}\\
1a & Main & Maximum & Nonpartisan & Abs. Rep. seat diff. from sims. & \ttfamily{log(1 + corrupt)} & \ttfamily{-0.15150} & \ttfamily{0.10380}\\
1a & Main & Maximum & Nonpartisan & Abs. Rep. seat diff. from sims. & \ttfamily{initiative} & \ttfamily{0.14910} & \ttfamily{0.10740}\\
\end{longtable}

\normalsize


\end{document}